\documentclass{aastex63}       \newif\ifdraft \drafttrue
\ifdraft \else \newcommand{\submitjournal}[1]{\relax} 
               \newcommand{\correspondingauthor}[1]{\relax} 
\fi
\submitjournal{AJ}
\usepackage{graphicx}
\usepackage{xcolor}
\usepackage{upgreek}
\usepackage{amsmath}	
\ifdraft
   \usepackage{hyperref}
   \hypersetup{
               colorlinks=true,
               linkcolor=black,
               filecolor=black,
               citecolor=black,
               urlcolor=blue
              }
     \newcommand{\web}[1]{\Blb{\url{#1}}}
\else
     \newcommand{\web}[1]{#1}
\fi

\newcommand{\PIMA}{$\cal P\hspace{-0.067em}I\hspace{-0.067em}M\hspace{-0.067em}A$ }

\newcommand{\ntab}[2]{ \multicolumn{1}{#1}{#2} }
\newcommand{\nntab}[2]{ \multicolumn{2}{#1}{#2} }
\newcommand{\nnntab}[2]{ \multicolumn{3}{#1}{#2} }
\newcommand{\nnnntab}[2]{ \multicolumn{4}{#1}{#2} }
\newcommand{\nnnnnntab}[2]{ \multicolumn{6}{#1}{#2} }

\definecolor{Dred}{rgb}{0.312,0.070,0.070}
\definecolor{Dblue}{rgb}{0.070,0.070,0.312}
\definecolor{Dgreen}{rgb}{0.070,0.312,0.070}
\definecolor{Db}{rgb}    {0.050,0.0,0.320}

\newcommand{\Blb}[1]{\textcolor{Dblue}{\bf #1}}

\newcommand{\Gaia}{{\it Gaia}}
\newcommand{\beq}{ \begin{eqnarray} }
\newcommand{\eeq}[1]{\label{#1}\end{eqnarray}}
\newcommand{\eeqn}{ \nonumber \end{eqnarray} }
\newcommand{\Frac}[2]{\frac{\displaystyle\strut #1}{\displaystyle\strut #2} }
\newcommand{\vex}{\vspace{1ex}}
\newcommand{\dss}{\displaystyle}
\renewcommand{\tau}{\uptau}
\renewcommand{\vec}[1]{{\mathbf #1}}

\newcommand{\vce}[1]{\underline{\vec{#1}}}
\newcommand{\lp}{ \left(  }
\newcommand{\rp}{ \right) }

\newcommand{\mat}[1]{\widehat {\mathstrut \cal #1}}

\newcommand{\hp}{\phantom{+}}
\newcommand{\hm}{\phantom{-}}

\newcounter{note}
\let\oldmarginpar\marginpar
\renewcommand\marginpar[1]{\-\oldmarginpar[\raggedleft\footnotesize #1]%
{\raggedright\footnotesize #1}}
\newcommand{\Note}[1]{\Rdb{#1}{\addtocounter{note}{1}%
\marginpar{\small\underline{\Rdb{\footnotesize Corr \arabic{note}}}}}}
\newcommand{\note}[1]{\Rdb{#1}}
\renewcommand{\Note}[1]{#1}
\renewcommand{\note}[1]{#1}

\received{August 20, 2020}
\revised{October 05, 2020}
\accepted{October 19, 2020}

\begin{document}

\title{The wide-field VLBA calibrator survey --- WFCS}
\ifdraft \relax \else \correspondingauthor{Leonid Petrov} \fi
\email{Leonid.Petrov@nasa.gov}

\author[0000-0001-9737-9667]{Leonid Petrov}
\affil{NASA Goddard Space Flight Center \\
Code 61A1, 8800 Greenbelt Rd, Greenbelt, 20771 MD, USA}

\begin{abstract}

  The paper presents the results of the largest to date VLBI absolute 
astrometry campaign of observations of 13,645 radio sources with 
the Very Long Baseline Array (VLBA). Of them, 7220 have been detected, 
including 6755 target sources that have never been observed with VLBI 
before. This makes the present VLBI catalogue the largest ever published. 
Positions of the target sources have been determined with the median 
uncertainty 1.7~mas, and \note{15,542 images of 7171} sources have been generated. 
Unlike to previous absolute radio astrometry campaigns, observations were made
at 4.3 and 7.6~GHz simultaneously using a single wide-band receiver. 
Because of the fine spectral and time resolutions, the field of view was 
4 to 8 arcminutes --- much greater than 10 to 20 arcseconds in previous 
surveys. This made possible to use input catalogues with low position 
accuracy and detect a compact component in extended sources. Unlike to 
previous absolute astrometry campaigns, both steep and flat spectrum 
sources were observed. The observations were scheduled in the so-called 
\note{filler} mode to fill the gaps between other high priority programs. 
That was achieved by development of the totally automatic scheduling 
procedure. 

\end{abstract}


\section{Introduction}

   The method of very long baseline interferometry (VLBI) first proposed
by \citet{r:mat65} provides very high angular resolution. It was quickly
realized that VLBI is a powerful tool for geodesy and astronomy. The first 
catalogue of source coordinates determined with VLBI contained 35~objects 
\citep{r:first-cat}. Since then, continuous efforts were put in the
extension of the source list and improvement of accuracy. Absolute 
astrometry and geodesy programs in the 20th century at 8.6 and 2.3~GHz 
(X and S bands) using the Mark3 recording system resulted in the ICRF1 
catalogue of 608~sources \citep{r:icrf1}. Later, thousands sources 
were observed with the Very Long Baseline Array (VLBA), the Long Baseline 
Array (LBA) in the southern hemisphere, and the Chinese VLBI Network (CVN)
in a number of dedicated absolute astrometry programs: the VLBI Calibrator 
Survey (VCS) \citep{r:vcs1,r:vcs2,r:vcs3,r:vcs4,r:vcs5,r:vcs6,r:vcs-ii}, 
the VLBA Imaging and Polarimetry Survey (VIPS) \citep{r:vips,r:astro_vips}, 
the VLBA Galactic plane Survey (VGaPS) \citep{r:vgaps}, the Long Baseline 
Array Calibrator Survey (LCS) \citep{r:lcs1,r:lcs2}, the Ecliptic Plane 
Survey \citep{r:shu17}, the VLBA regular geodesy RDV program \citep{r:rdv}, 
and several other programs \citep{r:obrs1,r:egaps,r:obrs2,r:bessel,r:npcs}.

  The goal of these programs was to build a catalogue of positions of 
the most compact components in active galactic nuclea (AGNs) with a nanoradian 
level of accuracy (1 \Note{nrad} $\approx$ 0.2~mas). Such a catalogue is used for 
for imaging of weak sources with phase-referencing, as targets for geodetic 
VLBI observations, for space navigation, for associations of $\gamma$-ray 
sources \citep{r:aofus1,r:aofus2,r:aofus3}, and for fundamental 
physics \citep{r:gamma_lambert}. Until recently, the method of VLBI was the 
most accurate. \Gaia\ Data release~2 \citep{r:gaia_dr2} demonstrated the 
accuracy in par or better than VLBI. However, a detailed analysis of the 
differences \citep{r:gaia1,r:gaia4} showed that a fraction of matching 
sources has statistically significant position offsets along the AGN 
jet directions. \citet{r:gaia1,r:gaia2,r:gaia5} 
presented convincing argumentation in support of a claim that such offsets 
are not due to errors in VLBI or optical \Gaia\ catalogue, but is 
a manifestation of mas-scale optical jets that shift the centroid position. 
As a result, \citet{r:gaia3} concluded that high accuracy of optical 
catalogues cannot be transferred to the radio range beyond the 1--10~mas level, 
and positions derived from analysis of dedicated VLBI observations are 
necessary for applications that require higher position accuracy.

  Despite the total number of compact radio sources with positions derived 
from VLBI observations surpassed 7000 by January 2013, the density of 
calibrator sources was not high enough to ensure there is a good calibrator
within 2--$3^\circ$ of any direction. Therefore, a program for densification
of the grid of compact radio sources with precisely determined coordinates was
proposed. The goals of the program were

\begin{itemize}\setlength{\itemsep}{0.2ex}
   \item To increase the density of calibrator sources in the areas at 
         $\delta > -40^\circ$ where their density was lower than on average.
         In particular, to have at least one calibrator within any field 
         of view of PanSTARRS \citep{r:ps_dr1} (disk of $1.5^\circ$ radius) or 
         LSST (disk of $1.75^\circ$ radius). The program was formulated and 
         observed before the \Gaia\ data release. It was not known that time 
         how useful \Gaia\ astrometry of faint sources of 15--20 magnitude 
         can be. In the absence of \Gaia\ astrometry, the presence of several 
         objects with positions known with the accuracy better 1~mas could be 
         used as calibrators and boost the accuracy of PanSTARRS source 
         position catalogues.

   \item To reach the completeness over 95\% level at 150~mJy level to perform 
         a study of the population of compact radio sources.

   \item To study the population of steep spectrum sources. The population of 
         steep spectrum sources is poorly studied due to a heavy selection 
         bias in prior surveys towards the so-called flat spectrum 
         sources, i.e. the sources with spectral index $\alpha > -0.5$
         defined as $S \sim f^{+\alpha}, $\Note{where $S$ is flux density and
         $f$ is frequency}.

   \item To re-observe the sources with large radio-to-optical position offsets.
\end{itemize}

  We consider a source as a calibrator if it can be detected with the signal to
noise ratio (SNR) $>$ 10 at baselines longer 5000~km for 30~s integration. 
The SNR is defined as the ratio of the fringe amplitude to the mean amplitude
of the noise. Sources brighter than 15--20~mJy satisfy this criteria, 
provided they are observed at 4~Gbps recording mode at the network with the 
sensitivity similar to the VLBA, i.e. with the system \Note{effective} flux 
density (SEFD) in a 250--400~Jy range.

   The objectives of the program were 1)~to determine coordinates of target 
sources with a milliarcsecond level of accuracy and 2)~to synthesize images of 
all detected sources. The catalogue was available on-line \Note{since April 19,
2013} even before the campaign was completed. Since the campaign used a number 
of novel techniques, there is a necessity to describe them in depth. 
The technology of VLBI surveys is the main focus of this article. The design 
of the campaign, results of the pilot programs, the source selection, and the 
scheduling algorithm are discussed in section~2. The post-correlator, 
astrometric, and imaging data analyses are described in section~3 and 4. 
The catalogue is presented in section~5 followed by the summary.

\section{Observations}

\subsection{Selection of frequencies}

  The objectives of the program determine a choice of a receiver and a 
recording mode. To get a high position accuracy, observations should be done 
at two bands simultaneously. The combination of group delay observables 
at upper and lower frequencies 
\beq
  \tau_{\rm if} = \frac{f^2_u}{f^2_u -f^2_l} \tau_u \: - \: \frac{f^2_l}{f^2_u -f^2_l} \tau_l
\eeq{e:e1}
  is not affected by the ionosphere, since the ionospheric path delay
is reciprocal to the square of frequency. Here $\tau_u$ and $\tau_l$ are 
group delays at the upper and lower frequency respectively, $f_u$ and
$f_l$ are the effective frequencies that are close to the central frequencies 
of recorded bands. In 2013--2016, when the experiments were observed, the VLBA 
supported two options of dual-band single polarization observations at the 
2~Gbps recording rate: 1)~simultaneous 2.2--2.4~GHz (S-band) and 8.4--8.9~GHz 
(X-band) observations using a dichroic plate that sends the signal to two 
receivers and 2)~recording at two remote wings of the broadband 
4--8~GHz receiver. 

  In 2013 when the program started, there were no technical capability
to record the entire band. The band is split into 16 sub-bands, 32~MHz wide,
hereafter called intermediate frequencies (IF). The hardware imposes certain
restrictions on frequency selection. In particular, the sub-bands should be 
assigned to two groups, and each group should have the spanned bandwidth 
not exceeding 480~MHz. The placement of IFs within sub-groups affects the 
accuracy of group delay computation and the probability of picking up 
a secondary maximum during the fringe fitting processes. The frequency setup 
used in this campaign is presented in Table~\ref{t:fs}. The highest secondary 
maximum of the delay resolution function for this sequence is 0.678 at 2.7~ns 
and the uncertainty of group delay is 90.9~ps at the SNR=10 .

  I ran several pilot projects. In the first one I examined the VLBA 
performance at different frequency setups using the 4--8~GHz receiver
and in the second I examined the differences in the ionosphere total electron
contents (TEC) derived from quasi-simultaneous 4.3/7.6~GHz and 2.2/8.4~GHz 
observations. Test observations have confirmed no noticeable degradation of 
sensitivity with respect to the more frequently used 4.9--6.6~GHz part 
of the band.

  The use of dual-band observations increases the uncertainty of the 
ionosphere-free combination of lower and upper band observables:
\beq
  \iftwocolstyle\hspace{-2em}\fi 
  \sigma(\tau_{\rm if}) = \sqrt{ \frac{f^4_u}{(f^2_u -f^2_l)^2} \tau_u^2 \: + \: 
                                 \frac{f^4_l}{(f^2_u -f^2_l)^2} \tau_l^2}.
\eeq{e:e2}
  The wider frequency separation, the lower an increase in the uncertainty with 
respect to a single-band observation at the upper band. Therefore, at a given SNR
the group delay uncertainty from the data collected with the broadband C-band 
receiver is worse than the group delay uncertainty from the data collected with 
the S/X band receiver. However, the sensitivity of the C-band receiver is higher 
than the sensitivity of the S/X receiver. According to the National Radio Astronomy 
Observatory (NRAO) gain monitoring 
program\footnote{\web{https://science.nrao.edu/facilities/vla/docs/manuals/obsguide/calibration}}, 
the SEFD in the zenith direction of {\sc pietown} antenna in 2015 was 190, 250 
and 280~Jy at 4.9, 6.6, and 8.4~GHz respectively. If to take this into account, 
the group delay uncertainty from the data collected with the broadband C-band is 
worse than from the S/X receiver by a factor of 1.26 for flat spectrum sources
and by a factor 1.22 for sources with the spectral index of -0.5 that is typical for 
the program sources. However, the sensitivity at 4.3~GHz is a factor of 1.47 
better than at 8.4~GHz for a flat spectrum source and is a factor of 2.06 better 
for a source with spectral index -0.5. If a source is weak, it may not be 
detected at all if the antenna sensitivity is not high enough.

  Another factor that affects the choice of used frequencies is the presence of 
radio interference (RFI). The RFI is the worst at 2.2--2.4~GHz. It reduces the
usable band to less than 140~MHz, requires considerable efforts to edit the 
data, and poses the risk of losing some observations. At the same time, no 
serious RFI were reported within 4--8~GHz in 2013, except of the presence of
narrow-band signals (bandwidth less than 100~kHz) at 4.2 and 7.8~GHz frequencies 
that are due to a leakage from synthesizers. 

  I consider an improvement of the detection limit by a factor of 1.5--2.1 and 
the RFI alleviation is more important than a 22--26\% improvement in source 
position accuracy for strong objects.
\begin{table}
   \caption{The frequency sequence of the low edges of 32~MHz wide IFs 
            used in the campaign.
           }
   \begin{center}
      \begin{tabular}{cc}
         \hline
         Low band & Upper band \\
          GHz  &  GHz  \\
         \hline
         4.128 & 7.392 \\
         4.160 & 7.424 \\
         4.192 & 7.456 \\
         4.224 & 7.552 \\
         4.416 & 7.744 \\
         4.512 & 7.776 \\
         4.544 & 7.808 \\
         4.576 & 7.840 \\
         \hline
      \end{tabular}
   \end{center}
   \label{t:fs}
\end{table}

\subsection{The field of view of the survey}

\begin{figure}
  \includegraphics[width=0.50\textwidth]{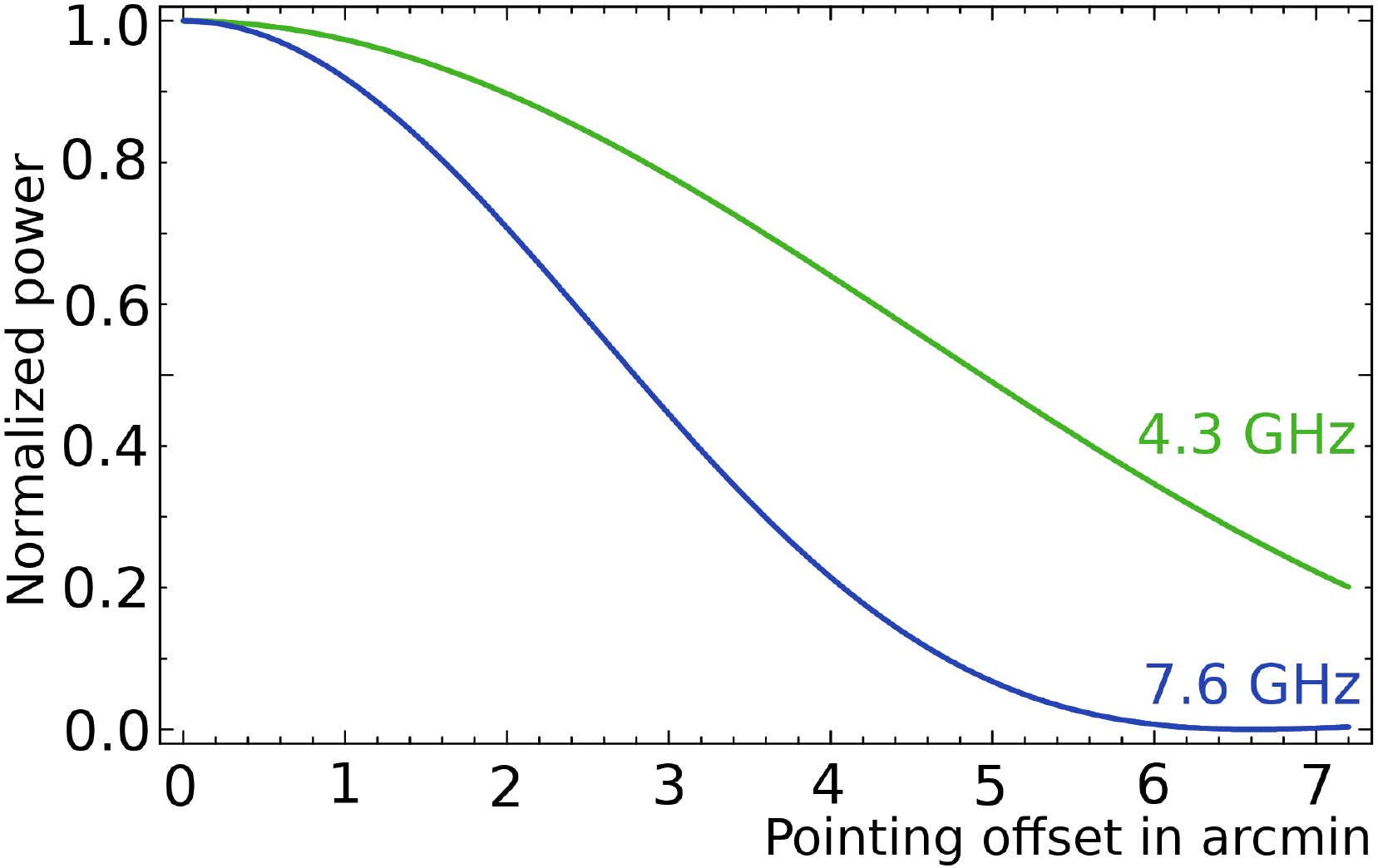}
  \caption{The primary beam attenuation of a 25~m VLBA antenna at 4.3
         (upper green curve) and 7.6~GHz (lower blue curve).}
  \label{f:beam}
\end{figure}

  The full width half maximum (FWHM) of VLBA antennas is $10'$ at 4.3~GHz and 
$5.8'$ at 7.6~GHz (See Figure~\ref{f:beam}). However, for traditional observations 
with accumulation period lengths 2~s and spectral resolutions of 128 channels 
per IF that are very often used as default for processing VLBI observation, 
the field of view of a radio interferometer is significantly narrower. The 
field of view of the VLBA with these settings is limited to 10--$20''$.
I define the field of view as the area of the sensitivity reduction at a level 
not exceeding 50\% with respected to the pointing direction.

  There are four factors that affect the field of view:

\begin{enumerate}
   \item {\it Antenna primary beam}. All VLA and VLBA antennas are identical 
          25~m dishes. The primary beam power diagram of an ideal antenna 
          is described by the Airy pattern \citep[e.g.,][]{r:bw99}
         \beq
           B(x) = \lp 2 J_1(x)/x \rp^2,
         \eeq{e:e3}      
         where $J_1(x)$ is the Bessel function of the first order, $x$ is the 
         normalized distance from the center of the field 
         ${\pi D}/{\lambda} \, \theta$, $D$ is the antenna diameter, 
         $\lambda$ is the wavelength, and $\theta$ is the offset with respect
         to the pointing direction.

         The presence of the obstructing secondary mirror, the \Note{quadrapod}, 
         and the deviation of the antenna surface from the parabaloid causes 
         a departure of the beam pattern from expression \ref{e:e3}. Even when
         the beam power diagram is known precisely, pointing errors cause 
         errors in computation of the primary beam attenuation.
         
         This effect cannot be mitigated for an antenna of a given size, but 
         the amplitude loss can be calibrated and taken into account during 
         data analysis.

   \item {\it Tapering.} The DiFX correlator \citep{r:difx1,r:difx2} used for data 
         analysis, is of the FX type. Input data stream segments 
         are shifted according to the a~priori station-based delays, Fourier 
         transformed, cross-multiplied, and accumulated. If the a~posteriori
         delay is the same as the a~priori delay, all data in the input 
         segments are used for cross-multiplication and accumulation. 
         If the a~posteriori delay differs from the a~priori delay, one 
         input data stream is shifted with respect to another and therefore, 
         only a fraction of the data is cross-multiplied and accumulated.
         If the shift exceeds the segment length, no data can be accumulated 
         at all. Since the data were recorded at the Nyquist frequency 
         without overlapping, the share of the accumulated data is
         \beq
             L_{\rm t}(\Delta \tau) = 1 - |\Delta \tau | \, \Frac{2 \,B_{\rm sr}}{N},
         \eeq{e:e4}
         where $\Delta \tau$ is the residual delay, $N$ is the segment
         length (1024 for this survey) and $B_{\rm sr}$ is the sampling rate 
         (64~megasamples/s). 

         The antennas are pointed to the direction where a source is expected,
         and the correlator uses the a~priori path delay computed for 
         these directions. If a source is located $\Delta \alpha$, $\Delta \delta$,
         off the a~priori position, the path delay is incremented by
         $\partial \tau/\partial \alpha \, \Delta \alpha + \partial \tau / \partial \delta \, \Delta \delta$,
         and therefore, the fringe amplitude is reduced according to 
         equation~\ref{e:e4}. The array loses its ability to detect a source 
         even if $L_{\rm t} > 0$ when the signal to noise ratio (SNR) is 
         reduced by $L_{\rm t}$ to a level below the detection threshold.

         An obvious way to mitigate tapering is to increase $N$ and therefore, 
         to increase of the spectral resolution of visibilities that is 
         $2B_{\rm IF}/N$, where $B_{\rm IF}$ is the IF bandwidth. This results
         in a growth of the correlator output size that was considered 
         undesirable in the past and made the wide-field VLBI unpopular. 
         Advances in computer hardware made wide-field VLBI affordable.

%

   \item {\it Time smearing}. Although the correlator used 15.625~$\mu$s long 
         segments for this campaign, the visibilities are averaged 
         over longer accumulation periods. Averaging visibilities over a finite 
         time causes decorrelation at the edges of the time intervals. It can 
         be easily shown \citep[e.g.,][]{r:tms} that the fringe amplitude 
         loss factor due to time smearing is 
         \begin{widetext}
         \beq
            L_{\rm ts} (\Delta \dot{\tau})
                       = \left| \Frac{1}{\Delta t} 
                                \dss \int\limits^{t+\Delta t/2}_{t-\Delta t/2} 
                                \cos ( 2\pi f_0 \dot{\tau} t)  \; dt \; \right|
                       =
                         \cos ( 2\pi f_0 \dot{\tau} t ) 
                         \left| \Frac{\sin{\pi f_0 \dot{\tau} \, \Delta t}}
                                          {\pi f_0 \dot{\tau} \, \Delta t} \right|,
                     \qquad
         \eeq{e:e5}
         \end{widetext}
         where $\Delta t$ is the accumulation period during correlation and $f_0$
         is the reference frequency.

         Though, amplitude losses due to time smearing can be mitigated by reducing
         accumulation period lengths. This also increases the output dataset size.
  
   \item {\it Non-linearity of fringe phase}. The fringe search procedure 
         assumes the fringe phase varies linearly over a scan with 
         both time and frequency \citep{r:vgaps}. The fringe phase at the 
         accumulation period $i$ and the frequency channel $j$ is expressed as
         \begin{widetext}
         \beq
            \phi_{ij} = \omega_o \tau_p + \omega_o (t_i - t_o) \dot{\tau_p} + 
                        (\omega_j - \omega_o) \; \tau_g + 
                        (\omega_j - \omega_o) \; (t_i - t_o) \; \dot{\tau_g},
                        \qquad
         \eeq{e:e6}
         \end{widetext}
        where $\omega$ is an angular frequency, $\tau_p$ and $\tau_g$ 
        are phase delay and group delay. The contribution of the third mixed
        delay derivatives, namely $\partial^3 \tau/\partial^2 t \partial \alpha$ 
        and $\partial^3 \tau/\partial^2 t \partial \delta$, causes a quadratic 
        term in the dependence of the fringe phase on time. It may become 
        significant if the position offset is large and a scan is long. In the 
        first approximation, the time delay is expressed via the baseline 
        vector $\vec{r_1} - \vec{r_2}$ and the unit source position vector 
        $\vce{S}$ up to terms $O(1/c^2)$ as: 
\beq
        \tau = \Frac{1}{c} \, \mat{E} \, ( \vec{r_1} - \vec{r_2} ) \cdot \vce{S},
\eeq{e:e7}
        where $\mat{E}$ is the Earth rotation matrix. Differentiating it over 
time twice, we get an expression for $\ddot{\tau}$:
\beq
   \ddot{\tau} = \Frac{1}{c} \, \ddot{\mat{E}} \, 
       ( \vec{r_1} - \vec{r_2} ) \cdot \vce{S}.
\eeq{e:e8}
\end{enumerate}
   The maximum value of $\ddot{\tau}$ for a baseline with the Earth's radius is
$1/c \, R_\oplus \Omega_\oplus^2$, or about $10^{-10}$ 1/s. The maximum values 
of $\partial^3 \tau/\partial^2 t \partial \delta$ and 
$\partial^3 \tau/\partial^2 t \partial \alpha$ are close to $\ddot{\tau}$. 
Let us consider a source observed at 7.6 GHz in a 60~s long scan that is 
$10^{-3}$ rad ($3.3'$) off the pointing direction at a baseline with the 
length of the Earth's radius. Then the maximum magnitude of the phase curvature 
over the scan will be $ 2\pi f/c \, R_\oplus \, \Omega_\oplus^2 \, (t/2)^2/2 \approx 2$
radian. The contribution for a 2~minute long scan will be 8~radians, i.e. over 
one phase turn.

  The fringe amplitude loss due to the  non-linear phase variation can be mitigated
by iterations when a preliminary source position is determined at the first
iteration, and then that position is used for computation of the non-linear
phase variation that is subtracted before the next iteration of fringe fitting. 
Since the non-linear phase variation due to the error in the a~priori source 
position is proportional to the baseline length, a source still can be 
detected at short baselines. \Note{The accuracy of its positions derived from
these observations is sufficient to compute precisely the quadratic term in
fringe phases.}

  There are two approaches for an increase of the field of view: 1)~to increase 
the spectral and time resolutions during correlation and 2)~to perform multiple 
correlation passes with different phase centers at expected source positions 
\citep{r:mid13}. The first approach is more straightforward, but generates 
a large amount of data that requires significant computing resources after 
correlation. The second approach requires more resources during correlation. 
The DiFX correlator implements this approach very efficiently. This approach 
has an advantage if we have the a~priori knowledge that one or several sources
are located within 10--$20''$ of the specific positions within the 
primary beam of an antenna and other areas are {\it excluded} from the search. 
However, if to cover the primary beam with a mosaic of many phase 
centers as \citet{r:mor11} and \citet{r:mid13} did, the output dataset size
and the amount of required resources for post-processing are not decreasing, 
and it becomes more complex. The dataset size can be decreased substantially if 
to correlate at phase centers of the sources known 
from low-resolution connected element interferometers, such as NVSS 
\citep{r:nvss}. \citet{r:mjive20} used this approach for detection of weak 
sources within the primary beam at 1.4~GHz. However, the downside of this 
approach is that we can find only those sources that are known before and 
lose others. There are instances when a compact component is located far away 
from the brightest extended component that dominates at low resolution images 
\citep[See Figures 1--5 in][]{r:obrs2}. In contrast, the first approach 
relaxes the requirement to the accuracy of a~priori positions: a source will 
be detected anywhere in the field of view. I used the first approach in 
this campaign.

\subsection{Design of the wide-field VLBI campaign}

  We see in the previous section that a finite spectral and time 
resolution reduces the field of view with respect to the individual 
antenna beam size. The old hardware correlators limited the output rate 
and restricted the choice of time and spectral resolution. Advances in 
computing hardware made if feasible to correlate VLBI experiments at 
general purpose computers using flexible software correlators, such 
as DiFX or SFXC \citep{r:sfxc}, and these limitations were lifted. 
However, post-processing of large datasets was considered impractical 
until recently.

\begin{figure*}
  \begin{minipage}[t]{0.32\textwidth}\centerline{$\delta = 70^\circ$}\end{minipage}
  \hspace{0.01\textwidth}
  \begin{minipage}[t]{0.32\textwidth}\centerline{$\delta = 20^\circ$}\end{minipage}
  \hspace{0.01\textwidth}
  \begin{minipage}[t]{0.32\textwidth}\centerline{$\delta = -30^\circ$}\end{minipage}
  \par\vspace{0.0ex}\par
  \includegraphics[width=0.32\textwidth]{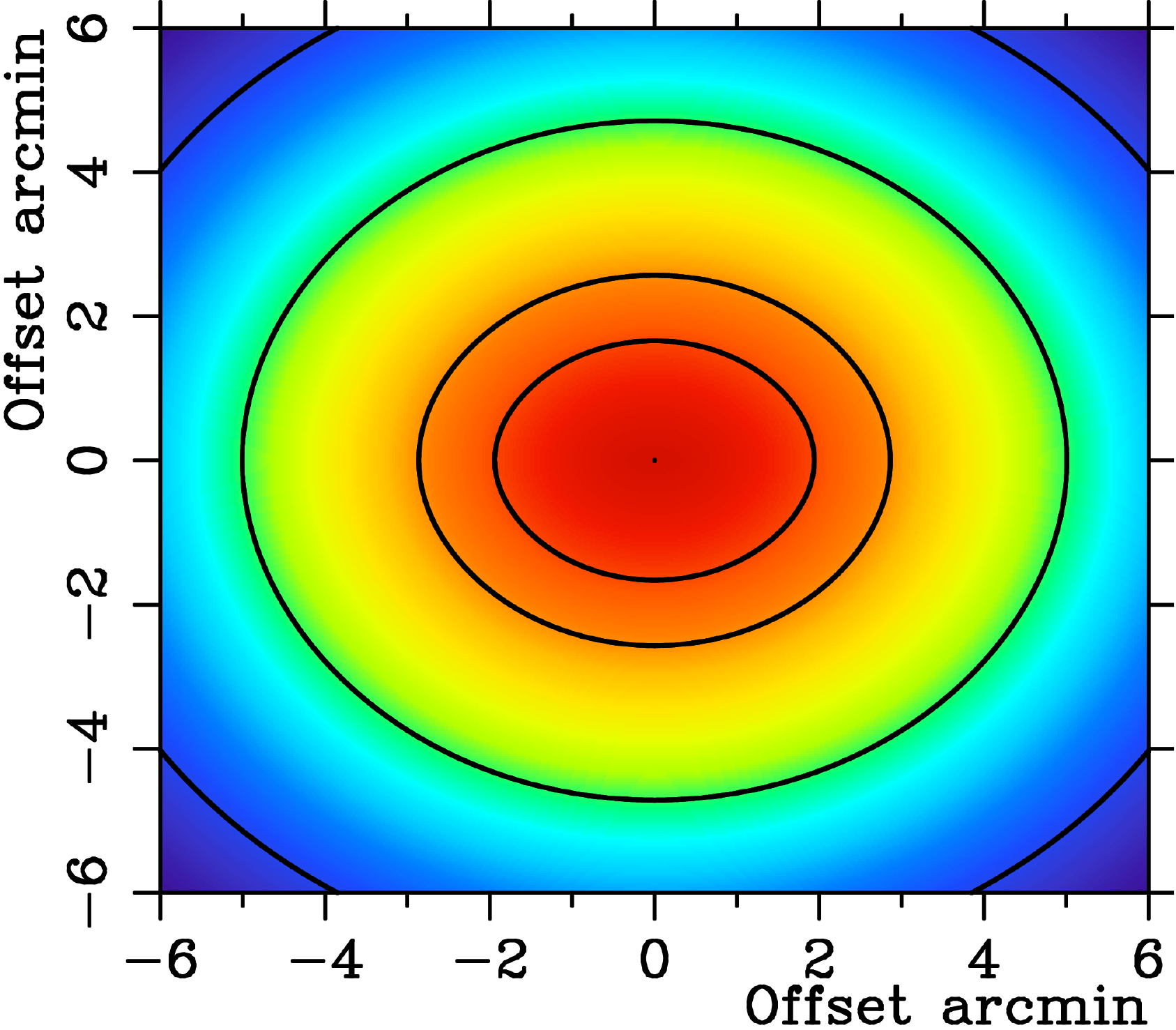}
  \hspace{0.01\textwidth}
  \includegraphics[width=0.32\textwidth]{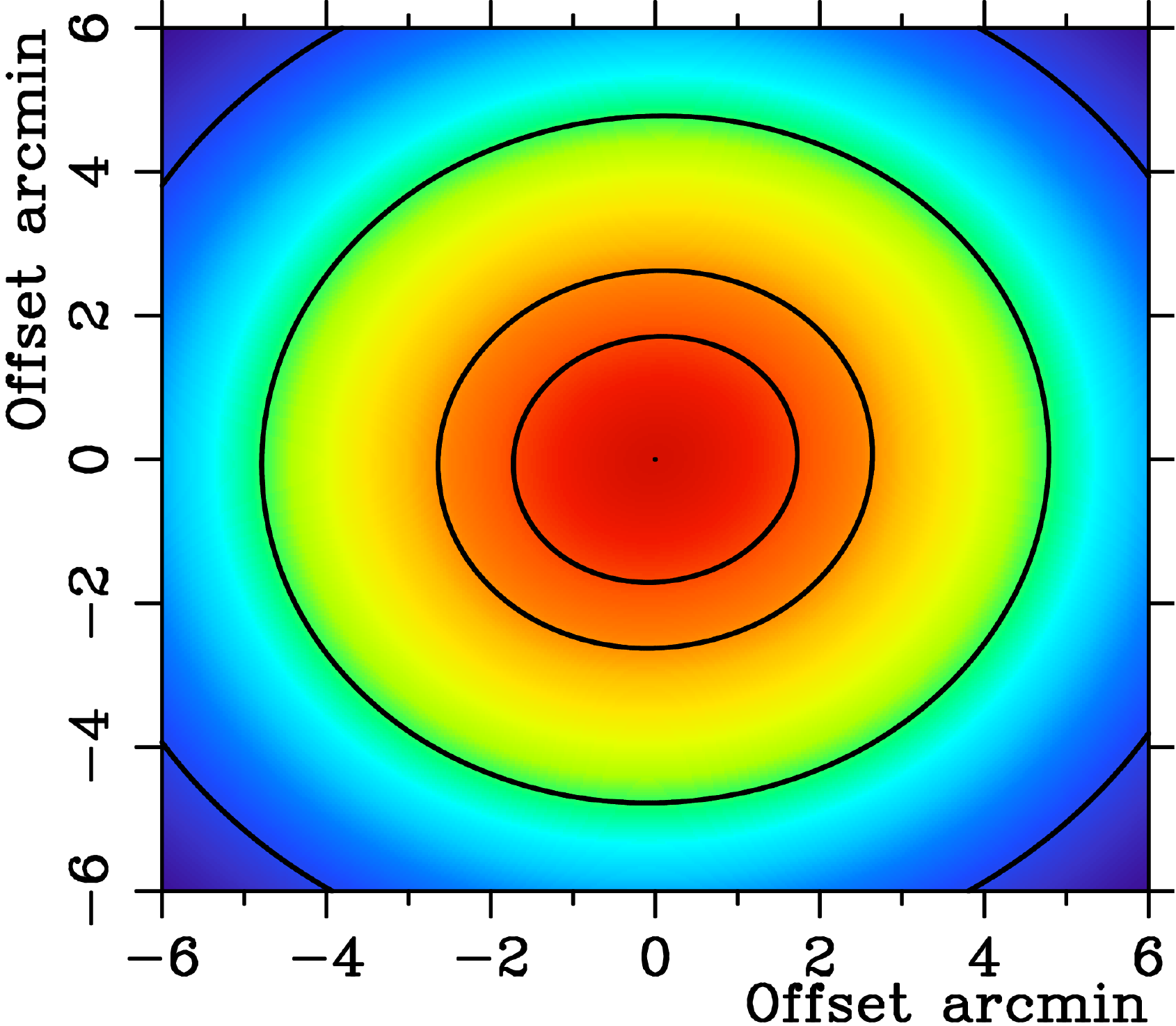}
  \hspace{0.01\textwidth}
  \includegraphics[width=0.32\textwidth]{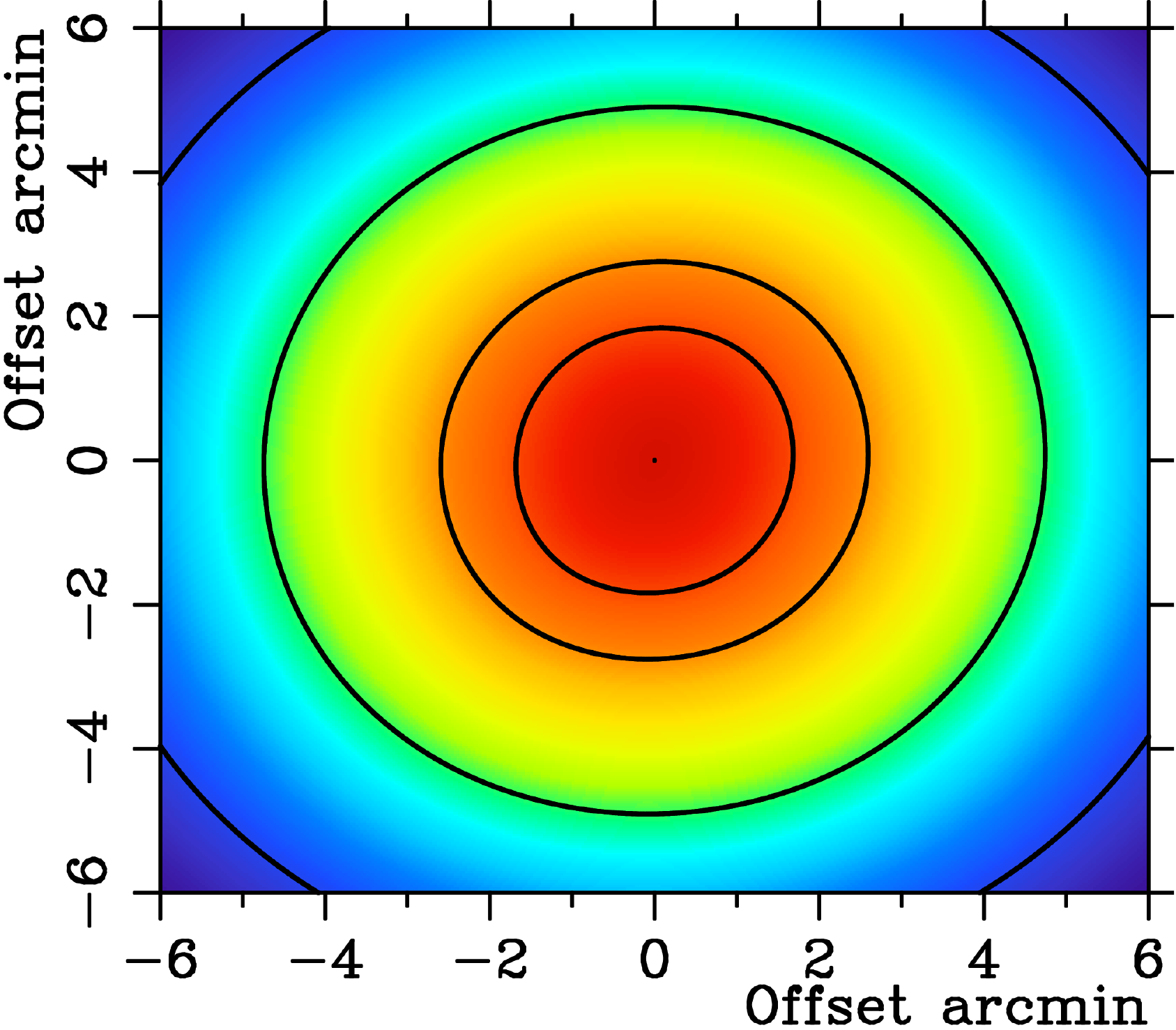}
  \par\vspace{0.5ex}\par
  \includegraphics[width=0.32\textwidth]{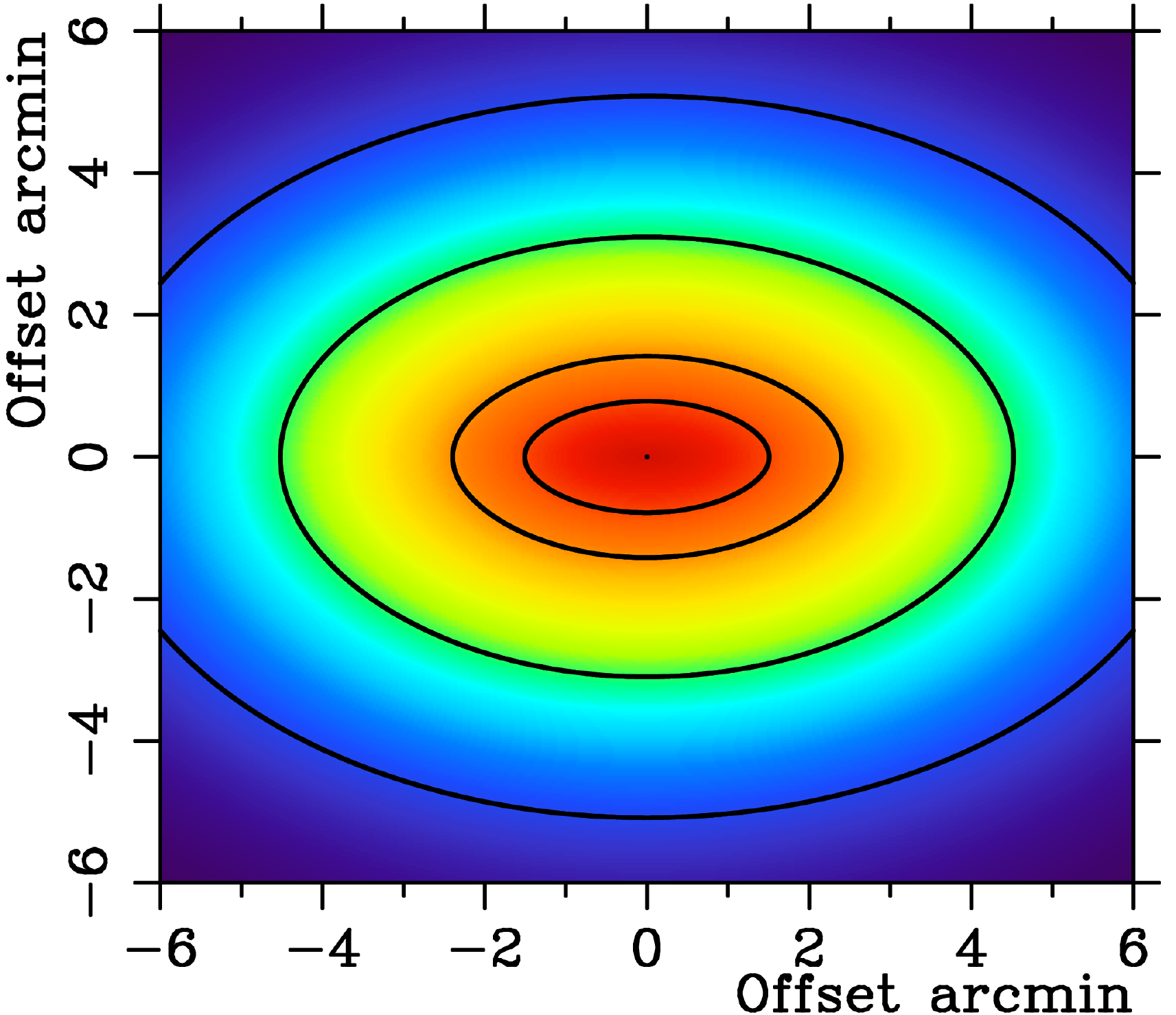}
  \hspace{0.01\textwidth}
  \includegraphics[width=0.32\textwidth]{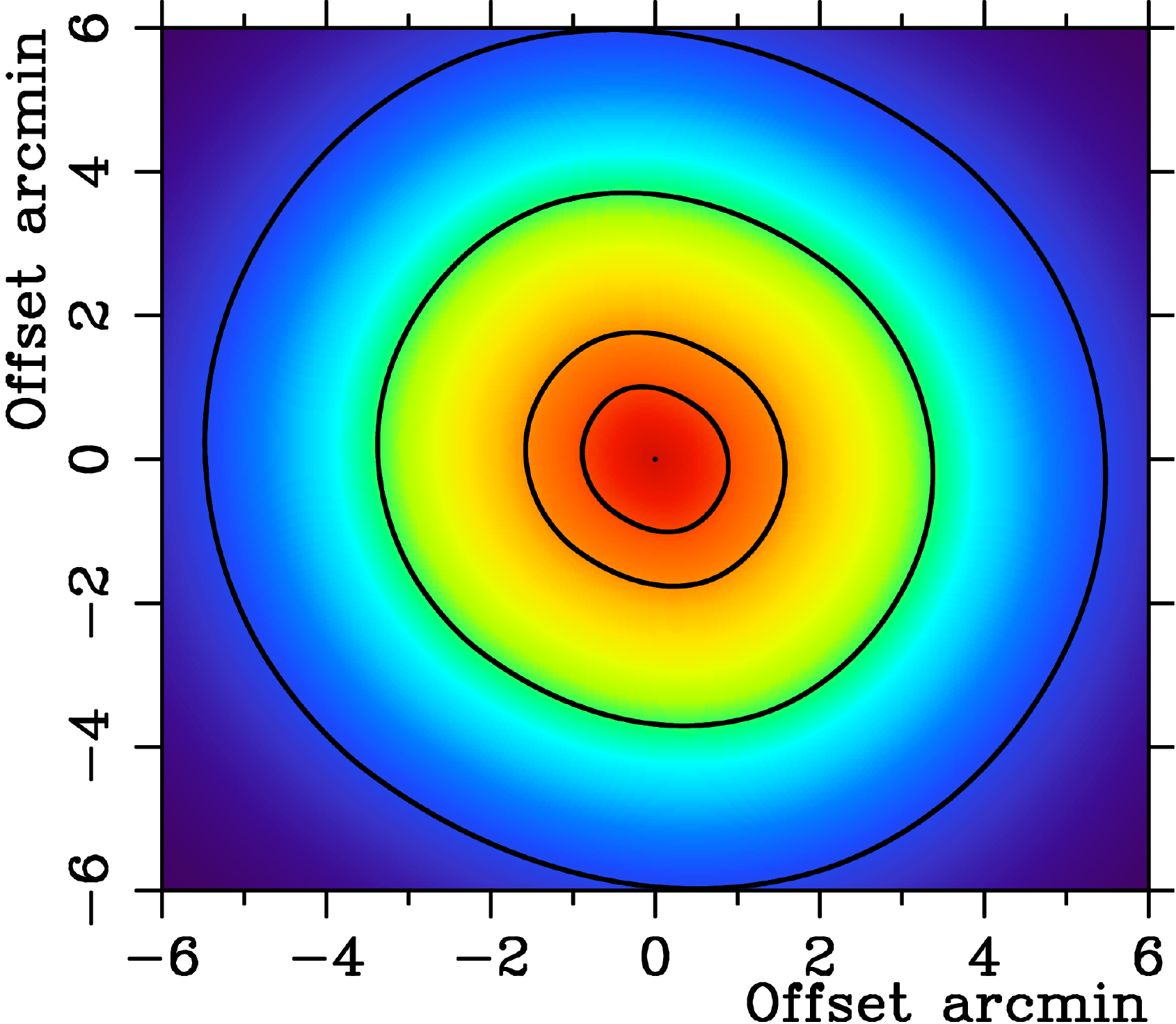}
  \hspace{0.01\textwidth}
  \includegraphics[width=0.32\textwidth]{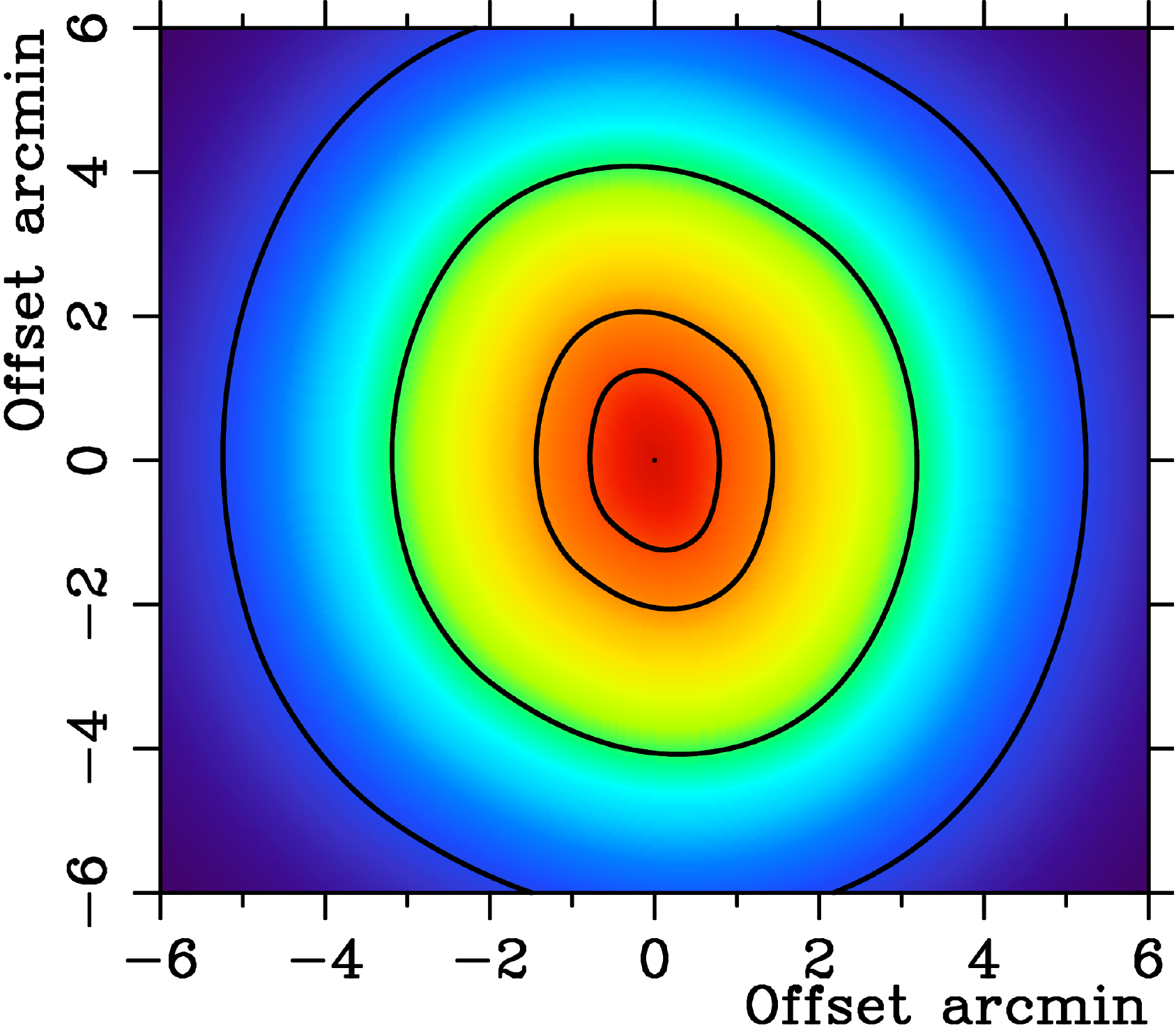}
  \par\vspace{0.5ex}\par
  \includegraphics[width=0.32\textwidth]{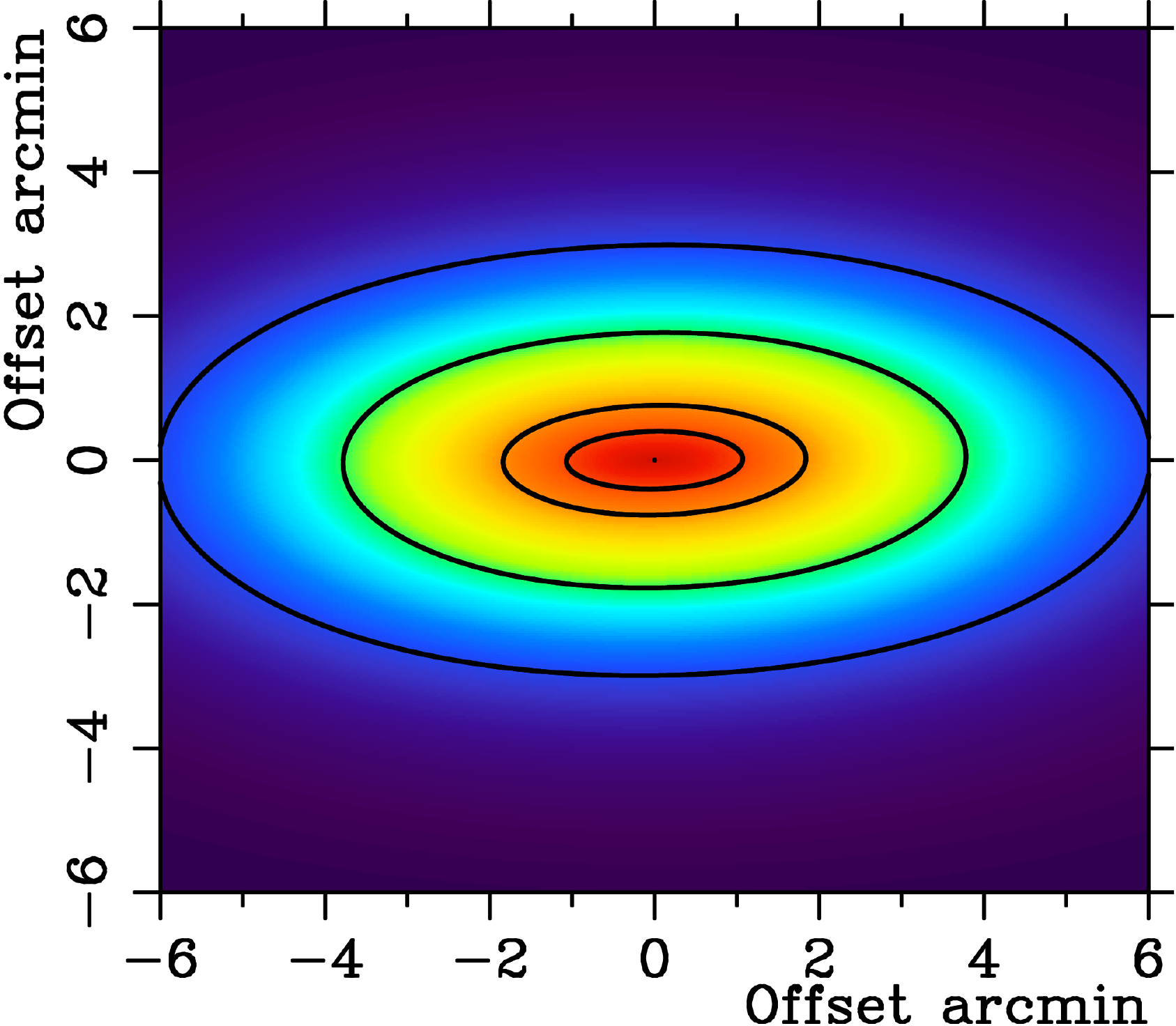}
  \hspace{0.01\textwidth}
  \includegraphics[width=0.32\textwidth]{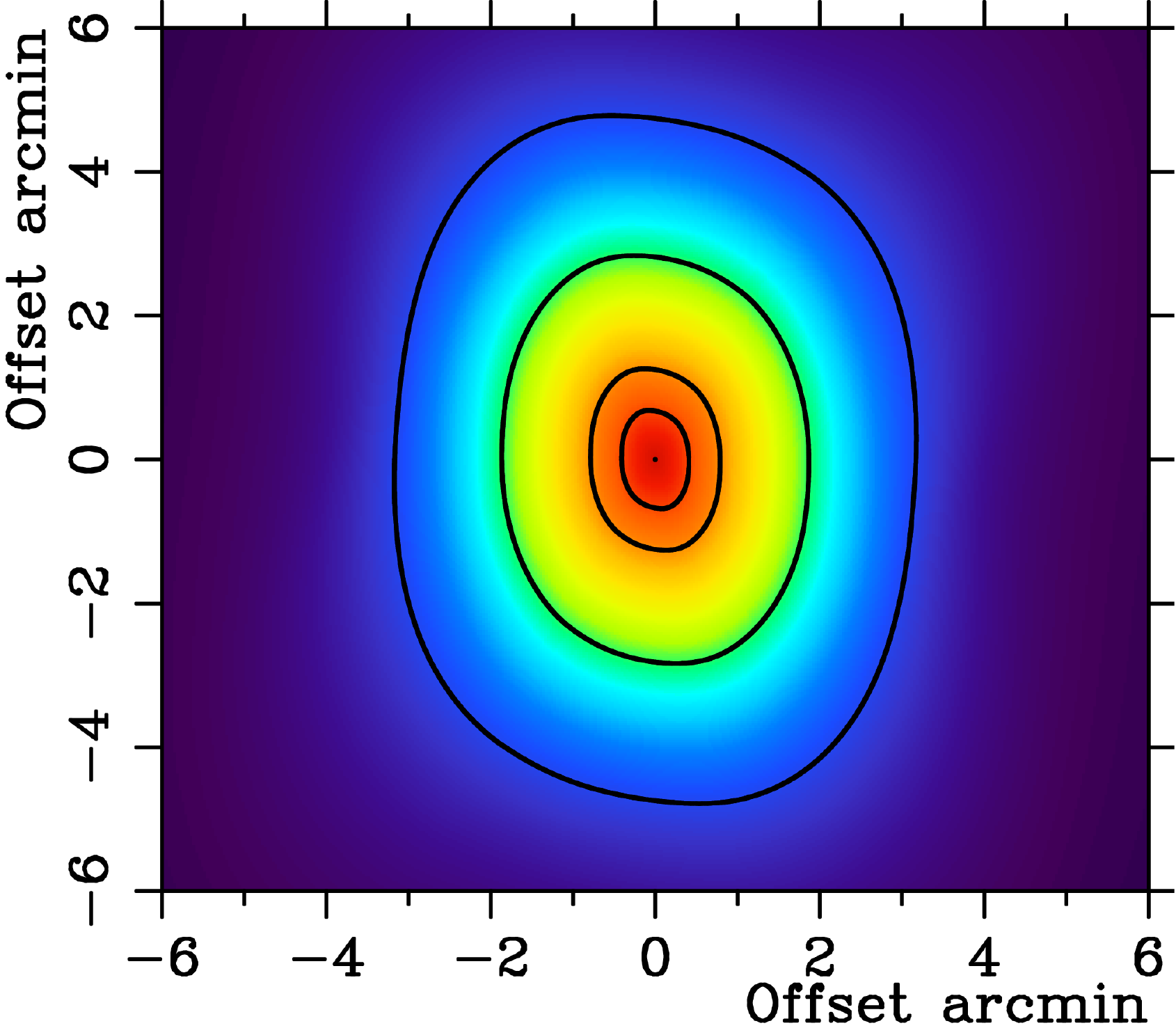}
  \hspace{0.01\textwidth}
  \includegraphics[width=0.32\textwidth]{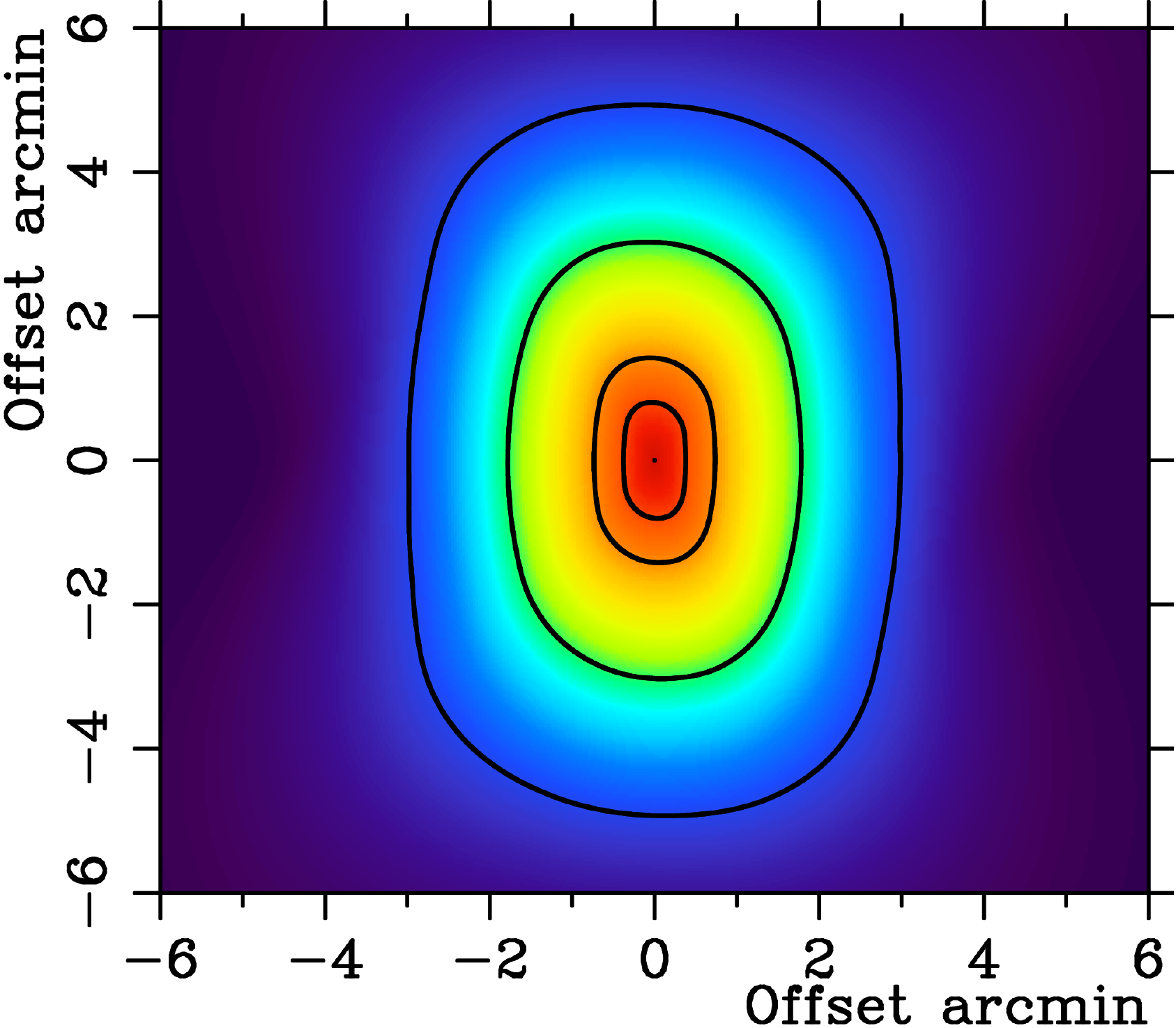}
  \par\vspace{1.0ex}\par
  \includegraphics[width=0.32\textwidth]{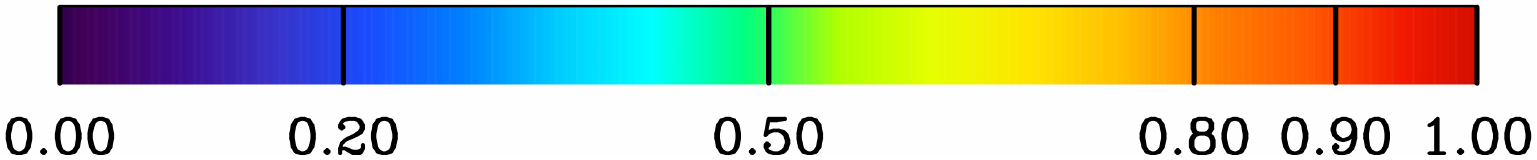}
  \caption{The typical fields of view of the VLBA network at 4.3~GHz when correlated
           with accumulation periods of 100~ms and with the spectral resolutions 
           of 62500~Hz. Color shows a reduction of the fringe amplitude as 
           a function of the source position offset with respect to the pointing
           direction in a range from 0 to to 1.
           The first row provides the averaged field of view for the inner
           part of VLBA at baselines shorter 1000~km, the second row provides
           the field of view at baseline lengths in a range of 2000--4000~km,
           and the third row provides the field of view at baselines longer 5000~km.
           Three columns correspond to observation of a source at declinations
           $70^\circ$, $20^\circ$, and $-30^\circ$ respectively.
          }
  \label{f:fov}
\end{figure*}

  I evaluated the size of the field of view at 4.3~GHz band achievable with
the wide-field correlation setup (0.1~s time resolution and 62500~Hz
frequency resolution). Tapering and time smearing depends on the 
baseline vector. I ran a simulated 24 hour schedule of observing three 
sources at declinations $-30^\circ$, $20^\circ$, and $70^\circ$ every 
5~minutes. Observations at elevations below $5^\circ$ were discarded. 
I averaged the amplitude losses due to the four factors discussed
in the previous section for three subarrays: 1)~10~short 
baselines in the inner part of the array with lengths shorter 1000~km; 
2)~13~medium baselines with lengths in a range of 2000--4000~km, and 
3)~6~baselines with lengths longer 5000~km. Figure~\ref{f:fov} shows 
the simulation results.

   We see that tapering and time smearing almost do not affect the field
of view at 4.3~GHz for short baselines: it is determined by the size of 
the primary beam. The field of view shrinks to 7--$8'$ at medium size 
baselines and to $3.5'$ at the longest baselines. The spectral resolution
should have been increased by a factor of 4 to avoid it. The field of 
view is smaller by the factor of 1.76 at short baselines at 7.6~GHz 
and by the factor of 1.1 at longest baselines. 

\subsection{Pilot observations to test elimination of the ionosphere contribution}

   Since astrometry at 4--8~GHz with the broadband C-band receiver was
new in 2013, I ran a pilot project (NRAO code BP175) in order to quantify 
possible systematic differences of 4.3/7.6~GHz astrometry with respect
to the traditional absolute astrometry at 2.3/8.6~GHz. Ten sessions 3--8 
hours long each were observed with the VLBA in October--December 2013 during 
the pilot project. Sources with the correlated flux density brighter 
200~mJy were observed in scans of 180~s long. The array observed at 
2.3/8.6~GHz for 50~s, then switched receivers within 20~s, observed at 
4.3/7.6 GHz for 50~s, then switch back to 2.3/8.6~GHz, and the array observed 
the same source for 50~s more. The same recording rate, 2~Gbps was used for 
all observations. Although the 256~MHz wide band was recorded within 
[2.188, 2.444]~GHz, only its 178~MHz wide fraction was used, remaining part 
being masked out due to the interference caused by the satellite radio and 
due to the front-end filters at some stations. The frequency setup at 
8.6~GHz differed from the 7.6~GHz setup only by shifting frequencies 
by 1~GHz up.

  The goal of the pilot project was to assess using real data 
1)~the sensitivity of 4.3/7.6~GHz absolute astrometry; 2)~the errors 
in determination of the ionosphere-free combinations of group delays 
at two frequencies; and 3)~the magnitude of the systematic differences 
in source positions with respect to the traditional 2.3/8.6~GHz 
absolute astrometry.

  I found that fringe phase and amplitude are oscillating within 
5~seconds at the beginning of each scan and after each receiver change. 
The first 5 seconds of data after receiver change were  masked out in 
further analysis. Therefore, the total integration time in each scan 
was 90~s at 2.3/8.6~GHz and 45~s at 4.3/7.6~GHz.

  First, 8.6~GHz data were processed using the fringe fitting 
procedure implemented in \PIMA software package\footnote{See documentation
at \web{http://astrogeo.org/pima}}. Residual group delays, 
phase delay rates, group delay rates, and fringe phases were computed 
at the fringe reference time that was set to the weighted mean epoch 
of 8.6~GHz data. Then the data from three other bands were processed,
and the fringe reference time for each scan was set to be the same 
as for 8.6~GHz data. Then total group delays, phase delay rates, group 
delay rates, and fringe phases at \Note{2.3/8.6~GHz} were computed to 
the scan reference time --- the common moment of time for all 
observations at all baselines of a given scan. After that,
similar quantities at 4.3/7.6~GHz were computed to the same scan 
reference time as \Note{2.3/8.6~GHz} data.

   These quantities were imported to the geodetic VLBI data analysis software
VTD/pSolve and were preprocessed the same way as all other VLBI observations 
under geodesy and absolute astronomy programs. Data analysis included editing, 
suppression of the outliers that exceed 3.5 times the normalized error, 
checking for clock breaks, and update of additive weight corrections. In total, 
3\% of group delays were flagged out. Careful analysis revealed sudden phase 
jumps at approximately 1~rad that affected about 2\% observations. This problem 
was traced to the digital baseband converter hardware and was fixed before the 
start of the main observing campaign.

\begin{figure}[h]
  \includegraphics[width=0.495\textwidth]{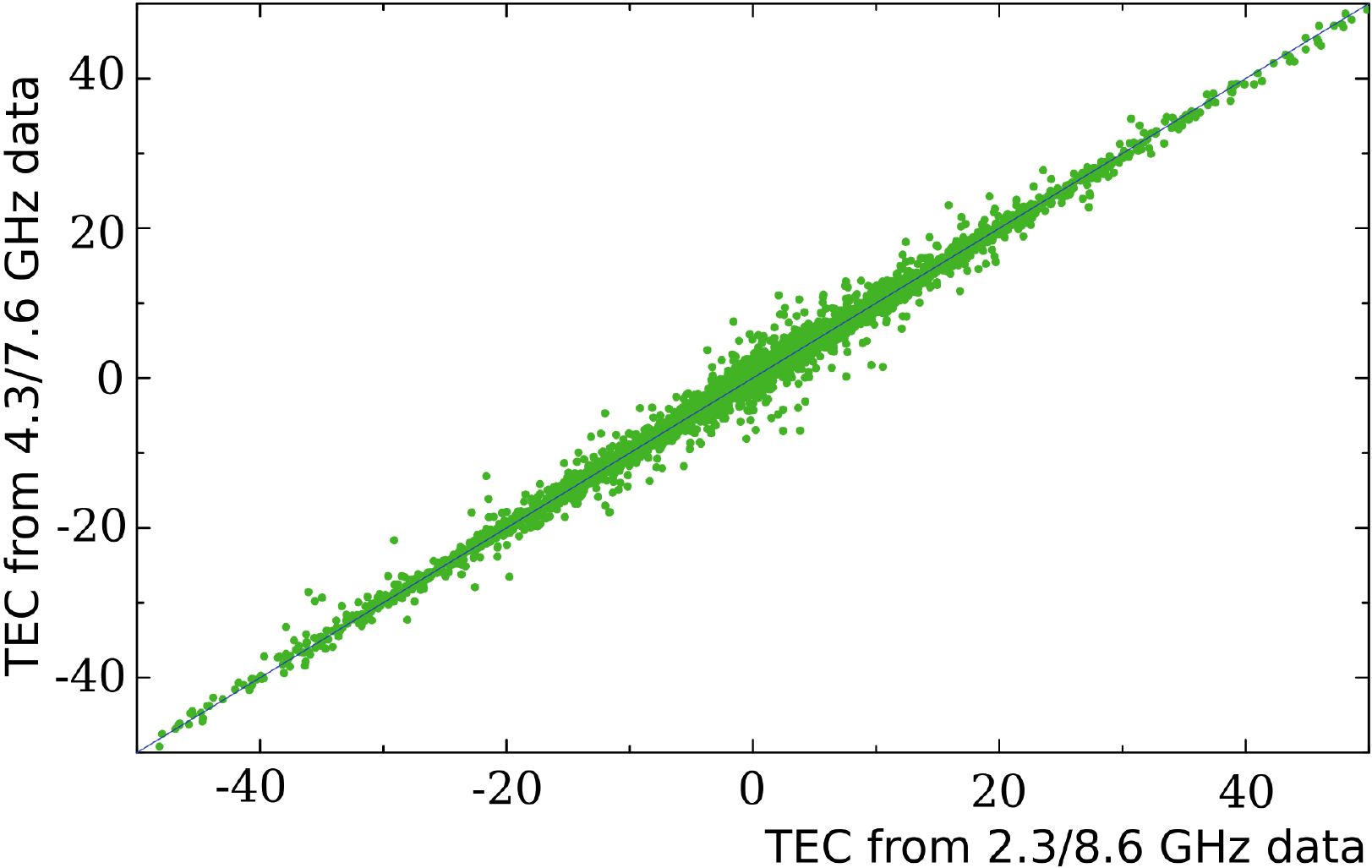}
  \caption{The estimates of the total electron contents from 
           quasi-simultaneous dual-band observations at 2.3/8.6 
           and 4.3/7.6 GHz.}
  \label{f:tec_tec}
\end{figure}

   Figure~\ref{f:tec_tec} shows the relationship between the TEC estimates
from \Note{2.3/8.6~GHz} and \Note{7.6/4.3~GHz} data. The correlation coefficient is 0.997. 
No systematic differences were found. Thus, we compelled to conclude 
that the ionospheric contribution derived from 2.3/8.6 and 4.3/7.6~GHz 
is practically the same.

\begin{figure*}
  \includegraphics[width=0.495\textwidth]{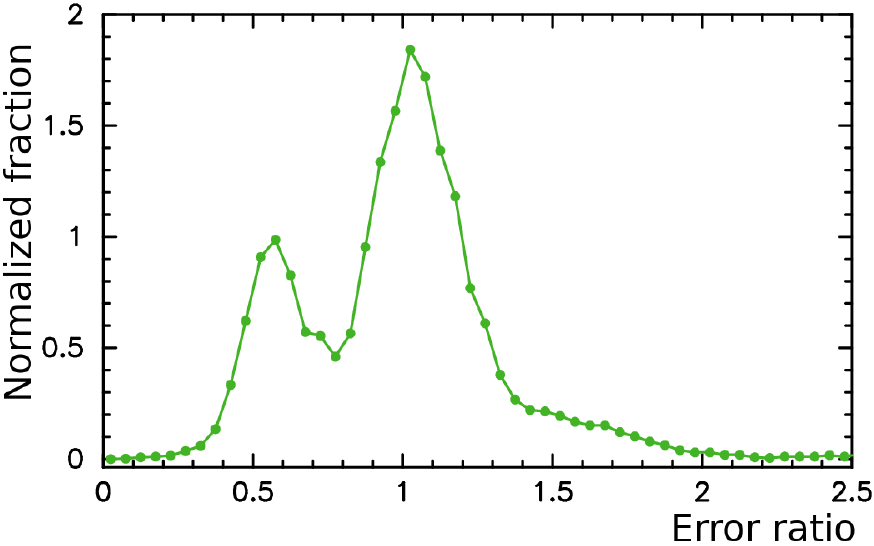}
  \hspace{0.005\textwidth}
  \includegraphics[width=0.495\textwidth]{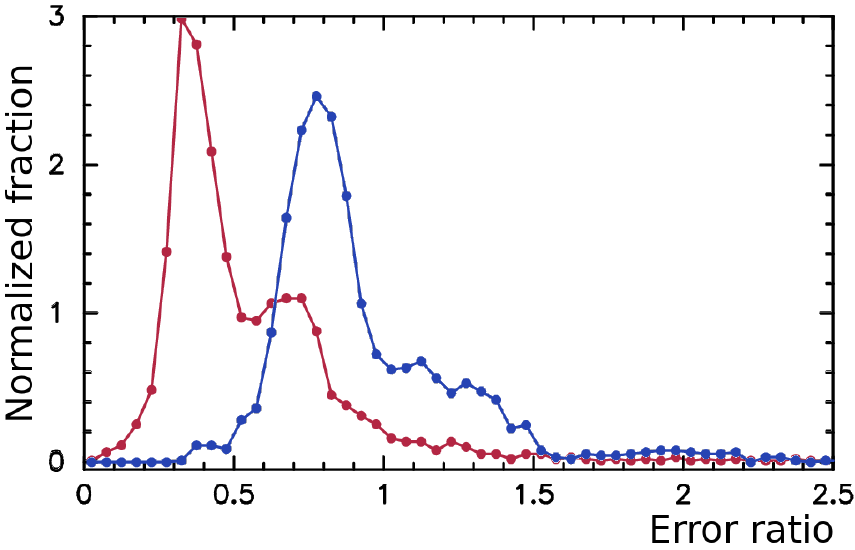}
  \caption{{\it Left} graph, green curve: the normalized histograms 
           of the ratio of the uncertainty of the ionosphere-free 
           combination of group delays derived using the 4.3/7.6~GHz
           data to the uncertainty of such a quantity derived from 
           simultaneous 2.3/8.6~GHz observations. {\it Right} graph, 
           red curve: a similar histogram but built using only the 
           group delays with uncertainties at 8.6~GHz greater 
           30~ps. {\it Right} graph, blue curve: similar histogram, 
           but using only group delays 8.6~GHz uncertainties less 
           than 5~ps.}
  \label{f:erat}
\end{figure*}

  Figure~\ref{f:erat} shows the histograms of the ratio of the uncertainty 
of the ionosphere-free combination of group delays derived from 4.3/7.6~GHz 
data to the uncertainty of such a quantity derived from simultaneous 
2.3/8.6~GHz observations. The 2.3/8.6 uncertainties were scaled by $1/\sqrt{2}$
because 4.3/7.6~GHz observations were twice shorter. The histogram shows 
two peaks: around 1.1, i.e., cases when the uncertainties are about the same,
and around 0.5, i.e., cases when 4.3/7.6~GHz uncertainties are twice smaller. 
I investigated that result further. When I kept only the observations with 
8.6~GHz group delay uncertainties greater than 30~ps, only the peak around 
0.35 remained (the red curve in the right graph of Figure~\ref{f:erat}). When 
I kept only the observations with strong SNR which provided uncertainty less 
than 5~ps at 8.6~GHz, the peak shifted to 0.8. This can be easily explained. 
\PIMA applies additive reweighting when computes uncertainties in group 
delay (see \citet{r:vgaps} for detail) that accounts for instrumental 
phase variations. This sets the floor in group delay uncertainties, 
typically 5--10~ps. When the floor is reached, effective sensitivity at 
7.6 and 8.6~GHz is about the same and wider frequency separation, 2.3 and 
8.6~GHz provides an advantage. That explains the peak around 1.1 in the 
left histogram of Figure~\ref{f:erat}. This error floor is not reached for 
weaker sources, and the higher sensitivity at 7.6~GHz explains 
the peak at 0.35 (red curve on the histogram in Figure~\ref{f:erat}).

  Analysis of the median in the cumulative distributions derived from the 
distribution presented in Figure~\ref{f:erat}, allows to us conclude that 
when a weak source is observed with the SNR $<$ 30 at X-band with VLBA, 
the estimates of the ionosphere-free combinations of group delays from 
4.3/7.6~GHz observations are a factor of 1.53 more precise than the 
combinations of delays derived from 2.3/8.6~GHz. When a strong source is 
observed, with SNR $>$ 200 at X-band, the 2.3/8.6~GHz ionosphere-free 
combinations of group delays \Note{have uncertainties lower} by a factor 
of 1.18. Thus, the tests have confirmed predictions.

  The 4.3/7.6 and 2.3/8.6~GHz dual-band observables were used for estimation 
of positions of 394 observed sources in two separate least square (LSQ) 
solutions. Prior to that, I ran a global reference solution that used all 
dual-band data since 1980 through 2020, except the data in the test campaign. 
That reference solution was made the same way as analysis of the main campaign 
was done. Then the variance-covariance matrix of the reference solutions 
was reduced by stripping the elements related to source positions. That 
reduced variance-covariance matrix was used as an input for analysis of the
test data. Therefore, only observations of the test data contributed to source 
positions. The test solutions used the same parameterization as the reference 
solution, except fixing the Earth orientation parameters to the IERS~C04 
series \citep{r:iersc04}. 

\begin{figure}
  \includegraphics[width=0.333\textwidth]{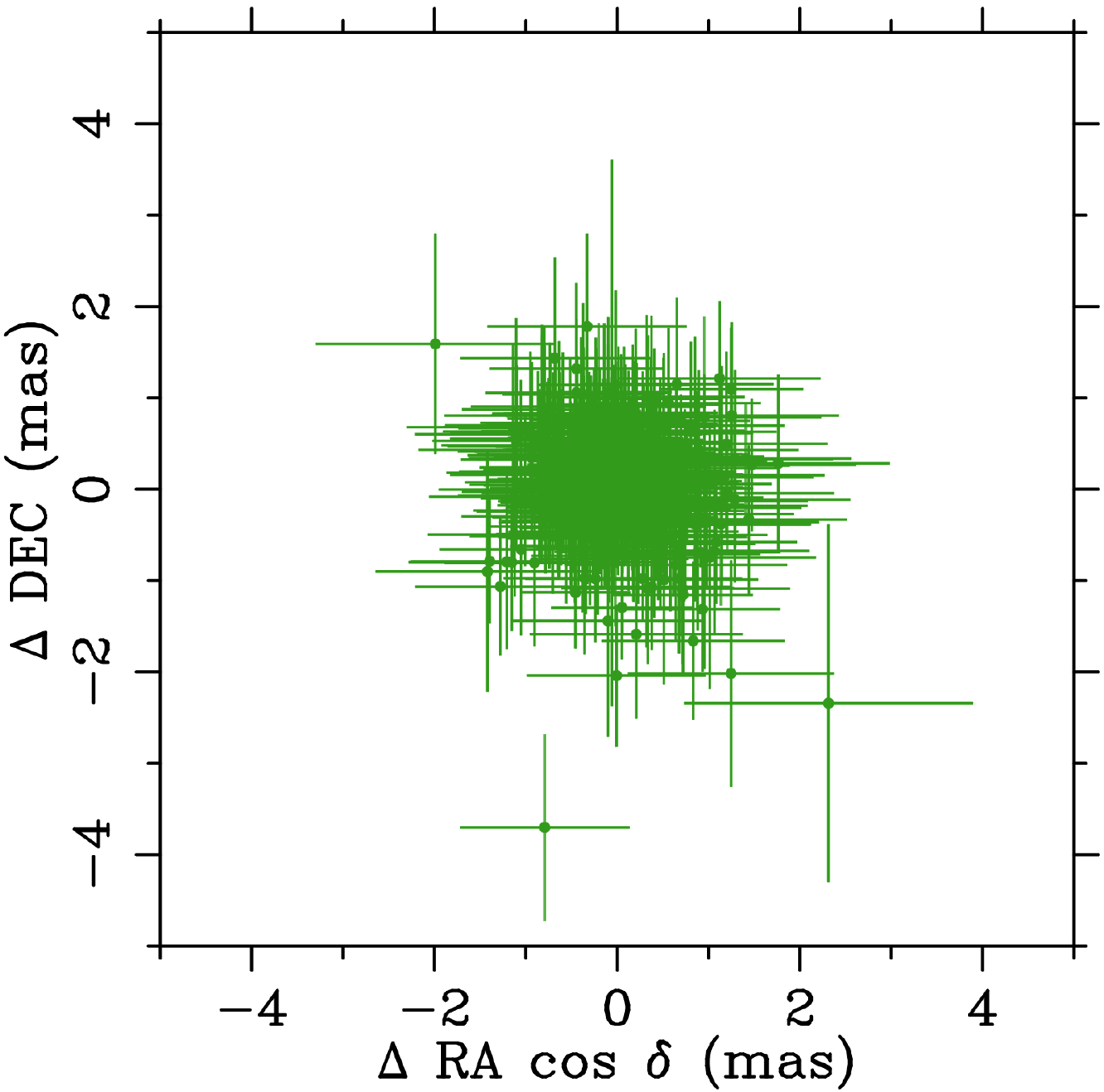}
%
%
%
  \caption{The differences in source position estimates from simultaneous 
           dual-band observations at 2.3/8.6 and 4.3/7.6 GHz.}
  \label{f:pos_sx_cf_diff}
\end{figure}

   Figure~\ref{f:pos_sx_cf_diff} shows the differences in estimates
of 394 source positions. The median position uncertainty is 0.62~mas
over right ascension scaled by $\cos\delta$ and 0.46~mas over declination.
The rms of position differences is 0.54~mas over right ascension scaled by 
$\cos\delta$ and 0.56~mas over declination. The mean weighted bias of 
4.3/7.6~GHz position estimates  versus 2.3/8.6~GHz position estimates is 
$-0.04 \pm 0.03$~mas over right ascension and $0.03  \pm 0.03$~mas over 
declination. The bias is statistically insignificant. This compels us to
draw a conclusion that switching from the traditional 2.3/8.6~GHz setup to 
4.3/7.6~GHz does not introduced measurable systematic errors.

\subsection{Pilot observations to test off-beam observations}

  To check whether the off-beam observations provide expected results, I ran
two tests. First, the same scan was re-correlated multiple times using 
different clock model that caused an increase of residual delay. The fringe
amplitude is expected to decrease linearly with an increase of the residual 
delay according to expression~\ref{e:e4} and fringe phase is expected to 
remain the same. The test showed no statistically significant deviation 
from the linear dependence. The phase difference remained statistically 
insignificant for $L_t$ above 0.1 and a systematic deviation reached 
0.2~rad at the lowest $L_t$ tested, 0.02. The source was not detected with
greater residual delays.

  A 25 minute long test was conducted on 2020 January 31 as a part of VLBI 
experiment BP245 that had the same frequency and correlation setup 
as the main campaign. A strong source 1741-038 was observed in 14 scans. 
In each scan, the antennas were pointed first at 1, 2, 3 \ldots 7 arcminutes 
away from its a~priori position along right ascension (hereafter, off-beam 
pointing) an then 1, 2, 3 \ldots 7 arcminutes along declination. Within each 
scan the antenna recorded for 30 seconds during off-beam pointing, then pointed 
to 1741-038, recorded for 30 seconds (hereafter, in-beam pointing) and 
then moved again to the same off-beam pointing. Then the procedure was 
repeated for the next off-beam pointing. 

   Fringe fitting was performed using both portions of the same off-beam 
pointing, and the fringe reference time was set the same for both in-beam 
and the off-beam pointings. The SNR around 1000 was achieved during in-beam 
pointing. The SNR was less for most of the off-beam pointings, and some 
observations at large offsets were not detected at all, which means the 
fringe amplitude was less than 0.006 with respect to the direct pointing.

\begin{table*}
   \caption{The differences in estimates of the 1741-038 off-beam position 
            with respect to its in-beam position derived from analysis of the 
            test experiment (columns 2--3 and 7--8) and the flux density
            estimates at 4.3 and 7.6~GHz integrated over the restored image. 
            The flux densities at zero offset are 3.84 and 4.26~Jy respectively.
            The left five columns show the position differences and the flux 
            densities when pointing was offset along the right ascension direction. 
            The right five columns show position differences and flux densities 
            when pointing was offset along the declination direction.
           }
   \hspace{-0.075\textwidth}   
   \begin{tabular}{crrrr @{\quad} |crrrr}
     \hline
    Off   & \nnnntab{c|}{Offset over $\alpha$}  & 
    Off   & \nnnntab{c} {Offset over $\delta$}  \\
    beam  & \ntab{c}{$\Delta \alpha$} & \ntab{c}{$\Delta \delta$} & C${}_{\rm tot}$ & X${}_{\rm tot}$  & 
    beam  & \ntab{c}{$\Delta \alpha$} & \ntab{c}{$\Delta \delta$} & C${}_{\rm tot}$ & X${}_{\rm tot}$  \\
    ($'$) & \ntab{c}{(mas)} & \ntab{c}{(mas)} & \ntab{c}{Jy} & \ntab{c|}{Jy} & 
    ($'$) & \ntab{c}{(mas)} & \ntab{c}{(mas)} & \ntab{c}{Jy} & \ntab{c }{Jy} \\
     \hline
      1 & -0.07 $\pm$  0.25 & -0.07 $\pm$  0.53 & 3.84 & 4.25 &  1 &  0.03 $\pm$  0.25 & -0.06 $\pm$  0.53 & 3.89 & 4.40 \\
      2 & -0.02 $\pm$  0.26 & -0.15 $\pm$  0.53 & 3.82 & 4.34 &  2 &  0.03 $\pm$  0.25 & -0.07 $\pm$  0.52 & 3.98 & 4.53 \\
      3 & -0.01 $\pm$  0.26 & -0.13 $\pm$  0.54 & 3.97 & 4.34 &  3 & -0.02 $\pm$  0.24 & -0.10 $\pm$  0.51 & 3.84 & 4.56 \\
      4 & -0.14 $\pm$  0.31 & -0.72 $\pm$  0.58 & 3.74 & 4.57 &  4 & -0.10 $\pm$  0.24 & -0.11 $\pm$  0.50 & 3.95 & 4.87 \\
      5 &  0.54 $\pm$  0.51 & -0.59 $\pm$  0.72 & 3.69 & 6.02 &  5 &  0.27 $\pm$  0.25 & -0.67 $\pm$  0.54 & 3.87 & 6.34 \\
      6 &  3.28 $\pm$  1.11 &  1.55 $\pm$  1.26 & 3.65 & n/a  &  6 &  0.90 $\pm$  0.36 & -0.97 $\pm$  1.00 & 3.70 & n/a  \\
      7 &  4.54 $\pm$  1.63 & 11.20 $\pm$  2.22 & 3.33 & n/a  &  7 & -1.97 $\pm$  0.49 &  2.64 $\pm$  1.58 & 3.66 & n/a  \\
     \hline
   \end{tabular} \hfill
   \label{t:pos_offbeam}
\end{table*}

   The in-beam and off-beam pointings were treated as different sources with 
different names. Both 4.3 and 7.6~GHz data were processed, and dual-band 
ionosphere-free combinations of group delays were used for estimation of 
1741-038 position from in-beam scans and 14 separate off-beam scans. The 
parameter estimation results are shown in Table~\ref{t:pos_offbeam}. 
Astrometry from off-beam pointings showed no systematic
errors above 1-$\sigma$ when a sources is observed up to $5'$ off-beam, and 
the fringe amplitude was above the 0.05 level achieved during in-beam pointing. 
There are systematic errors at 3-$\sigma$ only at pointings $6'$ and $7'$ off 
the beam when the amplitudes are less than 1\% of the in-beam pointing.

  I analyzed the normalized ratios $\eta(\theta_x,\theta_y)$ of the off-beam 
fringe amplitudes $a_{\rm ob}$ to $a_{\rm ib}$:
\beq
    \eta(\theta_x,\theta_y) \: = \: \Frac{a_{\rm ob}}{a_{\rm ib}} \, 
         \Frac{1}{L_t (\Delta \tau) \, 
                  L_{\rm ts}(\Delta \dot{\tau}) \, 
                  B(\theta_x,\theta_y)}.
\eeq{e:e9}
  These ratios should differ from 1 within the uncertainty of 
amplitude measurements. However, analysis revealed a significant scatter 
and a positive bias. \citet{r:mid13} investigated the primary beam of 
VLBA at 1.38~GHz. They adjusted the \Note{effective} antenna diameter as 
a single parameter for all the antennas. I tried to do the same, but this
resulted in a small reduction of variance. Then I extended the estimation 
model: I estimated the effective antenna diameter for each site separately
and included estimation of the antenna pointing corrections along
right ascension and declination. Considering the loss factors $L_t$ 
and $L_{\rm ts}$ are known precisely, we can present the unnormalized 
amplitude ratio $\rho (\theta_x,\theta_y)$ for a given baseline via 
the off-beam amplitude $a^c_{\rm ob}$ corrected for tapering and time 
smearing as 
\begin{widetext}
\beq
   \rho (\theta_x,\theta_y) = \Frac{a^c_{\rm ob}}{a_{\rm ib}} \, 
   \Frac{ \sqrt{B(\theta_x+ x_1, \theta_y+ y_1, D + d_1) \cdot
                B(\theta_x+ x_2, \theta_y+ y_2, D + d_2) }}
        { \sqrt{B(x_1, y_1, D + d_1) \cdot
                B(x_2, y_2, D + d_2) }},
\eeq{e:e10}
\end{widetext}
  where $x_i, y_i$ are the pointing corrections along right ascension and 
declination and $d_i$ is a correction to the a~priori antenna 
effective diameter $D$ with respect to the geometric diameter 25~m.
After taking logarithms from both sides, we get a system of non-linear
equations for $x_i$, $y_i$, and $d_i$. In order to avoid strong 
variations of fringe amplitude at large pointing offsets over 480~MHz 
bandwidth, I re-processed the data at a narrower bandwidth keeping 3~IFs 
around 4.54~GHz and 3~around 7.43~GHz. An iterative weighted LSQ 
was used to solve this system of equations. The following weights were 
used for parameter estimation:
\beq
   w = \Frac{1}{\sqrt{ \frac{2}{\pi} \rm{SNR}{}^2 + \sigma^2_0}},
\eeq{e:e11}
  where $\sigma_0 = 0.01$ is the floor of the amplitude variance
and a factor of $\sqrt{2/\pi}$ is due to the relationship between
the variance and the mean for the Rayleigh distribution.
Only observations at pointing offsets not exceeding $5'$ were used 
for processing 7~GHz data. Analysis was performed separately for 
4 and 7~GHz data.

\begin{figure}
  \includegraphics[width=0.495\textwidth]{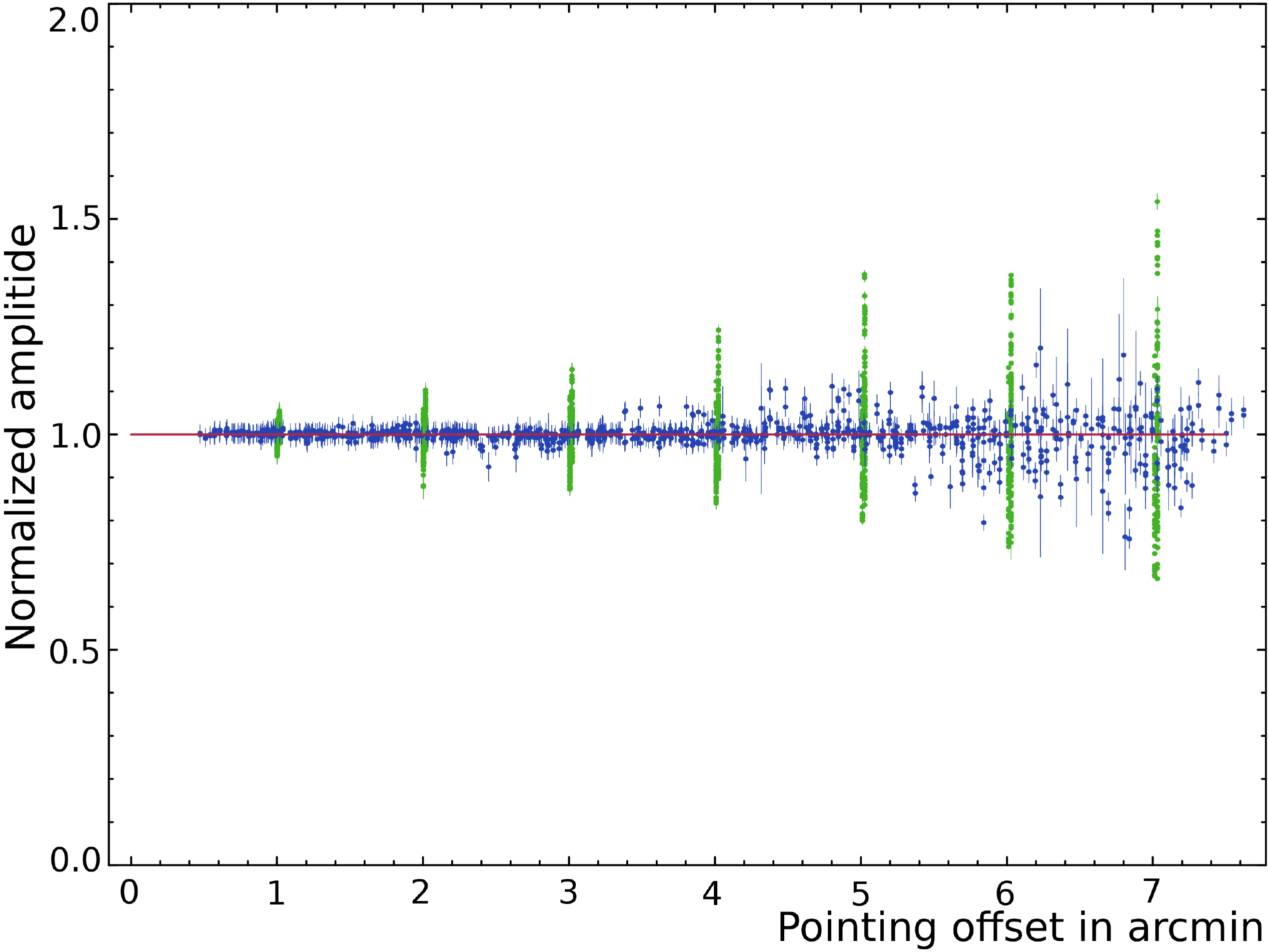}
  \includegraphics[width=0.495\textwidth]{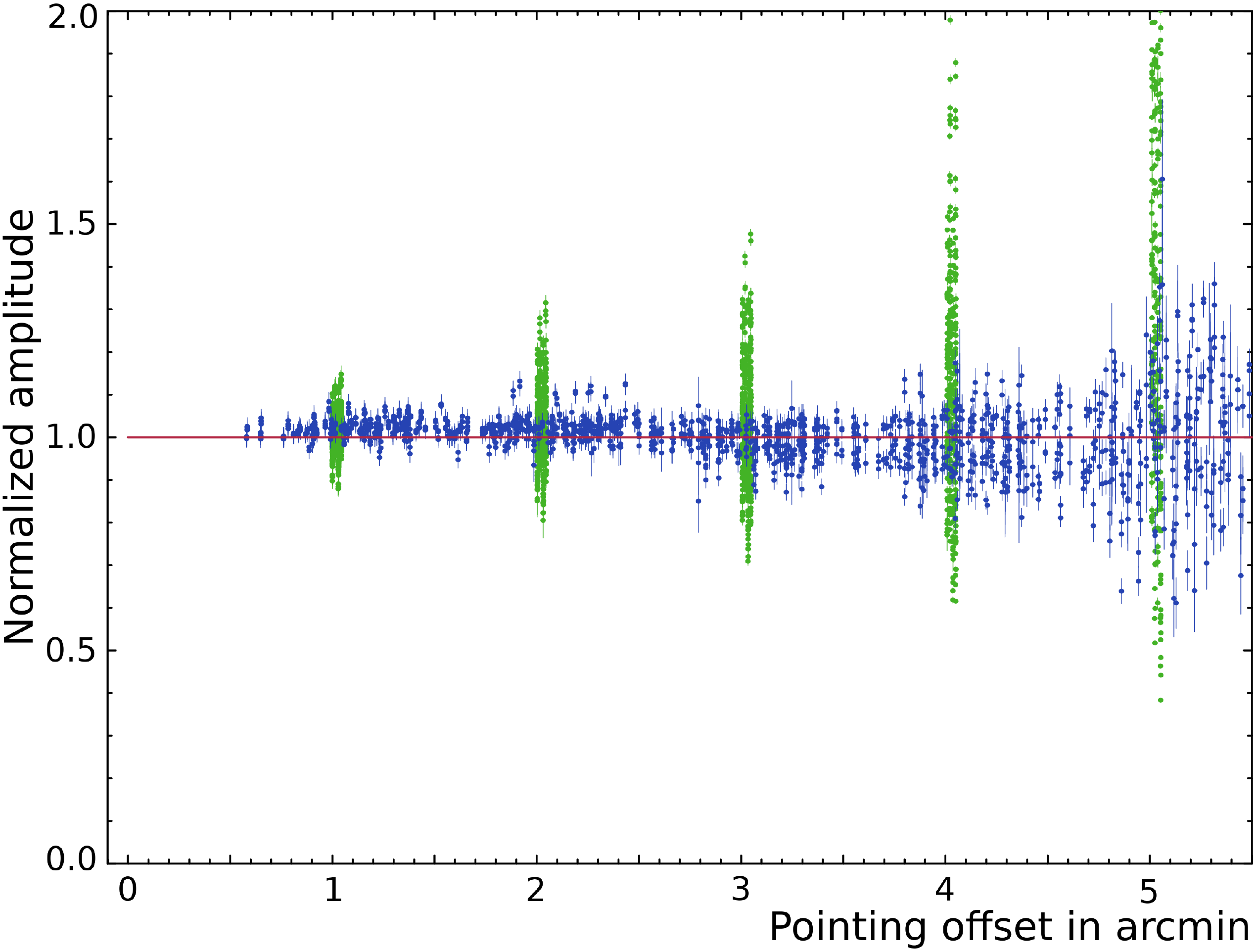}
  \caption{The normalized amplitude ratios $\eta({\theta_1,\theta_2})$
           for 4.5~GHz ({\it left}) and 7.4~GHz ({\it right}).
           The green points denote observed ratios as a function of
           nominal pointing direction. The blue points denote
           adjusted ratios when pointing corrections were estimated.}
  \label{f:norm_ampl}
\end{figure}

   Surprisingly large pointing corrections up to $55''$ were found. 
Pointing corrections at 4.5 and 7.4~GHz agree for most of the stations 
within $10''$. The estimates of the mean effective antenna diameters 
are 23.6~m and 25.4~m at 4.5 and 7.4~GHz respectively. Unfortunately, this 
test was not designed for determining pointing corrections since 
the antennas were pointed off the beam towards only one direction, which 
resulted in 0.95 correlations between estimates of pointing corrections 
and the effective diameters. Therefore, specific values of pointing 
corrections should be interpreted with a caution. However, estimation 
of pointing corrections and effective diameters resulted in a very 
substantial reduction of variance (see Figure~\ref{f:norm_ampl}) 
--- a factor of 4 at C-band and a factor of 3 at 7.4~GHz.

\begin{figure}
  \includegraphics[width=0.495\textwidth]{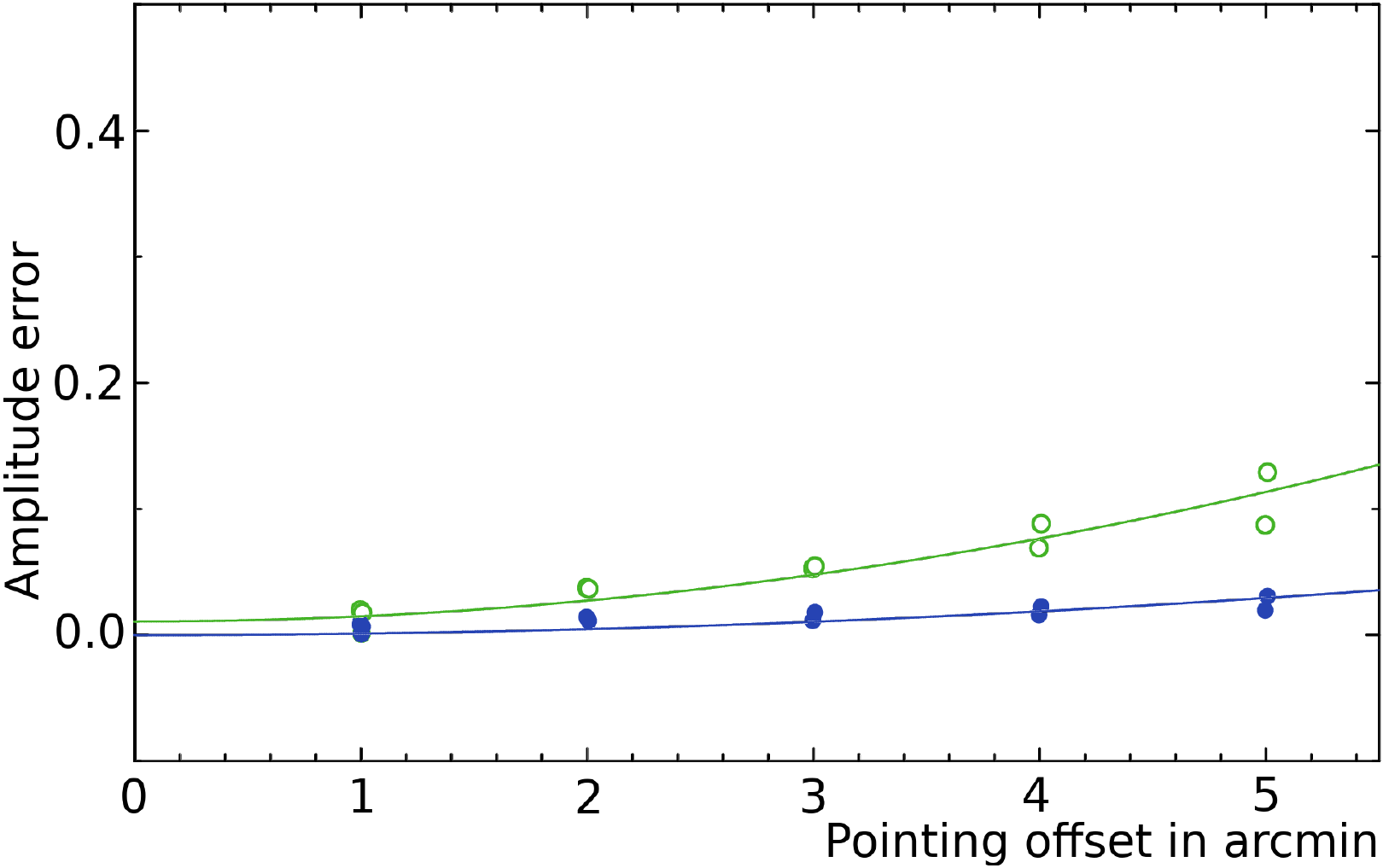}
  \includegraphics[width=0.495\textwidth]{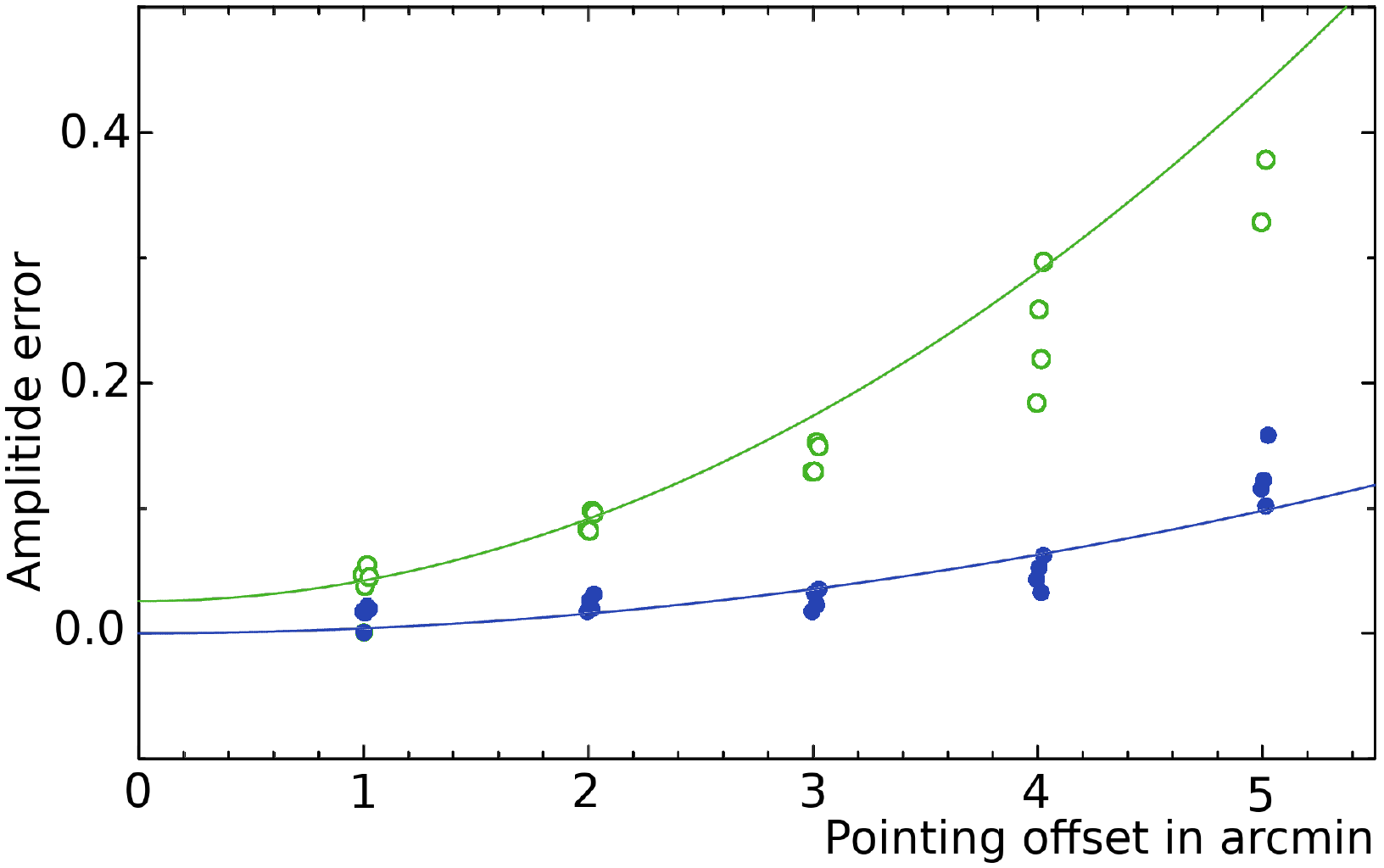}
  \caption{The amplitude errors as a function of pointing offset
           at 4.5~GHz ({\it left}) and 7.4~GHz ({\it right}).
           The green points denote observed ratios as a function of
           nominal pointing direction without adjustments. The blue 
           points denote adjusted ratios when pointing corrections 
           were estimated.}
  \label{f:ampl_err}
\end{figure}

   Adjusted ratios show no systematic bias with respect to 1 and their scatter
is greatly reduced. Figure~\ref{f:ampl_err} shows the root mean square (rms) 
of the scatter with respect to 1 as a function of the position offset 
before adjustment (green hollow circles) and after adjustment (blue filled 
points). Amplitude errors at the offsets that reduce the primary beam power 
by a factor of 2 ($5'$ at 4.5~GHz and $3'$ at 7.4~GHz) are 12\% before 
adjustment for pointing offsets. After adjustment for pointing corrections,
these errors are reduced to a level of 3\%. It should be noted that on average, 
pointing errors reduced the amplitude in the in-beam direction by 3\% at 
7.4~GHz. {A 30\% increase in the total flux density estimate at pointing
offset 
is consistent with expected amplitude errors shown in Figure~\ref{f:ampl_err},
and therefore, I consider pointing errors as its most probable cause .}
No correction for pointing offsets were made during the course of the WFCS 
campaign.

\subsection{Source selection}

   Even with the use of the wide-band VLBI technique, the VLBI field of
view is still narrow. With rare exceptions, we observe with VLBI the sources 
detected with low resolution single dish or connected interferometer
instruments. Scheduling algorithms ingest the input list of sources. 
A priority is assigned to a given source. Usually, an input source list has 
more objects than one can observe within the allotted time. The higher 
the over-subscription rate, i.e. the ratio of time required to observe every 
source in the list to the amount of the allotted time, the more efficient 
observing schedules the algorithm is able to generate because it has more 
candidates to choose from.

   There were three observing campaigns within the survey, VCS7, VCS8, and 
VCS9 that had different input source lists.

  The parent catalogue for VCS7 was generated using CATS 
database\footnote{\web{http://cats.sao.ru/}} \citep{r:cats} containing almost 
all radio catalogs known by 2013. Sources from these catalogues within $20''$ 
of their reported positions were matched against each other, and the spectral 
index was computed. The following sources  were included in the input list: 
a)~at declinations $>-45^\circ$; b)~with flux densities extrapolated at 8~GHz 
and greater 100~mJy; c)~with spectral index $> -0.55$; d)~that have no 
planetary nebulae or HII region with $2'$; e)~that were not observed with 
VLBI before. The sources in the zone of low density of known VLBI calibrators 
had higher priorities. In addition, 151 sources that are known to have 
intra-day variability and listed in MASIV catalogue \citep{r:masiv08} and 
27~sources previously observed with VLBI that showed significant differences 
between VLBI and \Note{optical} catalogue NOMAD \citep{r:zacharias14} were added 
(V. Makarov, private communication). In total, 6554 objects were selected 
as candidates and 1486 were observed.

  The parent catalogue for VCS8 was also generated with the use of CATS 
database. The selection criteria were similar to those in VCS7, except
the extrapolated flux density limit was raised to 150~mJy. In addition, 
23 sources detected in prior observations with position accuracy worse 
than 25~mas were added to the schedule. In total, 5712 objects were 
selected as candidates and 1233 were observed.

  The parent catalogues for VCS9 were GB6 \citep{r:gb6} and 
PMN \citep{r:pmn1,r:pmn2,r:pmn3,r:pmn4,r:pmn5,r:pmn6,r:pmn7} at 4.8~GHz.
The input list included 20641 sources with declinations $>-40^\circ$ 
brighter 70~mJy, excluding those that were previously observed. The list
also contained 174 previously observed sources with significant 
differences between VLBI and \Note{optical} catalogue NOMAD, 414 sources 
candidates with $\gamma$-ray associations, and 183 sources
with poor positions. Of 20641 input sources, 10575 were observed.

\subsection{Observing schedules}

  Scheduling was made with the use of software program sur\_sked
in a totally automatic fashion. The VLBI schedule is the sequence of 
start and stop time when the array or its part called a sub-array records 
radio emission. This time interval is called a scan, and data collected 
at a given baseline at this interval are called an observation. 
Considering VLBA has baselines over 2/3 of Earth's diameter, generation 
of the optimal sequence of scans that satisfies a number of constraints 
is a highly non-trivial problem \citep[for more detail see][]{r:viesched}.

  First, the minimum elevation angle and the minimum number of required 
stations was determined. It was set to $15^\circ$ and 10 stations,
i.e. full VLBA network, for sources with declinations above $-10^\circ$
and it was gradually lowered to $7^\circ$ and 7 stations for sources
with declinations below $-41^\circ$ and $5^\circ$ and 6 stations for sources
with declinations below $-46^\circ$, although very few sources were
scheduled at declinations below $-40^\circ$. Otherwise, sources with low 
declinations would have little chance to be observed. Then the interval
of time when a given source is above the specified elevation limits at 
the minimum number of required stations was computed, azimuth and 
elevation angles were calculated, and expanded into the B-spline basis 
as s function of time for fast interpolation. Such sources are called 
visible at a given moment of time. After that, the sequence of 
observed sources was generated. The list of sources that are visible
was determined for the moment of time at the end of integration time
of the previous source. Slewing time was calculated for each visible
source. The following score was computed for every visible source:
\begin{widetext}
\beq
    s = \lp 0.5 \cdot \lp30 - \max\lp20,\delta\rp\rp^2 + \Frac{10^6}{1+t_s} + 
            \Frac{10^7}{1+t_c^4} \rp p, \qquad
\eeq{e:e12}
\end{widetext}
  where $\delta$ --- declination in degrees, $t_s$ --- slewing time in seconds, 
$t_c$ --- time elapsed since the upper culmination at {\sc pietown} station 
in seconds. The first term up-weights low declination sources, because they have
shorter visible time, the second term up-weights sources with short slewing 
time, and the third term up-weights sources that are near the meridian
at {\sc pietown} station, and therefore, are observed at higher elevations. 
The last term, p --- a priority, is a number assigned to a source to quantify 
its preference. The source with the highest score is put in the schedule, and 
the process is repeated. The scheduled source is excluded from further 
consideration. With some exceptions, each source was scheduled in one scan of 
60~s long. The average schedule efficiency defined as the ratio of the total 
on-source time for target sources to the total duration of an experiment 
is~0.51. If to account observations of calibrator sources, the schedule 
efficiency is~0.63. Remaining time is used for slewing. 

  Some sources were scheduled in more than one scan of 120 to 300~s long. These 
are the sources with large \Note{radio-optical} offsets and with the excessive 
number of outliers from analysis of previous observing sessions.

  The process of generating the sequence of sources is interrupted every hour,
and four calibrators are inserted. Calibrator sources are selected in such 
a way that at least two of them are observed at the elevation range $[15^\circ, 
35^\circ]$ and two of them are observed at the elevation range $[45^\circ, 
88^\circ]$ at each station. The minimum number of stations for calibrator 
observations was 6. The calibrator list consists of 323 compact sources with 
the correlated flux density at 2 and 8~GHz greater 0.3~Jy and with the ratio 
of the correlated flux density at projected baseline lengths over 5,000 km to
the total flux density above 0.5. A brute force algorithm was used for finding 
4~calibrators at a given moment of time first selecting those that have 
the maximum number of participating stations in each scan and then choosing 
among them the variant that has minimum slewing time. Each calibrator was 
observed for 60~s.

  The calibrators were included in order to a)~improve estimation of the 
atmospheric path delay in the zenith direction by including the observations 
at low and high elevation angles, which helps to separate variables; 
b)~connect positions of the new sources, never observed with VLBI before,
with frequently observed objects with precisely known positions, 
c)~provide observations with high SNR to compute complex bandpasses; 
d)~provide observations of strong sources which images can be determined 
with the high dynamic range for gain calibration.

  The scheduling algorithm picks up the sources to observe from the input 
list automatically. Assigning source priorities increases a chance for 
the sources with high priorities being included in the schedule by an expense
of decreasing the overall schedule efficiency. The priorities were increased 
for a)~the sources in the area with a low density of known sources detected with 
VLBI; b)~flat spectrum sources, i.e. sources with spectral index flatter 
than -0.5; c)~brighter sources; d)~sources of a special interest, such 
as MASIV, or possible counterparts of $\gamma$-ray object detected
with {\it Fermi}, or sources with large \Note{radio-optical} offsets.

  All three campaigns were scheduled in the so-called \Note{filler} mode. That means
the start time of observing sessions and their duration is not known beforehand. 
The array operator finds a gap between high priority projects that require 
good weather, enters the start and stop dates using the web form and within 
1--3~minutes gets an automatically generated schedule file. The minimum 
duration of the observing session is 3.5 hours. The advantage 
of this approach is that the project can be observed almost for free: it takes 
time that the array would be idle otherwise. This is the only practical way to 
acquire hundreds of hours of VLBA observing time for projects like that.
The first disadvantage of this approach is that there is no direct control 
which source will be observed. The only leverage the principal investigator
has is raising or lowering source priorities. That increases or decreases 
a chance of a given source to be included in the schedule. The second 
disadvantage is that the distribution of scheduled sources over right 
ascension depends on the scheduling pressure of the array. There are ranges 
in the right ascension, specifically near the Galactic plane, that have 
a higher demand from competing projects. Since the \Note{filler} projects are 
scheduled at the lowest priority, a chance to observe the sources that 
culminate in the right ascension ranges that are in a high demand are 
less. As a result, the distribution of scheduled sources is not uniform over 
right ascensions, although the distribution of the input catalogue is uniform, 
and it inherits the footprint of the VLBA subscription pressure.

  Table~\ref{t:obs} summarizes the statistics of observing schedules.

\begin{table*}
   \caption{The summary of the observing campaigns: 
            (1) --- the campaign id;
            (2) --- the number of sources in the input catalogue;
            (3) --- the number of new observed sources observed in this campaign
                    that have never been observed with VLBI before;
            (4) --- the number of known sources in the input catalogue;
            (5) --- the number of observing sessions;
            (6) --- the amount of observed time allotted in hours;
            (7) --- time interval of observations. Since some sources were
                    listed in several campaigns, the total may be less
                    than a sum.
           }
   \begin{tabular}{lrrrrrl}
            & \nnntab{c}{Number of sources}    &          &                   \\
      camp  &  inp   &  new  & known & \# seg  & obs time & time interval     \\
            &        &       &       &         & in hours &                   \\
       (1)  &  (2)   &  (3)  &  (4)  & (5)     &  \ntab{c}{(6)} & \ntab{c}{(7)}     \\
      \hline      
      vcs7  &  6554  &  1486 &  140  &  17 &  71.7 & Apr 2013 --- Aug 2013 \\
      vcs8  &  5712  &  1233 &  153  &  10 &  47.7 & Jan 2014 --- Feb 2014 \\
      vcs9  & 20641  & 10575 &  433  &  99 & 536.0 & Aug 2015 --- Sep 2016 \\
      \hline      
      all   & 27609  & 13154 &  491  & 126 & 655.4 & Apr 2013 --- Sep 2016 \\
   \end{tabular}
   \label{t:obs}
\end{table*}

\section{Data analysis}

   Raw data were processed with DiFX \citep{r:difx1,r:difx2} correlator.
The spectral and time resolutions was 62.5 KHz and 100~ms respectively. These 
settings are 4 times finer for the spectral resolution and 20 times finer for 
the time resolution than typical values. That choice has increased the dataset 
size by a factor of 80 and increased computation time at least by a factor of
$4 \log_2 4 \times 20 \log_2 20 \approx 700$ with respect to  a commonly used 
correlation setup. The correlator is able to provide even finer resolution, and 
therefore, wider field of view. The choice of the resolutions was dictated 
but the available computer resources that the author was able to secure: 
a 20-core Xeon E5-2660-v3. Processing a 48~Tb dataset with 580,626,235 
visibilities required 12 years of CPU time per single core. Because some 
experiments have to be reprocessed due to errors discovered at latter 
stages during quality control, fringe fitting took one and half years 
to finish.

  The correlator provides the cross- and auto-spectrum of visibilities,
the spectrum of phase calibration signal, and the coefficients of the
polynomials for the a~priori model used during correlation. The output
dataset also contains system temperature and atmospheric parameters.

\subsection{Post-correlator analysis}

  I used a custom-designed software package \PIMA \citep{r:vgaps} for 
post-correlator processing. The goals of the post-correlator analysis 
are a)~to determine group delay and  phase delay rate for each observation 
as well as realistic uncertainties of these quantities for their subsequent 
use for absolute astrometry; b)~determine time and frequency averaged 
visibilities for their subsequent use for imaging.

  The data analysis pipeline has a number of steps.

\begin{enumerate}
  \item Automatic examination of the dataset, splitting the data into scans, 
        indexing, cross-referencing, \Note{discarding bad and orphaned
        data, i.e. visibilities without a header, time-tag, corresponding
        auto-correlation, or phase calibration.}

  \item Editing phase calibration signal. The VLBA injects narrow-band
        impulses in the feedhorn. Its spectrum in a form of a rail with
        a step of 1~MHz is computed by the correlator. Since the 
        phase-calibration unit and other parts of the VLBI hardware are
        fed by the signal from the same Hydrogen maser distributed at 1~MHz, 
        harmonics of the 1 MHz signal are spilled over and interfere 
        with the phase calibration signal. However their impact on a wide-band 
        signal from observed sources is negligible with some exceptions. 
        The phase-calibration signal has to be edited before use by
        removing spikes in the phase of the spectrum and masking out tones 
        at the IF edges. The VLBA IF filter has a significant spillover to 
        adjacent IFs. This reduces the fringe amplitude at the edges but also 
        causes an interference of the phase calibration signals from 
        adjacent IFs. Figure~\ref{f:pcal} demonstrates typical phase 
        calibration signal. 

\begin{figure}
  \includegraphics[width=0.495\textwidth]{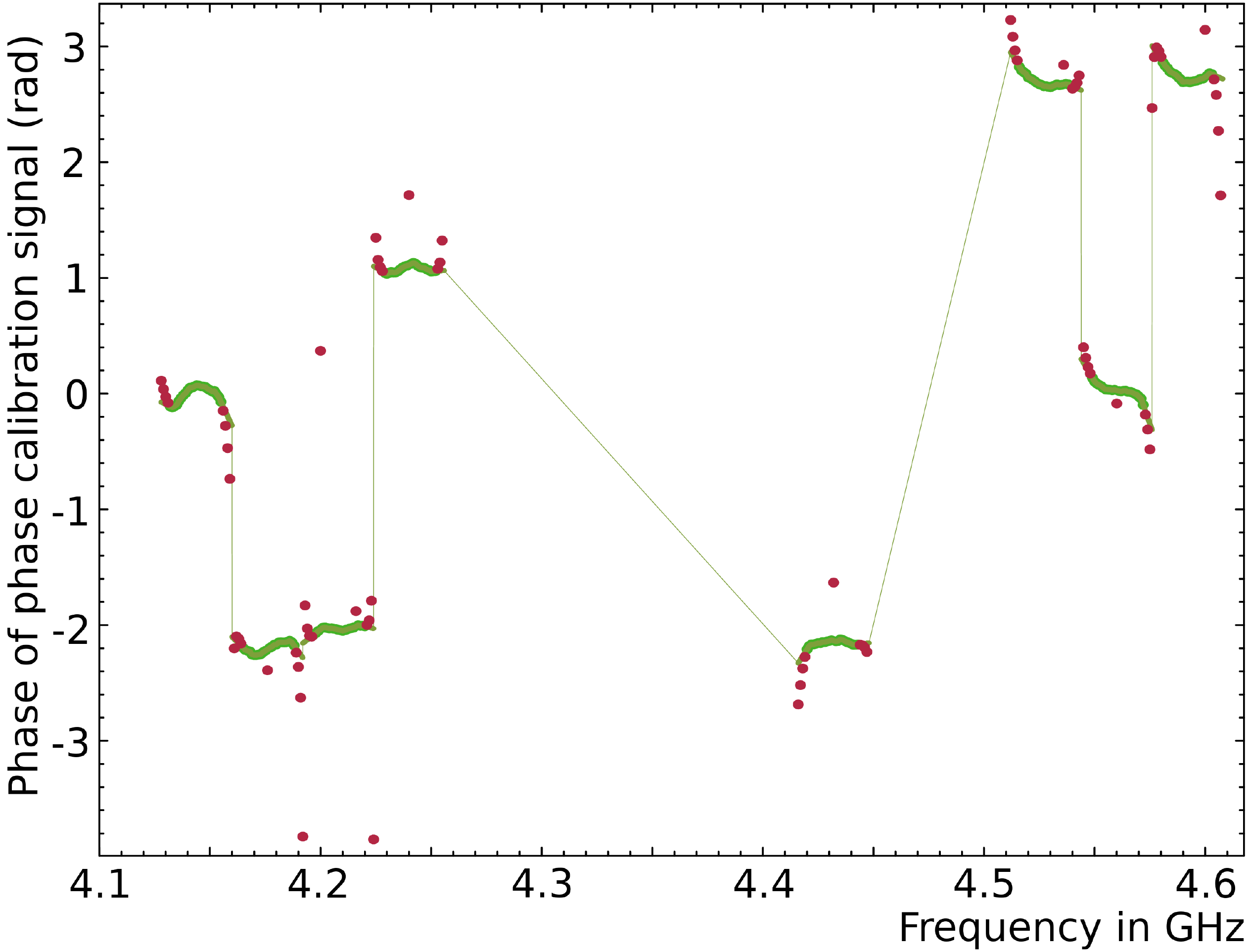}
  \caption{The phase of the phase calibration signal at station {\sc br-vlba} 
           during experiment BP192J8 on 2016 Sep 07 after fitting for 
           group delay, its removal, and unwrapping phase ambiguities. 
           Red points denote the tones that were masked out because they 
           are affected by internal interference.
           }
  \label{f:pcal}
\end{figure}

        After removal of phase calibration tones affected by the internal 
        interference, the phase of the phase calibration signal spectrum is 
        interpolated and/or extrapolated across each IF and applied to 
        data, i.e. subtracted from fringe phases.

  \item Coarse fringe fitting. A simplified procedure of fringe fitting
        is performed, without oversampling and without LSQ refinement. 
        The goal of the coarse fringe fitting is to get a set of 
        observations with high SNR at each baseline. Fringe fitting is 
        preformed for 4.3 and 7.6~GHz bands separately.

  \item Computation of the complex bandpass. Deficiencies in implementation
        of phase calibration does not allow us to restore precisely the 
        amplitude and phase characteristics of the VLBI signal chain. 
        Assuming the spectrum of observed sources within 480~MHz bandwidth 
        is flat, we can derive the phase and amplitude response by processing
        a number of strong sources with a high SNR. First, the algorithm runs
        fine fringe fitting for 12~sources with the highest SNR at all 
        baselines with a certain station taken as a reference. Then it 
        subtracts the contribution of phase delay and phase delay rate, 
        coherently averages over time and over 4~adjacent spectral frequency 
        channels in order to improve the SNR in each spectral bin, and then 
        normalizes the amplitude for the integral over frequency at each IF 
        to be equal to unity. Then the algorithm solves for the station-based
        complex bandpass $B_i(f)$ using LSQ by fitting the residual phases 
        and logarithm of the normalized amplitudes in a form of a B-spline 
        expansion with 9~knots uniformly distributed over frequency within 
        each IF that relates the observed visibility 
        $V^{\rm obs}_{ij}(f)$ and the ideal signal through
        $\Pi(f)$: $V^{\rm obs}_{ij}(f) = B_i(f) \, B_j^{\star}(f) \, \Pi(f)$ 
        relationship.
        
        Some spectral channels are masked out, i.e. corresponding visibilities are
        replaced with zero. These are the channels at the edge of IFs, channels 
        at 4.2 and 7.8~GHz that coincides with the synthesizer frequencies, and
        channels affected by the RFI. Occasionally, entire IFs have to be masked
        out due to hardware failures.

  \item Fine fringe fitting. The fringe fitting was repeated with the oversampling 
        factor 4, with the digitization calibration, phase calibration, and phase 
        bandpass applied. Group delay, phase delay rate, and group delay rate 
        were refined using LSQ in the vicinity of the main maximum of the 2D 
        Fourier transform with additive reweighting applied using all visibility 
        data of a given observation. See \citet{r:vgaps} for a detailed explanation 
        how this is done. 
        
  \item First astrometric solution. Results of fringe fitting are exported in 
        a database in geoVLBI format\footnote{See format specifications at 
        \web{http://astrogeo.org/gvh}} and ingested by the VLBI astrometry/geodesy 
        data analysis software pSolve\footnote{\web{http://astrogeo.org/psolve}}. 
        The first problem is to filter out 
        observations with no signal detected. Since on average, the signal was 
        detected only in 45\% observation, LSQ will not work unless the fraction 
        of observations with non-detections is reduced to a manageable level, 
        $< 10$\%. Figure~\ref{f:snr} shows the SNR distribution. The observed 
        distribution is a superposition of two very distinct partially 
        overlapping distributions: the distribution of detections (blue line) 
        that gradually increases towards low SNR because the number of weaker 
        sources is greater than the number of strong sources and the 
        distribution of non-detections (red line) that peaked at SNR=4.8. 
        These two distributions can be separated, fitted, and integrated. The 
        cumulative normalized distribution of non-detection immediately
        provides us an estimate of the probability of a false 
        detection \citep{r:vera_kcal}. the probability of false detection 
        at the SNR=6.0 is 0.0016. The probability reaches 0.1 at the 
        SNR=5.25 and 0.3 at the SNR=5.01. Therefore, we can consider that 
        observations with the SNR $<$ 5  are non-detections, with the SNR 
        $>6$ are mostly detections, and with the SNR from 5 to 6 are in 
        a transition zone. It should be noted that the false detection 
        probability depends on the search window size.

\begin{figure}
  \includegraphics[width=0.495\textwidth]{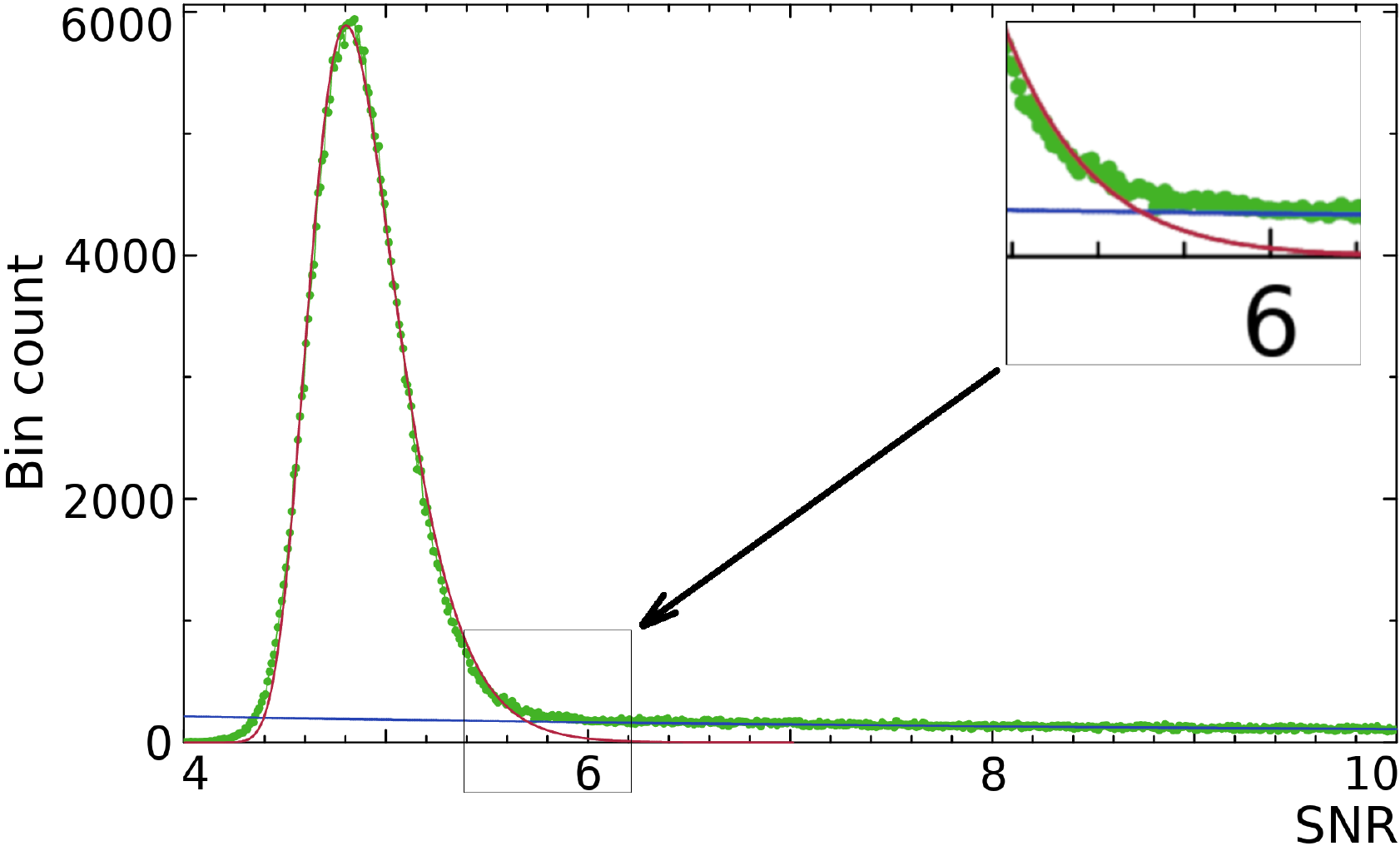}
  \caption{The SNR distribution. The green points show the observed 
           distribution. The blue line shows the distribution of the SNR 
           among detected observations. The line is extrapolated to 
           SNR=4.0. Red points show the distribution of non-detections.
           }
  \label{f:snr}
\end{figure}

        At the first step, the observations with the SNR $< 6.0$ are temporarily 
        suppressed, i.e. excluded from consideration. The positions of the 
        sources without prior astrometric VLBI observations are estimated using
        LSQ, as well as clock functions for all stations except the one taken as
        a reference. Then the automatic procedure for outlier elimination
        is executed: the observations with the largest normalized residuals are
        suppressed, the solution is updated, and the procedure is iterated 
        till no observations with normalized residuals greater $4.5 \sigma$ 
        remained. Although the mathematical expectation of the number of 
        non-detected sources with SNR $> 6.0$ among 5000 observations 
        (a typical number of sources in each individual observing session) 
        is 8, group delays may be wrong for a greater number of sources for 
        a variety of reasons: interference, phase instability due to hardware
        malfunction, etc. The sources with less than three remaining 
        observations are considered as non-detected, because one can always
        solve for two parameters using two observations and get zero residuals.
        Postfit group delays residuals of detected sources have the Gaussian 
        distribution with $\sigma$ 30--70~ps, while group delays derived from 
        the noise have the uniform distribution within the search window 
        [-8000, 8000]~ns. Considering that among detected observations the 
        typical rms of postfit residuals is 60~ps, the probability to find 
        residuals $< 4.5 \sigma$ among three observations with at least one 
        non-detection is $3 \times 4.5 \times 0.06/8000 \approx 10^{-4}$. 
        This estimate shows that a requirement for a source to have at least 
        3~observations with postfit residuals less than $< 4.5 \sigma$ is 
        a very powerful filter.

        The sources with more than three observations with SNR $>6$, but with
        less than three observations remaining after outlier elimination are 
        re-examined. It may happen that the outlier elimination  process 
        removed good observations but one or two bad observations affected
        by the RFI were kept. Examination of fringe phase residuals allows 
        us to identify the observations with a certain systematic pattern,
        flag them out, and the rerun the outlier elimination process for 
        the remaining sources from the very beginning. This usually fixes 
        the problem and allows us to restore the observations that were 
        incorrectly flagged out. Then the SNR limit is reduced to 5.9, 5.8, 
        and 5.7, and re-examination process is repeated. After that, 
        the parametric model is expanded, and estimation of residual 
        atmospheric path delay in the zenith direction at each station 
        is included.
     
        At the next step, a procedure reciprocal to the outlier elimination
        runs. It examines the suppressed observations with the SNR $> 4.8$, 
        finds the one that has the minimum normalized residual, flips the 
        suppression flag, updates the solution and repeats iterations till 
        the minimum normalized residual reached $4.5\sigma$. Since the presence
        of outliers distorts parameters estimates and residuals, the procedure 
        of the outlier elimination and restoration has to be repeated 2--3 
        times to reach convergence.

  \item The second fringe fitting run. The source positions determined in the 
        previous step are used as new a~priori. The non-linear term in phases
        due to large differences between the actual source positions and the
        positions used for computation of the correlator delay is evaluated.
        The visibilities are phase-rotated to cancel the contribution of the 
        quadratic term in phases, and the fringe fitting process is repeated.
        Re-fringing improves the SNR of the sources with large corrections to 
        their initial a~priori positions. The results of fringe fitting were 
        transformed to a database, but the previous flags were preserved.

  \item Second astrometric solution. The procedure for outlier elimination and
        restoration of the previously suppressed observations is repeated
        starting with the flags preserved in the previous astrometric solution.
        This time no limit on the SNR is imposed.
  
  \item Third fringe fitting run. The fringe fitting procedure is repeated for 
        all suppressed observations. Using results of the second astrometric 
        solution, expected group delays are computed. The fringe search window 
        has the semi-width 2~ns at 4.3~GHz and 1.3~ns at 7.6~GHz and is 
        centered at the expected values of group delay. The observations with
        the SNR $> 4.8$ are selected and added to the group delay dataset.
        This SNR limit is significantly lower since the search window is much 
        narrower and the probability that the peak in the narrow window with
        the SNR greater 4.8 could be found by chance is lower. Here information
        from another observations is utilized, what allows us to detect 
        sources with a weaker SNR. This procedure also often helps to recover
        observations affected by the RFI.
        
  \item Third astrometric solution. The procedure for restoration of previously 
        suppressed observations is repeated. The observations with normalized
        residuals less than $4.5\sigma$ are retained. Finally, the ionosphere-free 
        combinations of group delays of those observations that are detected 
        at both bands are formed. The algorithm runs a new LSQ solution using
        the ionosphere-free combinations. The procedure of outlier elimination 
        and restoration of previously suppressed observation is repeated till 
        no outliers exceeding $3.5\sigma$ or no suppressed observations with 
        normalized residuals less than $3.5\sigma$ remains. The $n \sigma$ 
        criteria is lowered for the dual-band solution since the contribution 
        of the additional noise due to the ionosphere is eliminated.
      
\end{enumerate}

\subsection{Detection statistics}

   Of 13,154 target sources, 6755 or 51\%, have been detected at at least
one band. Among 491 calibrators sources, 465 have been detected.
\Note{Table~\ref{t:det_dist} shows the cumulative distribution of the
distance of the detected sources from the a~priori position.}
  
\begin{table}[h]
   \caption{The share of sources detected at a given distance from the 
            a~priori position.}
   \begin{tabular}{rl}
      \hline
      50\%  & $>$ 10 arcsec \\
      20\%  & $>$ 23 arcsec \\
      10\%  & $>$ 36 arcsec \\
       5\%  & $>$ 59 arcsec \\
       1\%  & $>$ 207 arcsec \\
      \hline
   \end{tabular}
   \label{t:det_dist}
\end{table}

\begin{figure*}
  \includegraphics[width=0.495\textwidth]{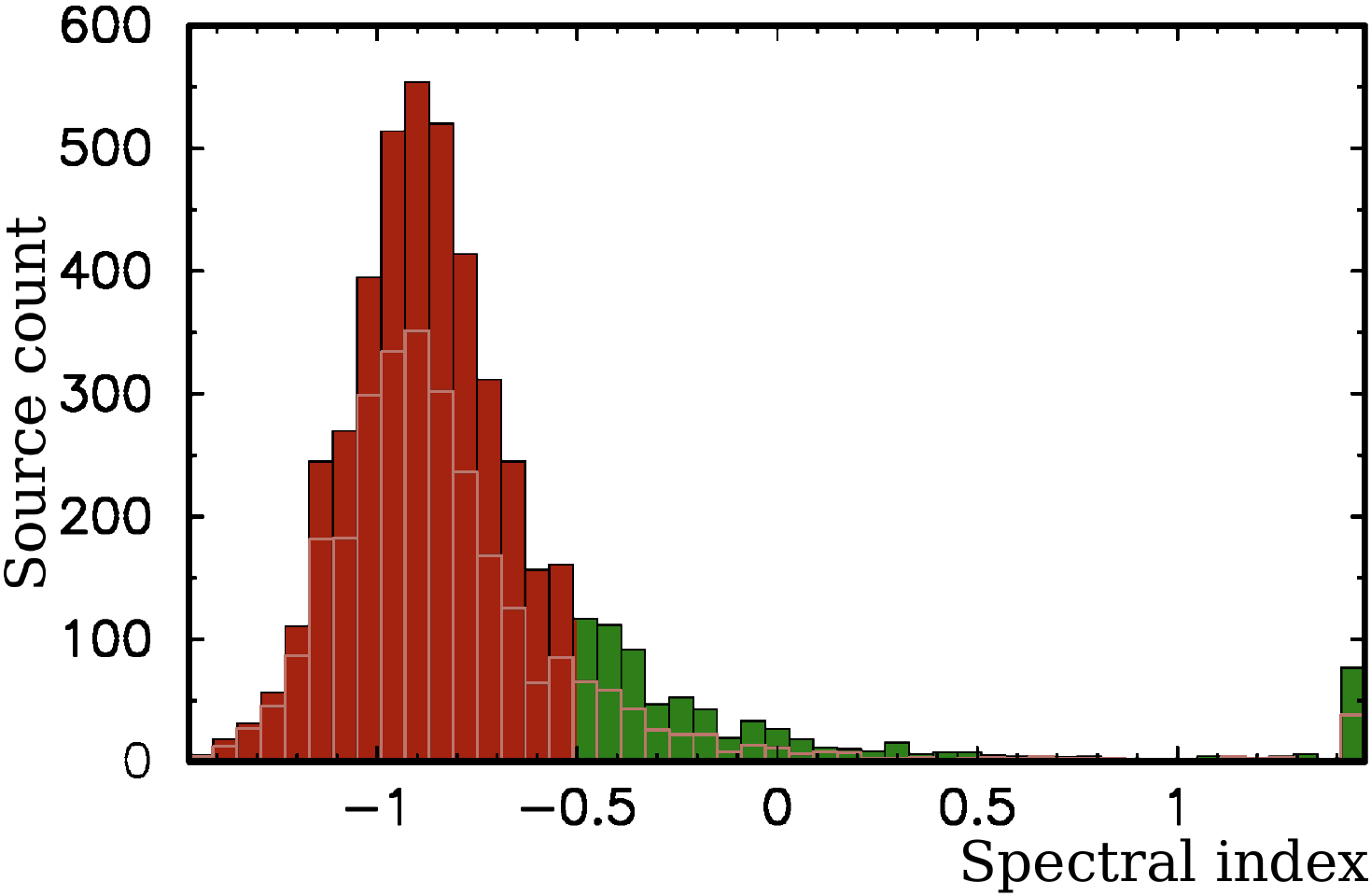}
  \includegraphics[width=0.495\textwidth]{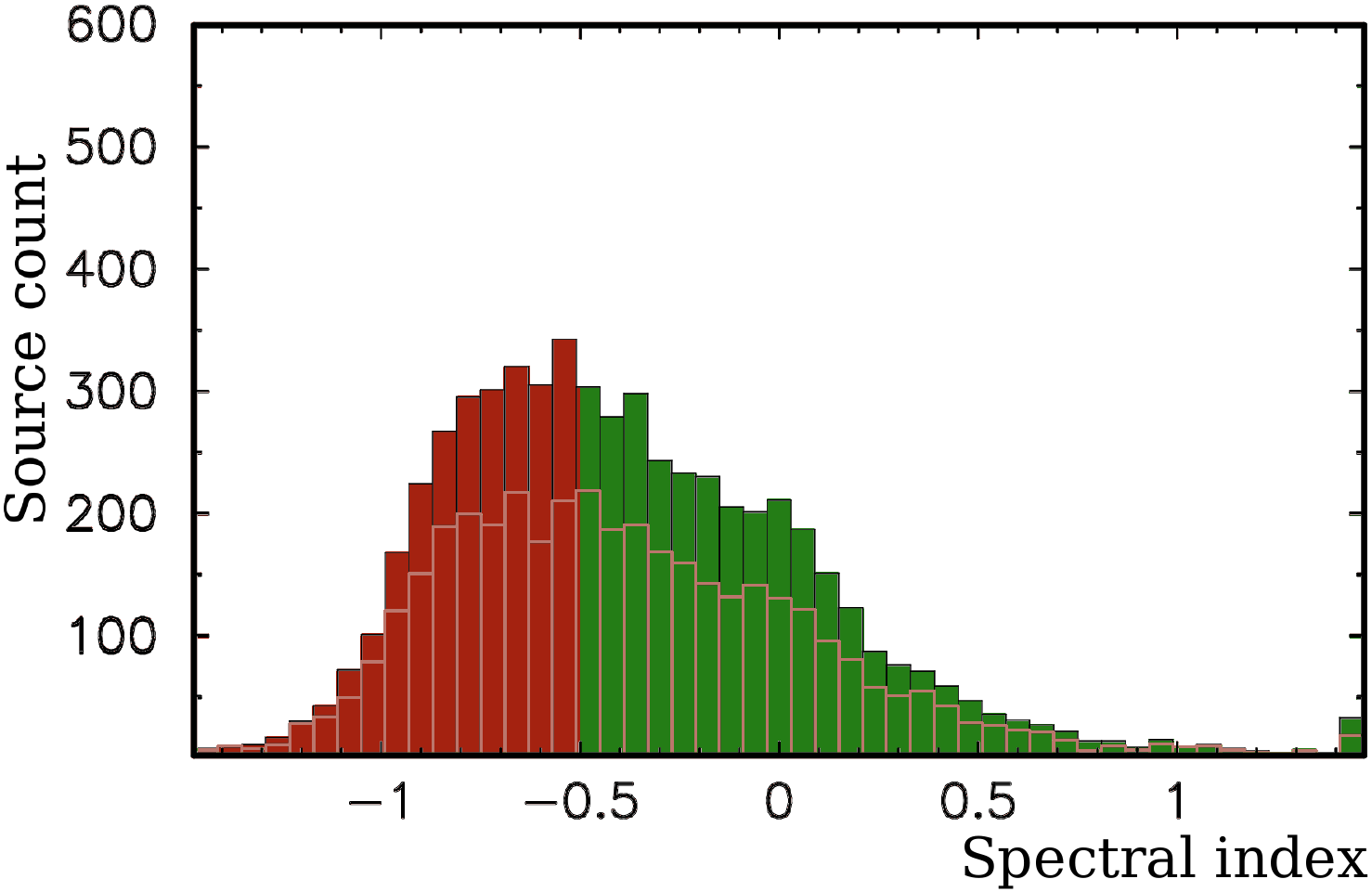}
  \caption{The distributions of observed sources over the spectral
           indices at kiloparsec scale among those which have not been
           detected (left) and among those that have been detected.
           Red color shows the steep spectrum source population 
           (spectral index $ < -0.5$). Green color shows the flat 
           spectrum population (spectral index $ > -0.5$).
           The filled boxes show the spectral histogram built
           using both PMN, GB6, and NVSS  catalogues. The light rose 
           line shows the histograms built using only GB6/NVSS data.
           }
  \label{f:spind_hist}
\end{figure*}

   I computed the spectral indices of the emission at kiloparsec scales 
between 4.85 and 1.4~GHz by cross-matching GB6 and PMN catalogues against 
NVSS catalogue. \Note{If a given source was observed in both GB6 and PMN, 
GB6 was used}. The distributions of spectral indices for detected 
and non-detected sources are shown in Figure~\ref{f:spind_hist}.

  Among detected sources, the share of steep spectrum objects is 42\%, 
while among non-detected is 83\%. The detection rate among observed flat 
spectrum sources is 79\% and among observed steep spectrum sources 
is 35\%. The detection rate drops to 20\% among the sources with the 
spectrum index steeper than -1.0. However, a caution should be taken 
in interpreting these numbers since the input sample is neither complete 
over the flux density nor over the spectral index.

  \Note{GB6, PMN, and NVSS catalogues were observed in different epochs:
in 1986--1987, 1990, and 1993--1996 respectively. Source variability 
affects spectral index estimates made from flux densities measured at
different epochs. Although as a comparison of 544 detected sources 
that are present in both GB6 and PMN catalogues show that the spectral
indices derived from GB6/NVSS and PMN/NVSS cross-matching indeed differ,
the impact of these differences on the distributions is negligible.
The Kolmogorov-Smirnov test shows the spectral index distributions 
formed from a subset of data GB6/NVSS and PMN/NVSS belong to the same
parent distribution (p-value = 0.53)}.

\subsection{Final astrometry analysis}

  I ran three independent global least square solutions. They differ
by the observables used. The experiment list consisted of two parts:
basic and specific. The basic part contains all observations collected 
under all VLBI geodesy programs, the VLBA regular geodesy program RDV, 
and VCS1-6. The specific part contains 4.3~GHz group delay observables for 
wfcs\_c solution, 7.6~GHz group delay observables for wfcs\_x solution, 
and ionosphere-free linear combinations of 4.3 C and 7.6~GHz observables 
for wfcs\_xc solution.

  Estimated parameters are split into three categories. Global parameters 
included station positions, station velocities, parameters of station non-linear 
motion, and source coordinates. They are estimated using all the data.
Session-wide parameters included pole coordinates, UT1 angle, their time 
derivatives, nutation angle offsets, baseline clock offsets, and clock 
breaks for some stations. They are estimated using data from each
observing session. Segment-wide parameters included the atmospheric path 
delay in zenith direction and the clock function for each station. They are 
modeled with a B-spline with the time span of 1 hour with constraints imposed 
on their rate of change with reciprocal weights 40~ps/s and $2 \cdot 10^{-14}$ 
for the atmospheric path delay rate and the clock rate respectively. 

  The estimated parameters include sine and cosine coefficients of 
the harmonic site position variations at diurnal, semi-diurnal, annual and 
semi-annual frequencies to take into account residual mass loading, thermal 
variations, and systematic errors in modeling the atmospheric path delay. 
In addition, parameters of B-spline with multiple nodes were estimated for 
some stations to account for co-seismic crustal deformation ({\sc mk-vlba} 
station) and local motion of the antenna foundation 
\Note({\sc pietown} station, see \citet{r:rdv} for details)).

  No-net-translation constraints were imposed on station positions and 
velocities and non-net-rotation constraints were imposed on station 
positions and velocities as well as source coordinates to find the solution
of the system of equations of the incomplete rank. In particular,
the new adjustments of the so-called 212 defining sources listed in the 
ICRF1 catalogue are required to have zero net rotation with respect to the 
positions reported in that catalogue.

  Weights associated with observations are computed the following way:
\beq
    w = \frac{1}{k \cdot \sqrt{\sigma_g^2 + a^2 + b^2(e)}},
\eeq{e:e13}
   where $\sigma_g$ is the group delay uncertainty, $k$ is the multiplicative 
factor, $a$ is the elevation-independent additive weight correction, and $b$ 
is the elevation-dependent weight correction. This form of weights accounts
for the contribution of systematic errors. I used $k=1.3$ based on the
analysis of VLBI-Gaia offsets \citep{r:gaia4}. Additive parameter $a$ was 
found by in iterative procedure that makes the ratio of the weighted sum 
of postfit residuals to their mathematical expectation close to unity. 
Parameters $b(e)$ were computed differently for dual-band and single-hand 
solutions. 

  I used $b(e)$ in a form of
$b(e)^2 = \beta \, (\tau(e_1)_{\rm atm,1}^2 + \tau(e_2)_{\rm atm,2}^2)$,
for processing dual-band observations, where $\tau(e_i)_{\rm i,atm}$ is the 
atmospheric path delay at the $i$th station. I set $\beta$ to  0.02 since it 
minimizes the baseline length repeatabilities (See \citet{r:rdv} on details 
how such tests are performed).

  I computed the contribution of the ionosphere to group delay utilizing the 
Total Electron Contents (TEC) maps derived from analysis of Global Navigation 
Satellite System (GNSS) observations and used them for processing single band 
observations. Specifically, CODE TEC time series 
\citep{r:schaer99}\footnote{Available at 
\href{ftp://ftp.aiub.unibe.ch/CODE}{ftp://ftp.aiub.unibe.ch/CODE}} with
a resolution of $5^\circ \times 2.5^\circ \times 2^h$ were used. However, the 
TEC maps account only partially for the ionospheric path delay due to the 
coarseness of their spatial and temporarily resolution. I have developed 
a simplified stochastic model in order to account for the impact of the 
mismodeled ionosphere contribution to positions derived from processing of 
single-band data.

  I collected statistics of the differences between the ionospheric 
contribution to group delays at 4.3~GHz derived from VLBI data, 
$\tau_{\rm v,iono}$, and derived from GNSS TEC maps, $\tau_{\rm g,iono}$. 
For each experiment and each baseline I computed the weighted mean 
$\tau_{\rm m,iono}$ and the root mean square (rms) of the zenith residual 
ionospheric contribution as 
$(\tau_{\rm g,iono} - \tau_{\rm v,iono} - \tau_{\rm m,iono})/{\tilde{M}(e)}$,
where $\tilde{M}(e)$ is the arithmetic mean of the ionospheric mapping 
functions over two stations at a given baseline. Then I computed the expected 
variance of the residual ionospheric contribution for each observation as
$\sigma_{\rm res,iono} = \mbox{rms} \cdot \tilde{M}(e)$. That variance was 
added in quadrature to parameter $a$ computed by the reweighting procedure. 
The additive variance was re-scaled by the $(4.3/7.6)^2$ factor when 
applied to group delay weights at 7.6~GHz.

  I compared 4.3~GHz and 7.6~GHz only solutions against the solution that used 
the ionosphere-free combinations of observables. First, I modified source 
position uncertainties $\sigma_0$ by adding in quadrature an ad~hoc term 
$l_r = 0.05$~mas to account for systematic errors. This quantity was derived 
from the decimation test: the dataset was divided into two equal subsets and 
the differences in source positions derived from the subsets were analyzed.

\begin{table*}[h]
  \caption{The parameters of the uncertainty re-scaling law and the position 
           offsets in right ascension and declination for the 
           4.3~GHz and 7.6~GHz solutions with respect to the dual-band
           solution. 
           }
  \begin{tabular}{l rrr @{\qquad} rrr}
      \hline
      Freq    & \nnntab{c}{RA}            & \nnntab{c}{DEC}  \\    
              & \ntab{c}{l}   & \ntab{c}{s} & \ntab{c}{off}      
              & \ntab{c}{l}   & \ntab{c}{s} & \ntab{c}{off}  \\
      GHz     & \ntab{c}{mas} &             & \ntab{c}{mas} 
              & \ntab{c}{mas} &             & \ntab{c}{mas}  \\
      \hline
      4.3 & 0.160 & 1.259 &  0.107 & 0.271 & 1.299 & 0.003    \\
      7.6 & 0.018 & 1.005 &  0.055 & 0.151 & 1.149 & 0.041    \\
      \hline
  \end{tabular}
  \label{t:sig_adj}
\end{table*}

\begin{figure*}
  \includegraphics[width=0.495\textwidth]{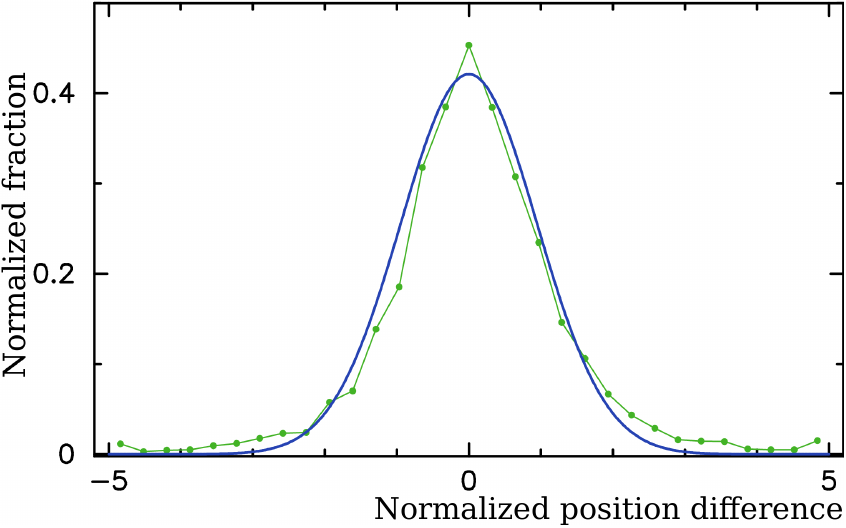}
  \includegraphics[width=0.495\textwidth]{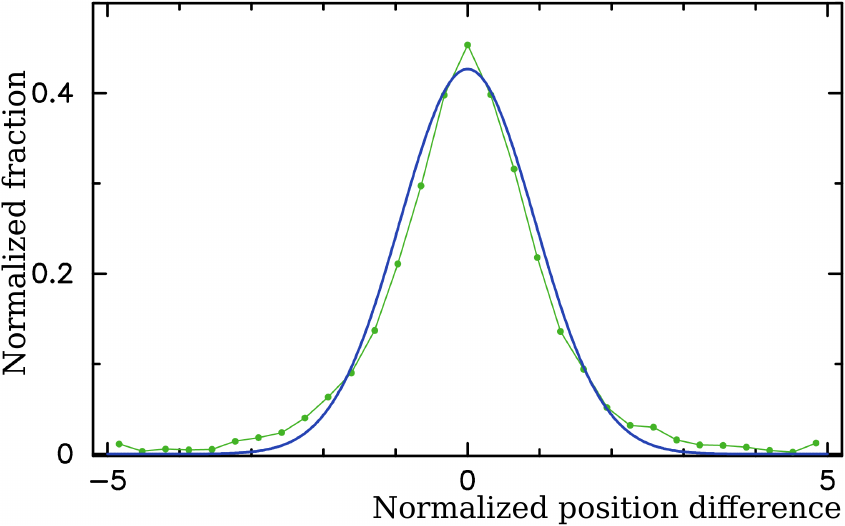}
  \caption{The distributions of the normalized position differences between
           the 4.3~GHz and the dual-band solutions after their
           re-scaling (green connected points). The blue lines show 
           the Gaussian distribution with $\sigma=1$ as a reference.
           \note{The left plot shows the differences over right ascensions 
           scaled by $\cos\delta$ and the right plot show the differences 
           over declinations.}
           }
  \label{f:pos_dif}
\end{figure*}

  I divided the position differences of 4.3 and 7.6~GHz solutions by the 
maximum between dual-band and single-band formal uncertainties $\sigma_m$ and 
computed their distributions. I fitted them into the Gaussian 
distribution and adjusted the mean position offset and two parameters of the 
formal uncertainty re-scaling law, $l$ and $s$, in a form 
$\sigma_a = \sqrt{l^2 + (s \, \sigma_m)^2}$. The results of the fit are shown 
in Table~\ref{t:sig_adj} \note{and the plots of normalized position 
uncertainties over right ascension and declinations from the 4.3~GHz solution 
is shown in Figure~\ref{f:pos_dif}}. The position offset was subtracted from 
all single-band position estimates. The original source position formal 
errors from the 4.3~GHz group delay solution were re-scaled as
\beq
     \sigma_c = \sqrt{ l_r^2 + l_c^2 + (s_c \, \sigma_0)^2}.
\eeq{e:14}
Formal errors from the 7.6~GHz group delay solution were re-scaled in 
a similar way. The final catalogue contains positions from the solution
that provided the least semi-major error ellipse axes for a given source.

\section{Imaging analysis}

   The same visibility data were used for imaging. Astronomers often spend
hours to get a VLBI image manually. Considering that the project involved
generating over 15,000 images, manual imaging was not an option. Therefore,
an automated pipeline has been developed. The imaging procedure includes 
a number of steps: cleaning system temperature, computing the a~priori 
gain, re-calibration, flagging visibilities when the antennas were off-source,
averaging the data over time and frequency, generating images using 
hybrid self-calibration, and computation of the total and median flux 
densities at long and short baselines.

\subsection{Amplitude calibration}

  The VLBA hardware measures system temperature ($T_{\rm sys}$) every 30~s.
It characterizes the antenna power in the absence of a signal. In addition, 
the NRAO periodically measures antenna gains and makes results of these
measurements available. The ratios of $T_{\rm sys}$ to gain would be sufficient 
for amplitude calibration if they were perfect. 
However, the measured $T_{\rm sys}$ often suffers from RFIs that manifest 
themselves as spikes in time series, overflows due to hardware problems, and 
misses values for some observations. Therefore, the raw $T_{\rm sys}$ data 
are to be processed in order to clean them. I developed two procedures called 
{\sf if-clean} and {\sf tmod-clean}. 

  The {\sf if-clean} procedure assumes the ratio of $T_{\rm sys}$ between IFs 
within a given band is kept the same since the data come from the same receiver 
and are transferred to the data acquisition system using the same cable. 
First, the algorithm determines the reference IF that has least missing data 
and spikes. Second, the time series of logarithms of $T_{\rm sys}$ ratios
with respect to the reference IF is computed, and an iterative procedure 
discards the values exceeding $3 \times\!\!$ rms. Then the ratios 
$r_{\rm \ij}$ between a given IF and the reference IF are computed for all 
IF combinations. Third, I compute a substitute for discarded values of 
$T_{\rm sys}$ if $T_{\rm sys}$ at at least one IF for the same moment of time 
is not flagged. The $T_{\rm sys,k}$ substitute at station $k$ is determined 
as the geometric mean of the products of $T_{\rm sys,i} \, r_{ik}$.

  The {\sf if-clean} procedure cannot recover $T_{\rm sys}$ if all IFs are 
affected. Since $T_{\rm sys}$ has a strong elevation dependence, and 
observations at adjacent scans are made at substantially different 
elevations, direct interpolation will not work, and a more sophisticated 
procedure is required.

\begin{figure*}
  \includegraphics[width=0.495\textwidth]{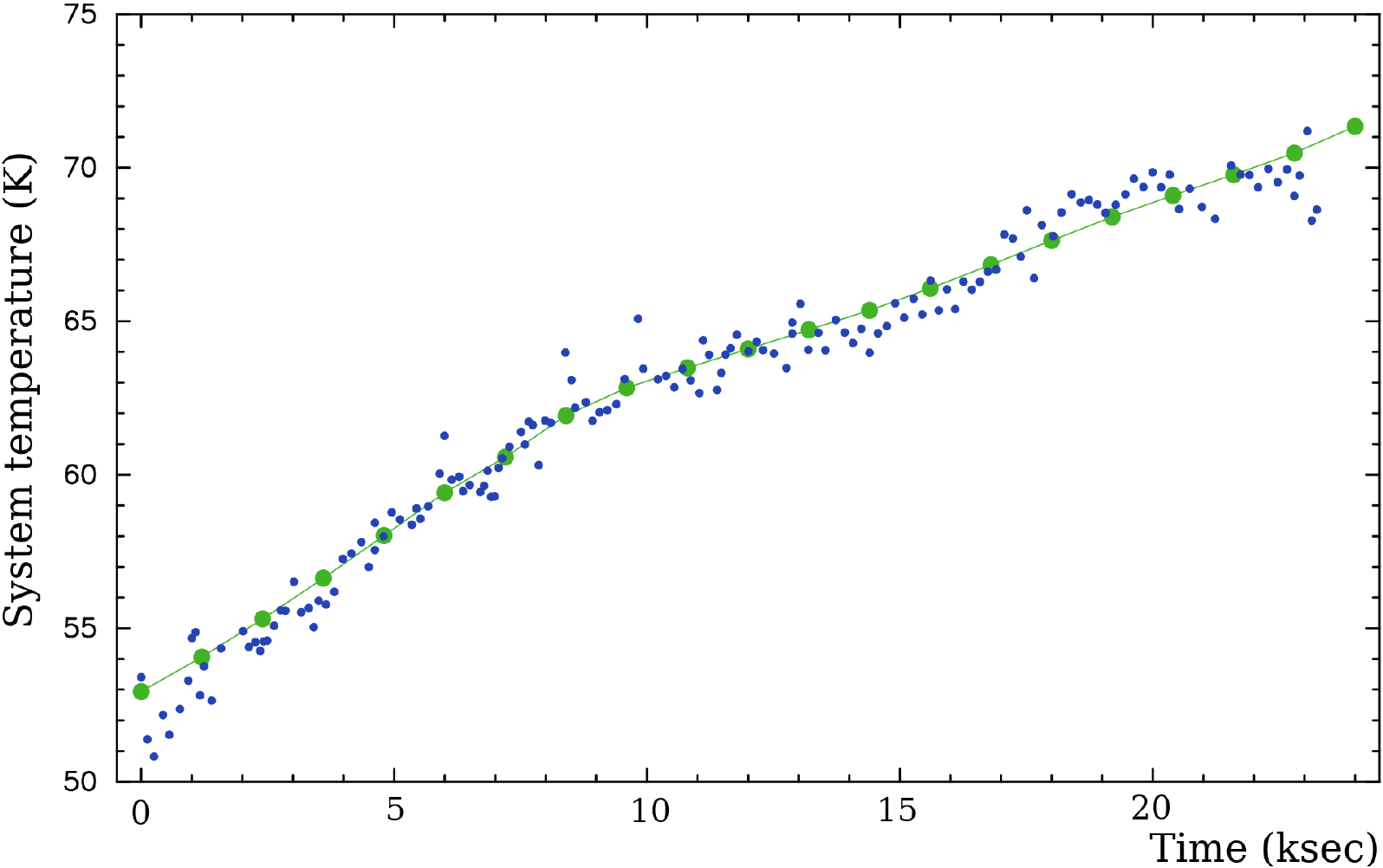}
  \includegraphics[width=0.495\textwidth]{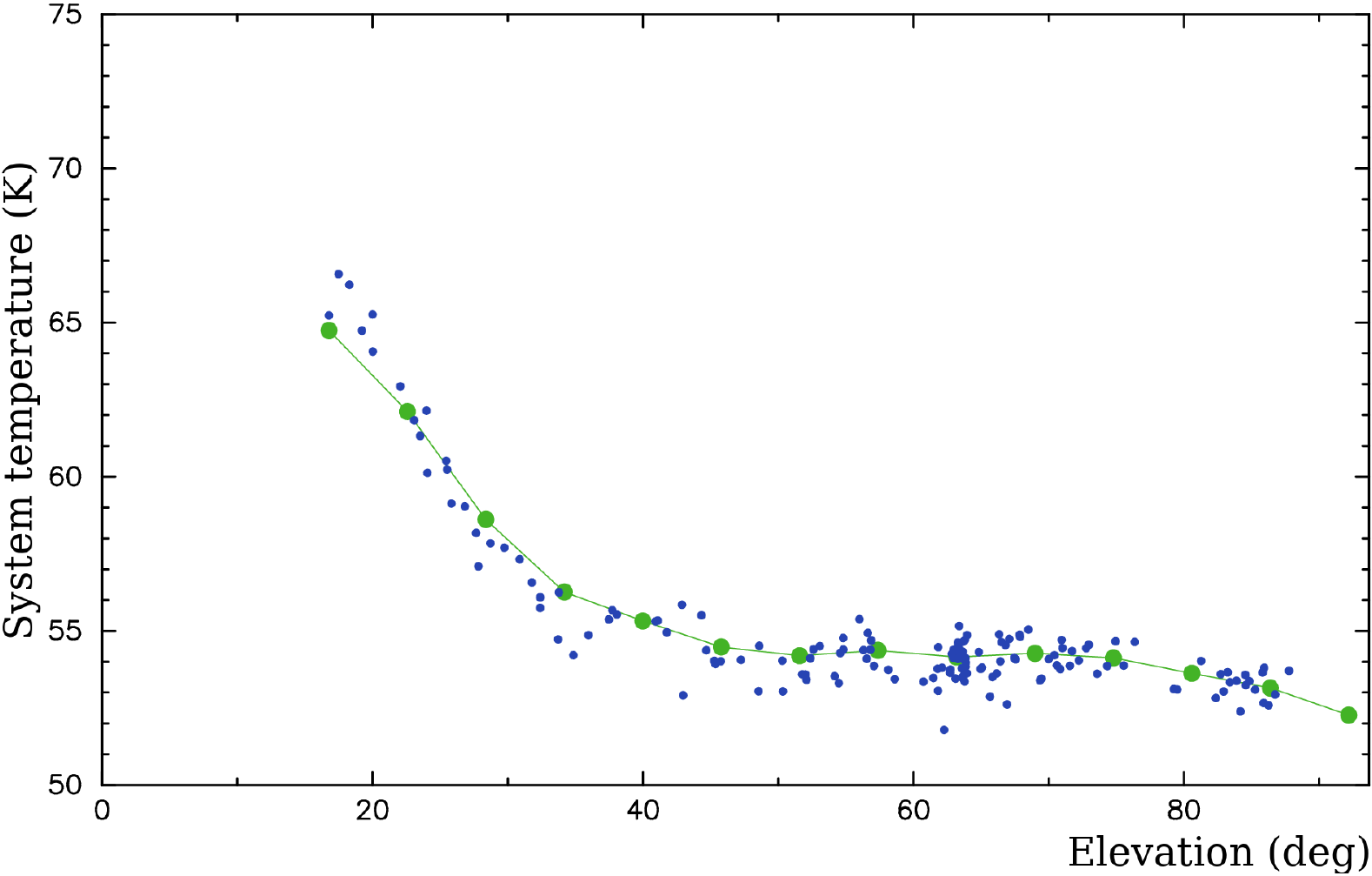}
  \caption{The result of the system temperature decomposition on time and 
           elevation dependence at station {\sc fd-vlba} for
           experiment BP192J8 that ran on 2016.09.07. {\it Left:} 
           time dependence. {\it Right:} elevation dependence.
           The green connected points show the model value. The blue 
           points show the sum of the model and residuals.
           }
  \label{f:tsys_decomp}
\end{figure*}

  The {\sf tmod-clean} procedure performs a decomposition of $T_{\rm sys}$ 
into time and elevation dependence:
\beq
   T_{\rm sys}(t,e) = T_0 \cdot  T_a(e) \cdot T_z(t).
\eeq{e:15}
  Both $T_a(e)$ and $T_z(t)$ are sought using iterative non-linear LSQ in 
a form of their expansion into the B-spline basis of the 1st degree normalized 
in such a way that their minimum value is 1.0. Iterations are started 
using $T_z(t) = 1$ and expanding $T_{\rm sys}$ cleaned by the {\sf if-clean} 
procedure into the B-splines basis over elevations with 16 knots in a range 
of $3^\circ$ to $92^\circ$. The outliers exceeding $n$ normalized deviations 
are discarded. This expansion is normalized for $T_a(e)$, postfit residuals 
are computed, and they are used for evaluation of the $T_z(t)$ B-spline expansion 
with the time span between knots 20~minutes. Again the outliers are eliminated, 
and the procedure is repeated for the residuals with respect to $T_z(t)$ 
and $T_a(e)$ taken from previous iterations. In total, 8 iterations are 
performed. The $n \sigma$ outlier elimination criterion is $8.0$ for the 
first iteration and it is consecutively reduced to 6.0, 5.0, 4.0, 3.5, 3.0, 
3.0, and 3.0 $\sigma$. Constraints on $T_{\rm sys}$ values as well as 
derivatives over time and elevation with reciprocal weights 9~K, 0.0001~K/s 
and 5~K/rad are imposed to stabilize the solution. The procedure may not 
converge if the number of observations at a given station is less than 
8--10 per hour. The $T_{\rm sys}$ substitute are computed according to 
the empirical model in equation \ref{e:15} for missing measurements or 
those flagged as outliers. Figure~\ref{f:tsys_decomp} shows the result of 
$T_{\rm sys}$ decomposition at station {\sc fd-vlba} as an example. 

Finally, the amplitude was multiplied by the a~priori SEFD: the ratio of 
cleaned $T_{\rm sys}$ and the antenna gains.

\subsection{Amplitude re-normalization}

\begin{figure}
  \includegraphics[width=0.495\textwidth]{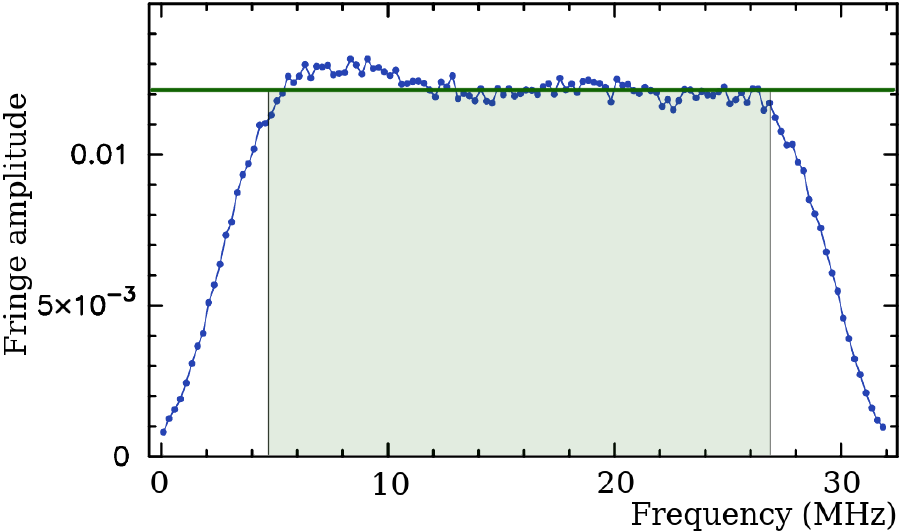}
  \caption{The amplitude of the visibility spectrum of WFCS J0745+1011 
           at baseline {\sc la-vlba/nl-vlba} within IF~1 at 4.128~GHz. 
           The amplitude is reduced at the band edges due to the interference with
           adjacent bands. The shadowed area shows the band kernel that
           was used for amplitude re-normalization.
           }
  \label{f:ampl_renorm}
\end{figure}

  A response of an ideal system to a source with the continuum flat spectrum 
at a given IF has $\Pi$-shape. The bandpass of the VLBA system deviates from 
the ideal and falls off at the edges. To compensate the signal loss due 
the bandpass non-rectangular shape, the visibilities are divided after fringe 
fitting by the function called the amplitude calibration bandpass that is 
the average of the normalized cross-spectrum of strong sources. The 
cross-correlation bandpass falls at the edges stronger than the product of 
auto-correlations. This suggests the signal incurs additional losses 
due to decorrelation at the edges. To account for these losses, I split the 
band into three areas: the central part called the band kernel and two 
edge parts (See Figure~\ref{f:ampl_renorm}). The amplitude bandpass 
is normalized in such a way that its integral over the kernel area is unity.

  The part of the band where the bandpass falls below 0.1 or has spikes
due to internal RFI is usually masked out before fringe fitting, since 
visibilities in this part of the spectrum are corrupted and bring more 
noise than signal. The auto-correlation spectrum was originally normalized 
to unity over the entire IF. Since the system temperature was recorded in 
the entire IF, removal of a part of the spectrum distorts normalization.
Therefore, the autocorrelation is renormalized to unity over the 
unmasked fraction of the spectra, and the cross-correlation amplitude is 
divided by the renormalization factor $R_m$
\beq
     R_m = \Frac{\sum_i \, A_i \, m_i}{\sum A_i} \cdot \Frac{n}{\sum_i m_i},
\eeq{e:16}
  where $A_i$ is the $i$th constituents of the auto-spectrum, $m_i$ is the 
mask, 0 or 1, and $n$ is the number of spectral channels.

\subsection{On-off flagging}

\begin{figure}
  \includegraphics[width=0.495\textwidth]{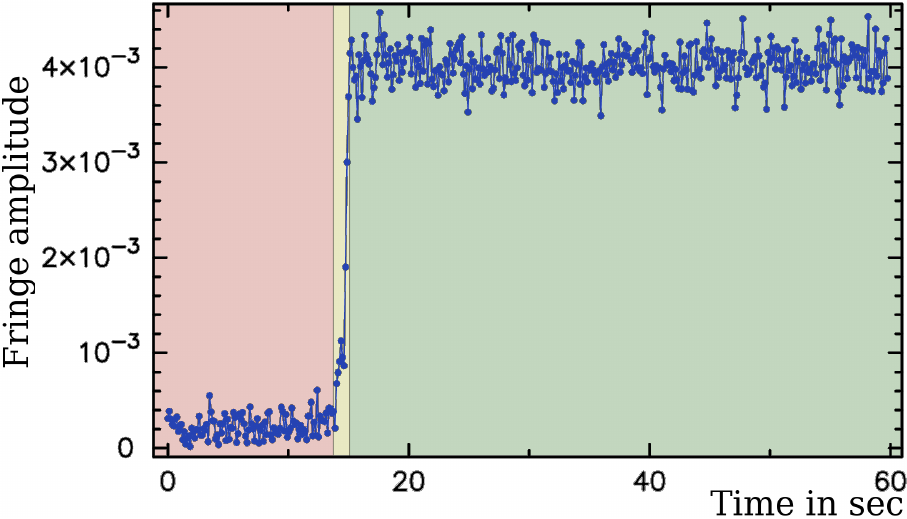}
  \caption{The amplitude of the visibility spectrum of the calibrator source 
           J1642+6856 at baseline {\sc br-vlba/nl-vlba} at 7.6~GHz. The 
           signal to noise ratio is 355. The antenna was off the source 
           for the first 13.9~s (the red area), on the source since 15.0~s 
           (the green area), and was moving on source in between 
           (the yellow area).
           }
  \label{f:onoff}
\end{figure}

  Although the VLBA antennas have a flagging mechanism to report events when 
the antenna finished slewing and reached the source, it is not uncommon 
to receive data recorded during slewing (See Figure~\ref{f:onoff}). 
If not flagged, the use of such data will seriously corrupt the image. 
I implemented an algorithm for detecting off-source data. It assumes
there is a period of time when the data are trusted, called a scan kernel, 
and a period of time when the data are questionable. The scan kernel 
was defined as a [0.4, 0.95] scan fraction.

   The first step is to compute the amplitude averaged over time and 
frequency over the scan kernel applying the results of fringe fitting. 
Then the data before the kernel are split into segments. The segment 
size measured in the number of accumulation periods is defined as 
\beq
   k_a = \max \left(  1, \biggl( n \, \Frac{\sigma_a}{a} \biggr)^2\right), 
\eeq{e:17}
where $a$ is the kernel fringe amplitude, $\sigma_a$ is the rms of the 
amplitude scatter with respect to $a$ over the kernel interval, and $n$ is 
a parameter. The algorithm coherently averages the visibilities within
each segment and checks their amplitudes in the reverse time order 
starting from the segment that is the closest to the kernel. If it finds 
the segment amplitude is less than the kernel amplitude by $n\sigma_s$,
where $\sigma_s$ is the segment amplitude uncertainty, all the visibilities
of that segment and the preceding segments are flagged out. Then a similar 
procedure is executed at the end of the scan. Although late on source is 
the most common reason of the amplitude drop, the fringe amplitude may have 
spikes or drops due to the internal RFI and hardware malfunction as well. 
Therefore, at the end, the kernel span is extended to all visibilities, 
and segments with the amplitude less than $n\sigma_s$ are flagged out.

  The parameter $n$ is selected depending on what is less desirable: 
to flag out good points or miss bad points. I selected $n=3$ which flags
the data more aggressively arguing that the presence of data with wrong 
amplitudes causes more damage than a removal of a fraction of good data 
points.

\subsection{Aggregation of visibilities}

   After flagging and calibration, visibility phases are rotated using 
group delays, phase delay rates, and group delay rates. However, several
complications have to be taken into account. First, since the fringe 
fitting procedure processed each observation individually, in general, 
the fringe reference time used for computation of the scan-averaged 
phases defined as the weighted mean epoch is different. Second, 
independently derived group delay, phase delay, and group delay rate 
estimates along any baseline $ij$, $ik$, and $jk$ do not preserve the 
closure relationship: $a_{ij} - a_{ik} + a_{jk} \neq 0$. If group delays 
or phase delay rates have non-zero closures, phase rotation will distort 
the closure relationship of original visibility phases. Third, if the 
a~priori quadratic term was added to phases before fringe fitting when 
processing a source with a large a~priori position error, it has to be 
subtracted.

   The common scan reference time $t_s$ is found as the weighted mean
over the baseline fringe reference time $t_f$. Then station-based 
group delays $\tau^g_i$, phase delay rates $\dot{\tau}^p_i$, and 
group delay rates $\dot{\tau}^g_i$ are computed from baseline quantities
using weighted least squares:
\begin{widetext}
\beq
   \left\{
   \begin{array}{l @{\enskip} 
                 l @{\enskip} 
                 l @{\enskip} 
                 l @{\enskip} 
                 l @{\enskip} 
                 l @{\enskip} 
                 c @{\enskip} 
                 l } 
               \dot{\tau}^p_i & 
             - \dot{\tau}^p_j & & & & 
               &=& \dot{\tau}^p_{ij} + \ddot{\tau}^p_{ij} \, (t_s - t_f)      \vex \\
                \dot{\tau}^p_i (t_s - t_f)  & 
              - \dot{\tau}^p_j (t_s - t_f)  & 
              + \tau^g_i                    & 
              - \tau^g_j                    & 
              + \dot{\tau}^g_i (t_s - t_f)  & 
              - \dot{\tau}^g_j (t_s - t_f) 
              &=& \tau^g_{ij} + \frac{1}{2} \ddot{\tau}^p_{ij} (t_s - t_{f})^2 \vex \\
              & & & & \hp \dot{\tau}^g_i & - \dot{\tau}^g_j 
              &=& \dot{\tau}^g_{ij}   \\
   \end{array}
   \right..
\eeq{e:e18}
\end{widetext}
  The system has $3 \, (n-1)$ equations, where $n$ is the number of stations. 
Only a portion of the system is shown. One of the stations is taken as 
a reference. Then using estimates of station-based $\dot{\tau}^p_i$, $\tau^g_i$, 
and $\dot{\tau}^g_i$, new baseline-based quantities are computed. Now they 
are referred to the same scan reference time and have a zero closure.

  The right-band side of system \ref{e:e18} contains $\dot{\tau}^p_{\rm ij}$, 
$\tau^g_{\rm ij}$, and $\dot{\tau}^g_{\rm ij}$ of detected observations. When 
the inverse transformation from the station-based to the baseline-based 
quantities is performed, a situation may occur that a given observation at 
a baseline $ij$ has not been detected, but the station-based quantities at 
stations $i$ and $j$ are available because there were detected observations 
at other baselines with these stations. Therefore, we can restore 
baseline-based $\dot{\tau}^p_{\rm ij}$, $\tau^g_{\rm ij}$, and 
$\dot{\tau}^g_{\rm ij}$ for an observation that has not been detected during 
the fringe fitting process. I compute the amplitude averaged over time and 
frequency for such observations and retain those with the amplitude a factor 
of 4.0 greater than the noise level. This criteria is more relaxed than the 
previously used detection thresholds. We are able to lower the detection 
threshold here because the visibilities at other baselines are utilized for 
processing a given observation.

  Finally, the visibility phases are rotated according to phase delay rates, 
group delays, group delay rates, and they are averaged over frequency within 
each IF and over time within 4~s long segments. The variance of the fringe 
amplitude within each segment is computed and used later for image restoration. 
All visibilities of a given source are combined and written into separate 
files, one file per source and per band. If a source was observed in several 
experiments, the data collected within 6 months are merged into a single file.

\note{Source variability within a 6 months period will cause some blurring
in the image made using merged data from two or more epochs, but for most of 
the sources this effect is small. In almost all cases changes of 
structure is associated with a change of the flux density of a compact 
component. Such changes manifest in a synchronous increase or decrease of
the correction factors determined during amplitude self-calibration.
Such a systematic change in correction factors was found in several sources.
These sources were re-imaged for each individual epochs.}

\subsection{Image restoration}

  I used Difmap \citep{r:difmap} for image restoration. First, for each session
I selected three strong calibrators that were observed at all baselines. 
These sources were imaged manually using the conventional technique described 
in Difmap manual. It includes iterations of phase self-calibration, selecting 
the areas on the image plane for CLEAN algorithm \citep{r:clean}, and amplitude 
self-calibration \citep{r:ampl_sc}. The empirical multiplicative gain
corrections adjusted during the amplitude self-calibration were extracted by 
differencing the original visibility data and the self-calibrated data saved by 
Difmap. The gain corrections are averaged over three calibrator sources. IFs 
with unstable or low amplitudes were flagged out by assigning them zero gain 
corrections in some cases. Then these gain corrections were applied to the 
remaining data.

  Further imaging was performed in an automatic fashion using the Difmap
script originally developed by M.~Shepherd and G.~Taylor and modified
by Y.Y.~Kovalev. The pipeline starts with phase self-calibration with
the solution interval 3600~s, then it runs a number of times the inner 
loop that consists of the CLEAN procedure over established windows 
that have the size a factor of $w$~greater than the clean beam size and 
the phase self-calibration with a solution interval~$s$. The inner loop of 
cleaning and phase calibration is repeated till no new peak above 
$d$~times the image rms is found. After each step, a set of CLEAN 
windows is accumulated. The first time the inner loop runs with 
s=3600~s, $w=4$, and $d=6$ with the uniform weights. Then it is repeated 
with natural weights with $w=6.4$, and $d=5.5$. After that the amplitude 
self-calibration is performed followed by the phase self-calibration with 
the solution interval 12~s. Then the inner loop with $s=12$~s, $w=6.4$, 
$d=5.0$ and natural weights is performed. Then the amplitude
self-calibration is performed with the solution interval 3600~s
followed by the phase self-calibration and the inner loop with $s=12$~s, 
$w=6.4$, $d=4.75$ and natural weighting. At that point the accumulated 
model is cleared, the CLEAN procedure with the uniform weights is 
performed over established CLEAN windows, followed by the CLEAN procedure 
with natural weighting and with the modified inner loop with the same 
parameters but without phase self-calibration. At the end, the CLEAN 
procedure over established windows runs once more followed by the phase 
self-calibration, and then the final CLEAN procedure runs over 
the entire map last time, and Difmap creates the final image.

\begin{figure*}
  \includegraphics[width=0.41\textwidth]{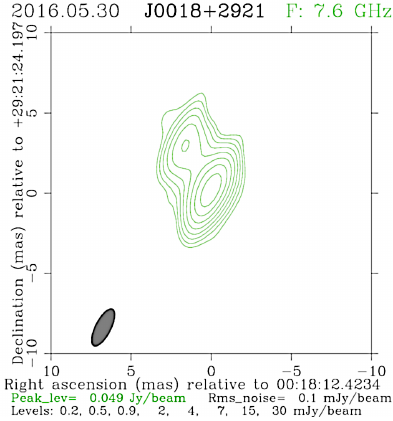}
  \hspace{0.049\textwidth}
  \includegraphics[width=0.54\textwidth]{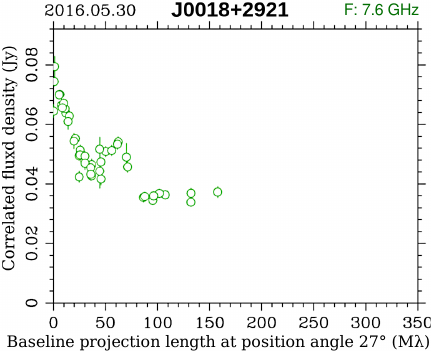}
  \caption{A sample image of WFCS J0018+2921 (left) and the self-calibrated
           visibilities averaged over time and frequency (right).
           The source has been detected at all baseline of the 10-station 
           network in one scan of 60~s long with the SNR in a range of 15 to 35.
           }
  \label{f:image}
\end{figure*}

  An example of an image derived from processing the survey data is shown
in Figure~\ref{f:image}. The self-calibrated visibilities are plotted against
the projection of the baseline vector into the jet direction. 
This source was detected at all 45 baselines, and no station failed. 
When the number of detections is less either because a source is weak,
or station failures, the image quality degrades. An image rarely provides 
useful information beyond the total flux density if the number of detections 
drops below 8--10. But anyway, the automatic procedure processed all the 
sources that have enough data to form closures, i.e. if the number of 
detections was at least 4--6.

  Upon completion of automatic imaging, I performed a manual inspection of 
all the images. If an image or a plot of self-calibrated visibilities showed
abnormalities, I reprocessed these images manually. Manual intervention required 
in 20--25\% cases. The most common problems are RFI, a sudden amplitude drop 
due to a hardware malfunction, and emission beyond the default mapping
area. The manual intervention solved the problem in most of the cases.

\subsection{Generation of median flux densities}

  An image is a two-dimensional array. To provide a more concise but coarse image 
characterization, I followed the practice used in prior VLBI Calibrator surveys
\citep[e.g.,][]{r:vcs5} and computed the total flux density by summing all 
CLEAN components, the flux density at short baselines --- the median flux 
density at baseline projection lengths  $< 900$~km, and the unresolved flux 
density --- the median flux density at baseline projection lengths  
$> 5,000$~km. For some sources no information about the unresolved flux 
density is available.

  The imaging process failed for 108 sources: 32 at C-band and 76 at X-band
because the number of observations, 3 to 6 was too small. The calibrated 
visibilities were coherently averaged for these sources over time and frequency
for each individual observation and then the amplitudes were incoherently 
averaged over all observations at different baselines. The mean visibilities 
produced that way were used as a substitute for the total flux density. 
No estimates of the flux density at short baselines and the unresolved flux 
densities are available for these sources.

  In total, images at least at one band are available for 6714 detected target 
sources out of 6755, and 6038 of them, or 90\%, were generated using only one 
scan of 60~s long. In addition, images for 465 calibrator sources were made, 
and many of them at more than at one epoch.

\begin{figure*}
  \includegraphics[width=0.492\textwidth]{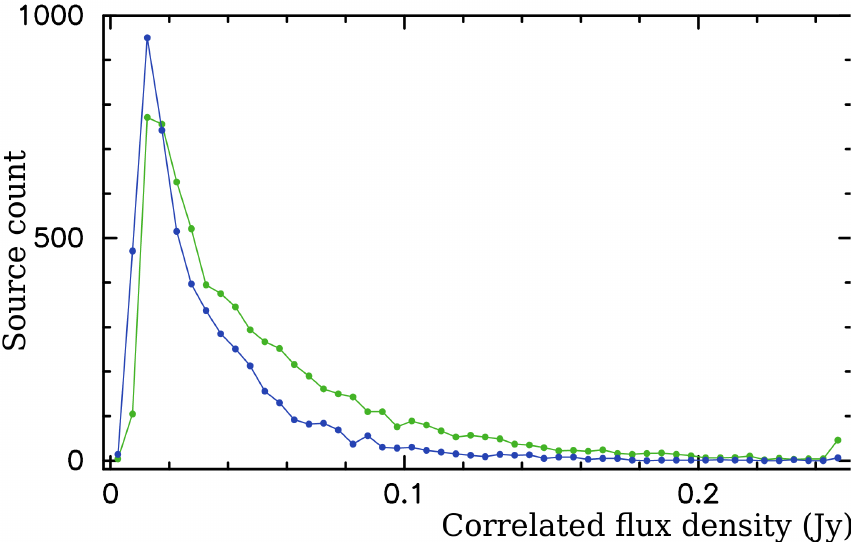}
  \hspace{0.009\textwidth}
  \includegraphics[width=0.495\textwidth]{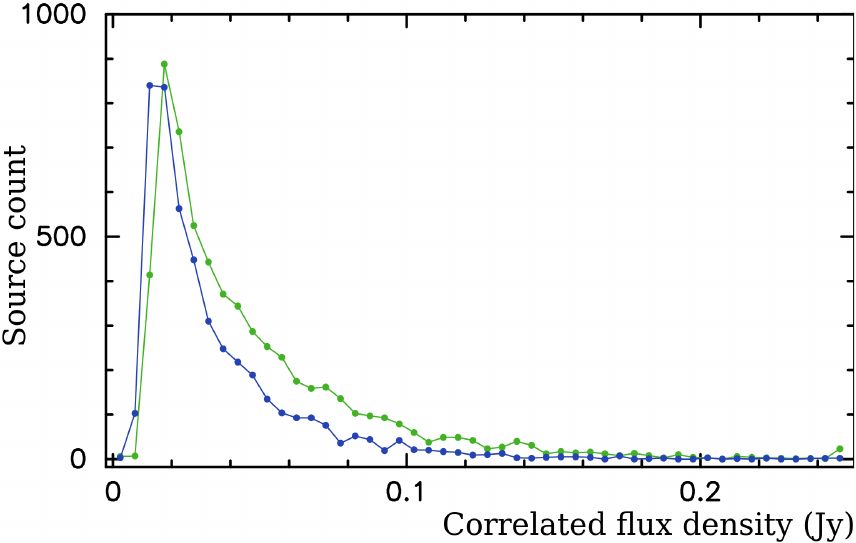}
  \caption{The distribution of the median flux densities at 4.3~GHz (left) 
           and at 7.6~GHz (right). The green points show the median flux 
           densities at short baseline projection lengths. The blue points 
           show the median flux densities at long baseline projection 
           lengths. 
           }
  \label{f:flux_distr}
\end{figure*}

  The distribution of flux densities at short and long baseline projection 
lengths of all target sources is shown in Figure~\ref{f:flux_distr}. The target 
sources are rather weak. Table~\ref{t:flux_median} shows the median flux 
densities of detected sources. Only 86 sources with flux density greater 
200~mJy at 7.6~GHZ have been detected among 13,154 observed. The detection 
limit was in a range of 10 to 15~mJy.

\begin{table*}
  \caption{The median total flux densities, the flux densities at short 
           baselines, and the unresolved flux densities at 4.3~GHz and 
           7.6~GHz. The table also shows the number of target sources 
           for which flux density information is available.}
  \begin{tabular}{l r r @{\qquad} r r}
       \hline
                                       & \nntab{c}{4.3 GHz} & \nntab{c}{7.6 GHz} \\
                                       & \ntab{c}{Flux} & \ntab{l}{\# Src} 
                                       & \ntab{c}{Flux} & \ntab{l}{\# Src}       \\
                                       & \ntab{c}{Jy} & 
                                       & \ntab{c}{Jy} &                     \\
       \hline
       Total flux density              & 0.041 & 6738       & 0.038 & 6115  \\
       Flux density at short baselines & 0.036 & 6688       & 0.034 & 6208  \\
       Unresolved flux density         & 0.023 & 5137       & 0.024 & 4601  \\
       \hline
  \end{tabular}
  \label{t:flux_median}
\end{table*}

\subsection{Multiple sources in the field of view}

\begin{figure*}
  \includegraphics[width=0.514\textwidth]{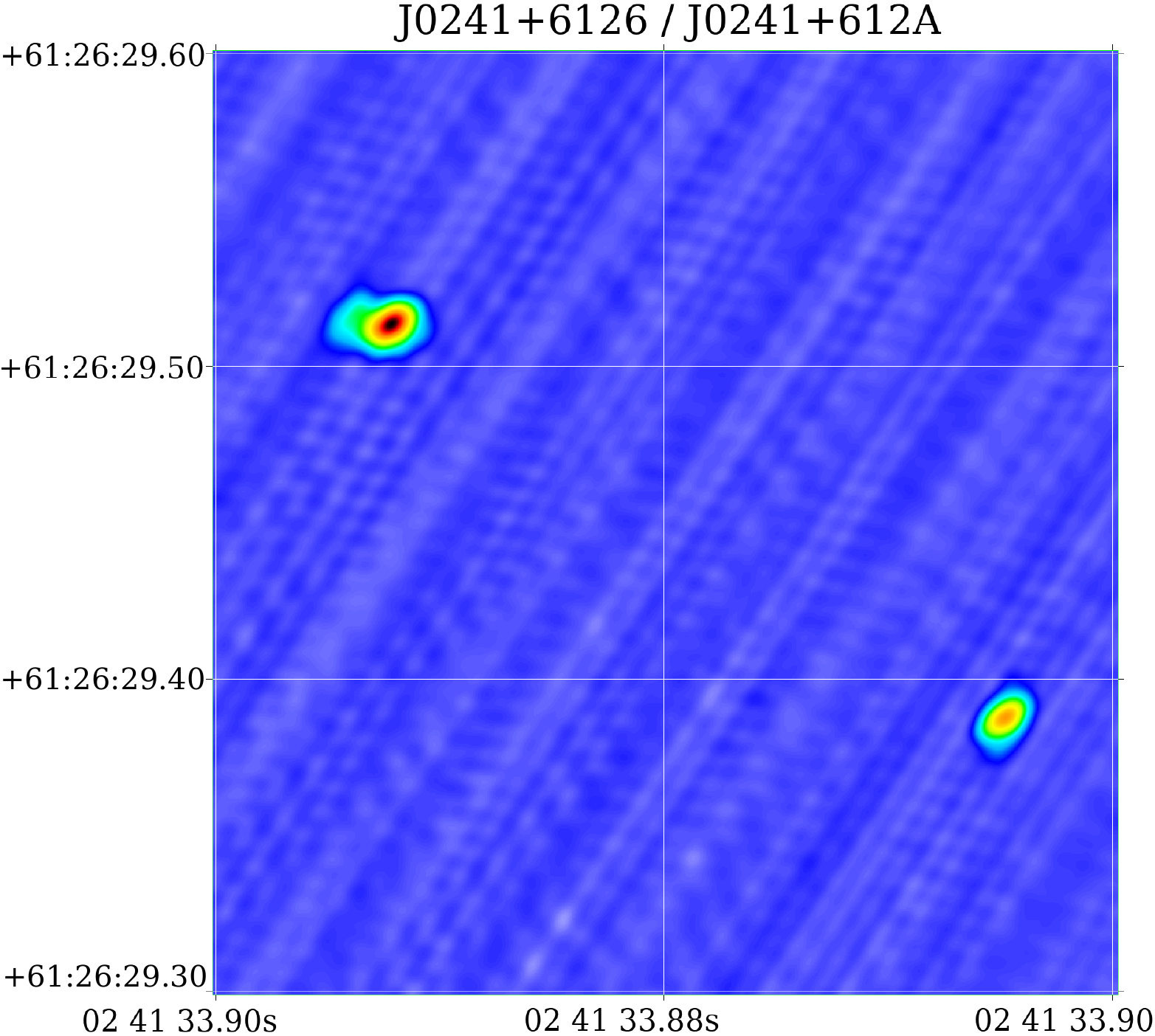}
  \hspace{0.049\textwidth}
  \includegraphics[width=0.41\textwidth]{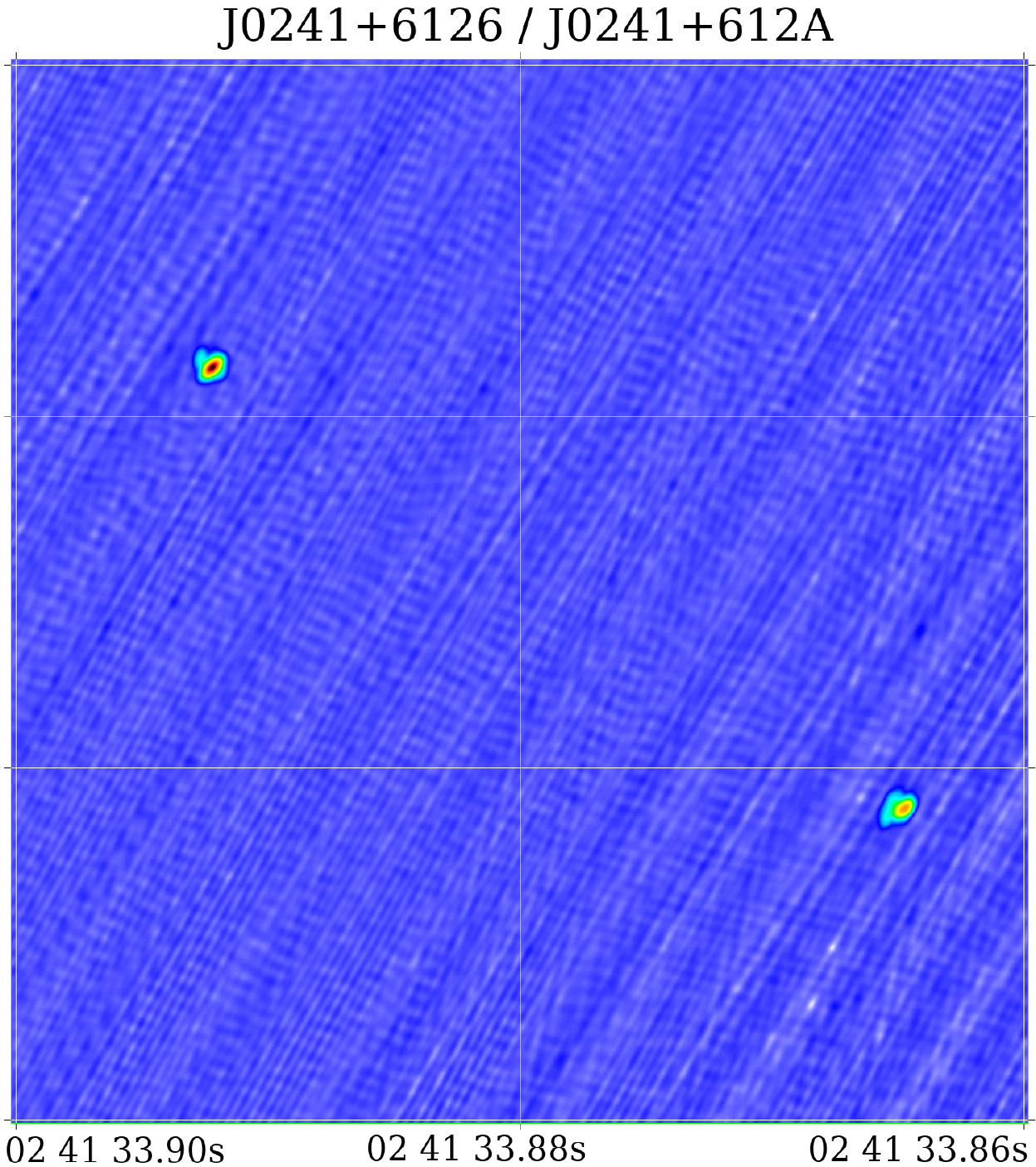}
  \caption{The images of the visually double source J0241+6126/J0241+612A 
           at 4.3~GHz (left) and at 7.6~GHz (right).
           }
  \label{f:pair}
\end{figure*}

  It is not uncommon to find several objects in the field when the field of 
view covers several arcminutes. If the field of view contains several strong 
objects, each of them at least a factor of 2 brighter than the detection 
limit, then the astrometric solution will have an unusually high number of 
outliers. I applied a technique of component separation for such cases. 
I examined fringe plots and flagged those that showed a pattern that is 
consistent with internal or external radio interference --- spikes, 
unusual spectrum, or sinc-shape pattern of fringe amplitude versus time. 
See Figures~3 and 4 in \citet{r:shu17} for examples.
If the outliers remained, I inverted the suppression status and tried to 
estimate positions using only those observations that were previously 
suppressed. Three outcomes of this procedure are possible. The solution may 
converge to a new position, and most of the group delays that were outliers
for the 1st component will be used in the solution for the 2nd component.
This happens if an observation was suppressed because the fringe-fitting 
procedure picked up one source component at some baselines and another at 
others. The solution may not converge to a new position. And the result may 
be inconclusive: a solution may converge to a new position, but a significant 
fraction of observations is included for estimation of both components, 
or is used in neither. If the distance between components exceeded
$1''$, its a~priori coordinates were updated, and a new iteration of fringe
fitting with accounting for the quadratic term in fringe phase followed. 
If the distance between components was below $2''$, the field was imaged
(See Figure~\ref{f:pair} as an example). If the second component did not 
to appear at the image, the hypothesis the source is double was rejected, 
and the observations were re-flagged. The component separation procedure 
works easily if the distance between components is above $0.3''$--$0.5''$. 
It does not work if the distance is less than $0.07''$--$0.15''$. The 
range $0.1''$--$0.4''$ is intermediate where the component separation is 
not always reliable. The second component may be missed if the flux 
density of the second component is a factor of 10~lower than the flux 
density of the primary component.

  In total, second components were found for 88 objects that resulted in 
separate catalogue entries. They are treated as different sources in the 
context of this paper although in many cases they may be a part of the 
same object. The source components are found at separations from several 
mas to several arcminutes. Separation of sources into several catalogue
entries was not done based on their physical properties. It was done 
\note{entirely} on the basis whether the position of the second component 
can be determined in the astrometric solution independently from the 
position of the first component. 

  Among 88 sources with several components found in the beam, 34 sources
have separations in a range of $0.08''$--$1.9''$ and 9 have separations in 
a range of $5''$ to $60''$. The first 8~pairs are shown in 
Table~\ref{t:pairs}. The entire table can be found in the electronic
attachment under name datafile2.txt. All second components (and for some 
sources the 3rd and 4th component) at distances less $2''$ were identified 
in images. Therefore, two or more components share the same image. 
Analysis of the nature of multiple structure is beyond the scope of this paper.

\begin{table}
   \caption{The first 8 rows of the table of 43 target sources 
            that have two or more components, which positions were
            independently determined.
           }
   \scriptsize\hspace{-3em}
   \begin{tabular}{l @{\quad} l @{\:} r}
      \hline
      Component 1     & Component 2     & Separation \\
      \hline
      WFCS J0403+702A & WFCS J0403+7026 & $0.0782''$ \\
      WFCS J0241+612A & WFCS J0241+6126 & $0.1140''$ \\
      WFCS J2108-210A & WFCS J2108-2101 & $0.1167''$ \\
      WFCS J0904+593A & WFCS J0904+5938 & $0.1401''$ \\
      WFCS J0031+540A & WFCS J0031+5401 & $0.1452''$ \\
      WFCS J0132+521A & WFCS J0132+5211 & $0.1481''$ \\
      WFCS J0023+273A & WFCS J0023+2734 & $0.1523''$ \\
      WFCS J0716+470A & WFCS J0716+470B & $0.1743''$ \\
      \ldots && \\
      \hline
   \end{tabular}
   \label{t:pairs}
\end{table}

\section{Source catalogue}

   The Wide-field VLBA Calibrator Survey catalogues has 6755 entries. 
The catalogue presents source positions at J2000.0 epoch, position
uncertainties, correlation between right ascension and declination,
the number of observations used in 4.3~GHz (C-band), 7.6~GHz (X-band),
and dual band solutions, the solution that was used for reporting 
positions; the total flux density, the median flux density at baseline 
projection lengths $< 900$~km, and the unresolved flux density defined 
as the median flux density at baseline projection lengths $> 5000$~km 
at 4.3 and 7.6~GHz. The first 8 rows of the catalogue are presented in 
Table~\ref{t:wfcs_cat}. The catalogue in full can be found in the 
electronic as attachment under name datafile1.txt.

\begin{figure*}
  \centerline{\includegraphics[width=0.99\textwidth]{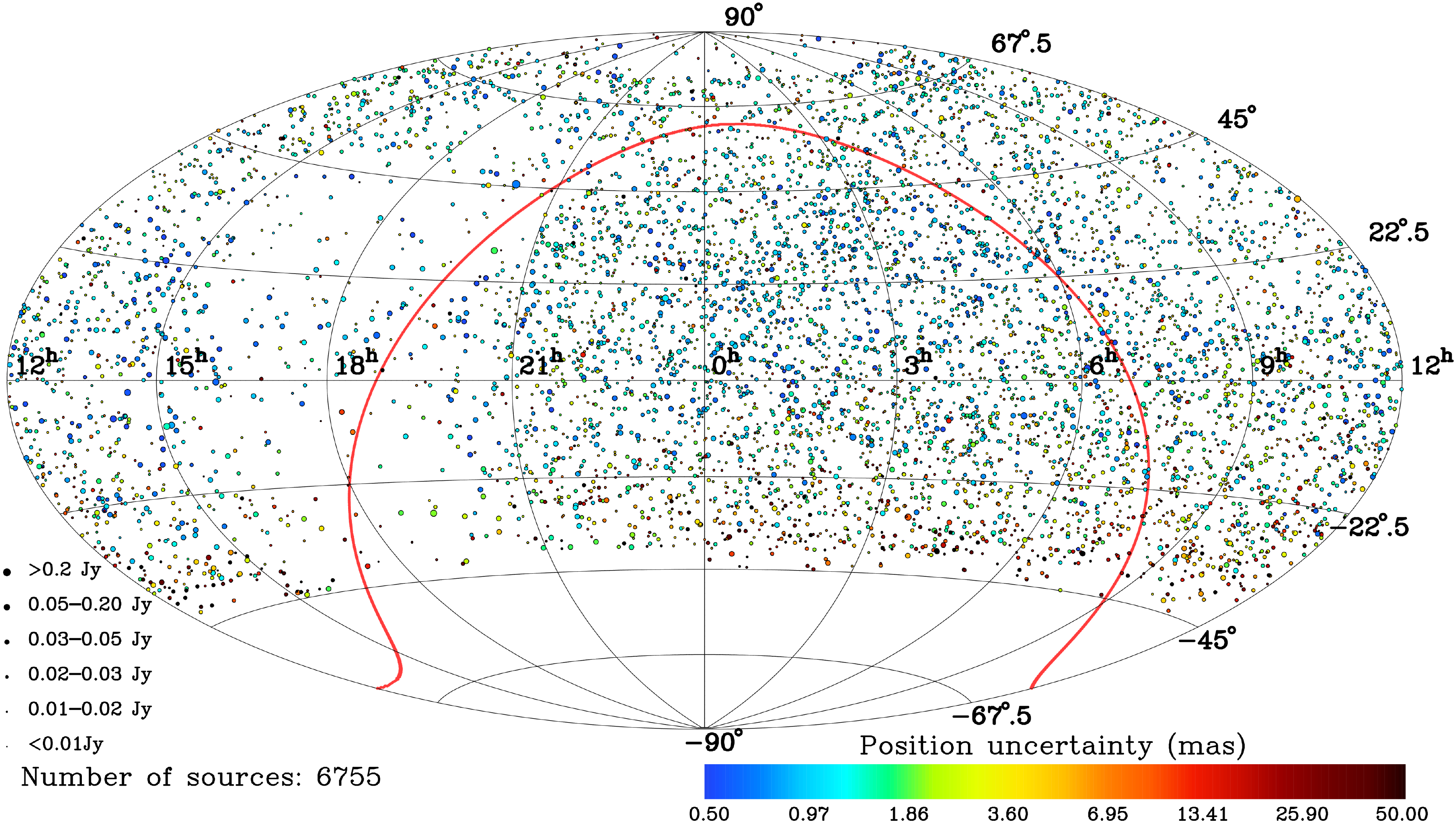}}
  \caption{The distribution of the detected sources over the sky. 
           The red line shows the Galactic plane.
           }
  \label{f:wfcs_distr}
  \par\vspace{-4ex}\par
\end{figure*}

\begin{table*}
   \caption{The first 8 records of the Wide-field VLBA Calibrator Survey 
            catalogue. The uncertainty in the right ascension $D_{\rm ra}$
            is given without the $\cos\delta$ factor. The column {\sf Corr} 
            contains the correlation between right ascension and declination. 
            The column {\sf Num Obs} contains the number of observations used in 
            the astrometric solutions at band X, C, and the linear combinations 
            of group delay observables. The column {\sf Band} contains a flag
            which of the solutions was used for reported positions. The six 
            columns {\sf Flux Density} contain the estimates of the total flux 
            density {\sf Tot}, the median flux density at baseline projection 
            lengths shorter 900~km {\sf Shr}, and the median flux density at 
            baseline projection lengths longer 5000~km {\sf Unr} for both bands, 
            4.3 and 7.6~GHz.
           }
   \scriptsize
   \iftwocolstyle \hspace{-0.32\textwidth} \else \hspace{-0.22\textwidth} \fi
   \begin{tabular}{ l @{\hspace{0.7em}} l @{\hspace{0.7em}} r @{\hspace{0.7em}} r @{\hspace{0.7em}} r @{\hspace{0.7em}} r @{\hspace{0.7em}}r @{\hspace{0.7em}} r @{\hspace{0.7em}} r @{\hspace{0.7em}}r @{\hspace{0.7em}}l @{\hspace{0.7em}}r @{\hspace{0.7em}}r @{\hspace{0.7em}}r @{\hspace{0.7em}}r@{\hspace{0.7em}}r @{\hspace{0.7em}}r}
       \hline
       J2000-name        & \hspace{-0.5em }B1950-name   & Right ascension & Declination  & $D_{\rm ra}$ & $D_{\rm dec}$ & Corr & \nnntab{c}{Num obs} & Band & \nnnnnntab{c}{Flux density} \\
                         &            &                 &                 &              &               &      & & & &  & \nnntab{c}{4.3 GHz} & \nnntab{c}{7.6 GHz} \\
                         &            &                 &                 &              &               & & X & C & \nntab{l}{\hspace{-0.8em}X/C} &  Tot & Shr & Unr & Tot & Shr & Unr \\
                         &            & hh mm ss.ffffff \hspace{4.0em}  & \hp dd mm ss.fffff \hspace{4.0em}  & mas   & mas    &       &     &     &     &     &     Jy &     Jy &     Jy &     Jy &    Jy &    Jy  \\
       \hline
       WFCS J0000$-$1352 & 2357$-$141 & 00 00 03.124493 & $-$13 52 00.75819 &   1.71 &   4.30 &  0.720 &  19 &  19 &  17 & X   &  0.021 & 0.021 & \hm 0.009 & 0.019 & 0.018 & \hm 0.015 \\
       WFCS J0000$-$3738 & 2357$-$379 & 00 00 08.414016 & $-$37 38 20.67746 &   3.90 &   6.96 &  0.921 &  32 &  32 &  29 & X/C &  0.019 & 0.018 & \hm 0.019 & 0.019 & 0.020 & \hm 0.020 \\
       WFCS J0000$+$0248 & 2357$+$025 & 00 00 19.282530 & $+$02 48 14.68956 &   0.56 &   1.35 &  0.025 &  29 &  28 &  28 & X   &  0.039 & 0.033 & \hm 0.018 & 0.038 & 0.037 & \hm 0.024 \\
       WFCS J0000$+$1139 & 2357$+$113 & 00 00 19.564227 & $+$11 39 20.72629 &   1.18 &   2.54 &  0.009 &  29 &  18 &  18 & C   &  0.022 & 0.025 & \hm 0.008 & 0.018 & 0.014 & \hm 0.011 \\
       WFCS J0000$+$0307 & 2357$+$028 & 00 00 27.022571 & $+$03 07 15.64962 &   0.47 &   1.15 & -0.080 &  36 &  36 &  36 & X/C &  0.085 & 0.085 & \hm 0.047 & 0.075 & 0.071 & \hm 0.027 \\
       WFCS J0000$+$3918 & 2358$+$390 & 00 00 41.527612 & $+$39 18 04.14826 &   0.43 &   0.67 & -0.142 &  45 &  45 &  45 & X/C &  0.070 & 0.066 & \hm 0.056 & 0.094 & 0.089 & \hm 0.076 \\
       WFCS J0000$+$5157 & 2358$+$516 & 00 00 51.385221 & $+$51 57 19.89483 &   7.30 &   7.88 &  0.216 &  11 &   6 &   6 & C   &  0.037 & 0.026 &    -9.9   & 0.015 & 0.017 &    -9.9   \\
       WFCS J0001$-$1741 & 2358$-$179 & 00 01 06.264989 & $-$17 41 26.61572 &  17.71 &  18.90 & -0.431 &   5 &   3 &   3 & C   &  0.078 & 0.025 &    -9.9   & 0.040 & 0.019 &    -9.9   \\
       \ldots &&&&&&&&&&&&&&&& \\
       \hline
   \end{tabular}
   \label{t:wfcs_cat}
\end{table*}

  The source distribution over the sky is shown in Figure~\ref{f:wfcs_distr}. 
The gap in right ascension 15 to $21^h$ is because this range was 
over-subscribed, and too few observing sessions covering it were scheduled.

  The catalogue is also accompanied with a dataset of 15,542 images. 
\note{Multiple sources with component separations less than $1.5''$ are shown
in one image file}. The dataset provides six files associated to each image: 
the image in FITS format, calibrated visibilities in FITS format, a table 
with estimates of the correlated flux densities, as well as pictures in 
Postscript format of the source maps, calibrated visibilities, and 
$uv$-coverage plots. The dataset is available at 
\web{http://astrogeo.org/wfcs/images}. \note{Two auxiliary tables with 
positions derived from all three solutions and the list of undetected 
sources are described in the appendix.}

\begin{figure}
  \includegraphics[width=0.495\textwidth]{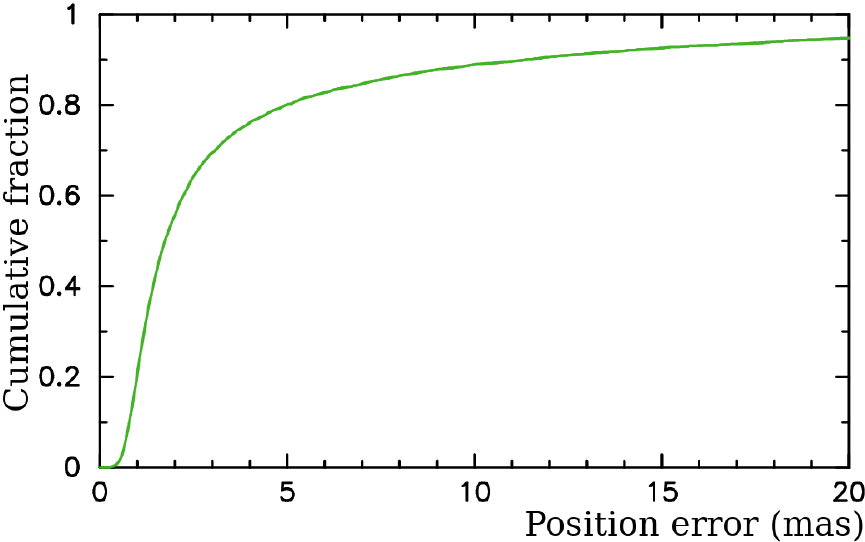}
  \caption{The cumulative distribution of the semi-major error ellipse axes.
          }
  \label{f:pos_err_distr}
\end{figure}

\begin{figure*}
  \includegraphics[width=0.495\textwidth]{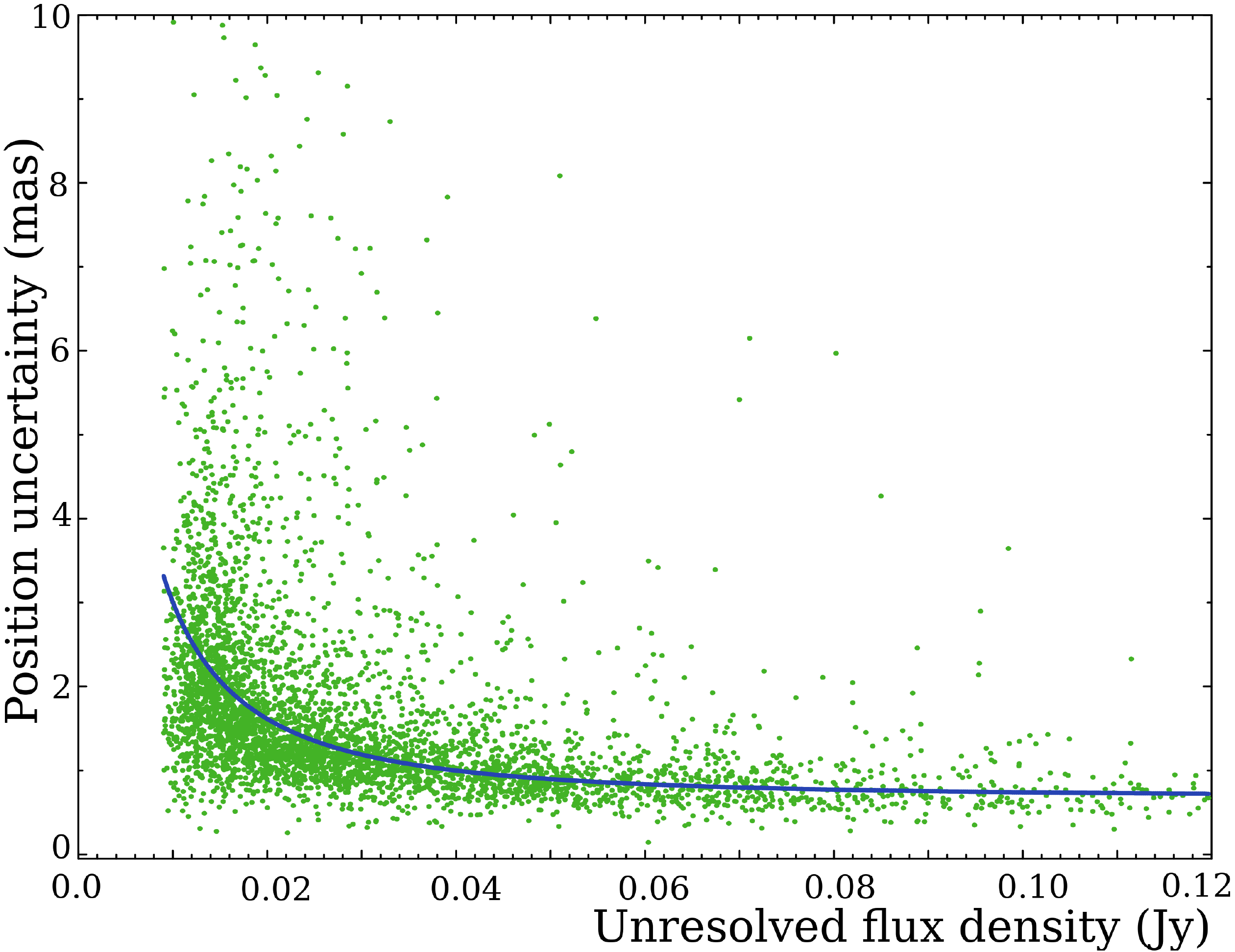}
  \hspace{0.009\textwidth}
  \includegraphics[width=0.495\textwidth]{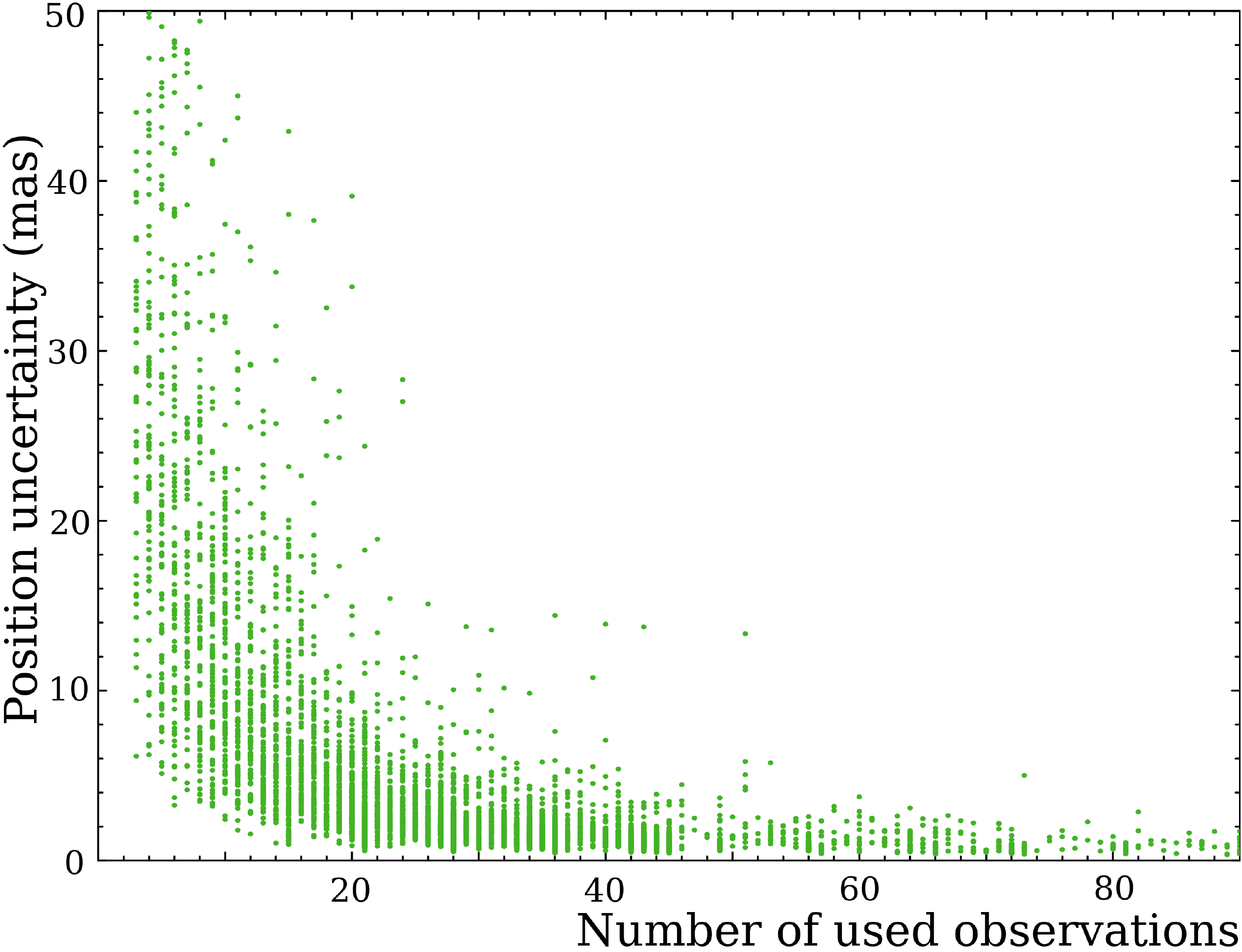}
  \caption{The dependence of the semi-major error ellipse axes as a function
           of the unresolved flux density at X-band (left) and the number
           of observations used in the solution (right).
           }
  \label{f:pos_err_dep}
\end{figure*}

  The position accuracy of the catalogue cannot be characterized by one 
number since it varies within 4 orders of magnitude, from 0.07 mas to $7''$. 
Figure~\ref{f:pos_err_distr} shows the cumulative distribution of the 
semi-major axes of the error ellipse. The median is 1.7~mas (0.9~mas
for right ascension scaled by $\cos\delta$, and 1.6~mas for declinations).
This is noticeably \note{higher} than in previous surveys VCS1--VCS6. 
Figure~\ref{f:pos_err_dep} shows the position uncertainties as a function of
the unresolved flux density at 7.6~GHz band and as a function of the number
of observations used in the solution. The mean position uncertainty is around
0.7~mas for the sources with the unresolved flux density 100~mJy. It drops to 
1.0~mas for the sources with flux density 50~mJy, 1.7~mas for the sources with 
flux density 20~mJy, and 3.0~mas for the sources with flux density 10~mJy.
Similarly, the position uncertainty steadily grows when the number of used 
observation decreases from 45 to 15, and then sharply increases.

  Thus, we see two factors that affected position accuracy. First, the WFCS
tapped the weak population --- stronger sources have already been observed in 
the previous surveys and the follow-up VCS-II campaign \citep{r:vcs-ii}.
Therefore, remaining sources are faint. Second, the amount of on-source time, one scan
of 60~s long is sufficient to detect a source to get its position with 
the milliarcsecond level of accuracy, get its coarse images, but not 
sufficient for reaching the 0.2~mas accuracy level and getting high fidelity,
high dynamic range images. In order to reach that level, approximately one 
order of magnitude more resources is required.

\section{Summary}

  A list of 13,645 sources has been observed in 2013--2016 with VLBA, 
including 13,154 target objects never before observed with VLBI. Of them,
more than one half, 6755 sources, have been detected. This numbers exceed
all prior published VLBI astrometry surveys combined.

  A novel technique of a wide-band survey has been successfully demonstrated.
The previous restrictions on the field of view caused by the limitations of 
hardware correlators and a lack of adequate computer resources are lifted.
VLBI with the field of view comparable with the antenna primary beam size is 
feasible and is expected to become routine in forthcoming surveys. Unlike to 
correlating with multiple phase centers, this approach does not involve the 
a~priori knowledge where a source should be. If one needs observe several 
sources with known positions, then the multiple phase center approach is 
preferable. If one needs to search for a source everywhere within the beam, 
a direct approach of correlation with high time and frequency resolution is 
preferable due to its simplicity. Tests showed that a precise calibration for
the amplitude loss as far from the center as at a 20\% level of the primary 
beam power pattern can be performed. The tests also highlighted the necessity 
of pointing offset monitoring for such observations.

  Historically, absolute astrometry surveys were conducted at S and X bands,
i.e. at 2.2 and 8.4~GHz. The choice was motivated by the availability of 
dual-band receivers that matched the NASA deep space navigation frequency
bands. It was demonstrated that simultaneous observations at remote wings 
of a single broadband C-band receiver provide a noticeable advantage 
in sensitivity and accuracy of derived source positions. Such observations 
are affected by the radio interference in a much lesser extent than 
observations at 2.2 GHz. It was demonstrated that transition from 
2.2/8.4~GHz to 4.3/7.6~GHz does not introduce any noticeable systematic 
differences related to the ionospheric contribution.

  In the past, the preference was given to VLBI observations of flat spectrum
sources. Since the pool of flat spectrum sources not yet observed with VLBI
is close to depleted, the spectral index selection criteria has to be lifted.
The detection rate among flat and steep spectrum sources was 79\% and 35\% 
respectively. We can find compact sources among steep spectrum sources, and 
vice versus, flat-spectrum sources can be resolved out. Since the WFCS 
catalogue does not form a complete flux limited sample, these numbers
should be taken with a caution. The recent paper of \citet{r:npcs} provides 
detailed statistics of the compactness of steep and flat spectrum sources. 
Two follow-up programs for reaching the completeness at a given flux density 
limit in some areas covered by the WFCS ran with VLBA in 2019--2020. 
Analysis of these campaigns that will be published soon, will investigate 
the properties of parsec-scale emission of flat and steep sources drawn 
from parent catalogues at different frequencies in detail.
   
  Although snapshot images do not have a high dynamic range, a number of 
sources with multiple components was spotted in the dataset. Analysis of 
their nature is beyond the scope of the present paper. It will require 
follow-up observations, and several such campaigns have already commenced.

  The WFCS median semi-major axis of the position uncertainty is 1.7~mas, 
which is noticeably lower than in a number of prior surveys. Two factors 
played the role: the median flux density was much lower, 35~mJy, and the 
sources were observed mostly only in one scan. The partly resolved sources 
were not detected at long baselines and this caused a significant position 
accuracy degradation. Allocation of substantially more resources is
required in order to reach a sub-milliarcsecond level of accuracy. It may 
not be practical to re-observe all the WFCS sources, although re-observation 
of a subset of sources, for instance those that are compact and suitable 
as calibrators, or sources exhibiting unusual morphology, is warranted.

  The catalogue presented in Table~\ref{t:wfcs_cat} was built using only the 
data from the three observing campaigns VCS7, VCS8, and VCS9 in 2013--2016.
Since then, a number of sources were re-observed. Up to date positions of
all these sources and many others can be found in the online Radio 
Fundamental Catalogue at \web{http://astrogeo.org/rfc} that is updated
on a quarterly basis.

\par\vspace{-4ex}\par
\acknowledgments

   This work was done with datasets BP171, BP175, BP177, BP192, and BP245 
collected with VLBA instrument of the NRAO and available at 
\web{https://archive.nrao.edu/archive}. The NRAO is a facility of the
National Science Foundation operated under cooperative agreement by
Associated Universities, Inc. This work made use of the Swinburne University 
of Technology software correlator, developed as part of the Australian Major 
National Research Facilities Programme and operated under license.

   It is my pleasure to thank Yuri Y.~Kovalev for lengthy and enlightening
discussions about imaging and amplitude normalization.

\appendix

\vspace{-3ex}\par

\note{
For completeness, two auxiliary tables are presented. 
Extended Table~\ref{t:det_ext} of detected sources has the following columns:
(1)~WFCS J2000 source name; 
(2)~B1950 source name;
(3)~right ascension from the C-band solution; 
(4)~declination from the C-band solution; 
(5)~error in right ascension (without cos(delta) factor) from the C-band solution in mas; 
(6)~error in declination from the C-band solution in mas; 
(7)~correlation between right ascension and declination from the C-band solution;
(8)~number of used observations in C-band solution; 
(9)~right ascension from the X-band solution; 
(10)~declination from the X-band solution; 
(11)~error in right ascension (without cos(delta) factor) from the X-band solution in mas; 
(12)~error in declination from the X-band solution in mas; 
(13)~correlation between right ascension and declination from the X-band solution;
(14)~number of used observations in X-band solution 
(15)~right ascension from the dual-band X/C solution; 
(16)~declination from the dual-band X/C solution; 
(17)~error in right ascension (without cos(delta) factor) from the dual-band solution in mas; 
(18)~error in declination from the X-band solution in mas; 
(19)~correlation between right ascension and declination from the dual-band solution;
(20)~number of used observations in the dual-band X/C solution;
(21)~flux density at 1.4 GHz from the cross-matched NVSS catalogue;
(22)~flux density at 4.85 GHz from the cross-matched GB6 or PMN catalogue and
(23)~Spectral index between 1.4 and 4.8 catalogues.
The entire table can be found in the electronic attachment under name datafile3.txt.}

\begin{table*}[h]
   \caption{The first 8 rows of the auxiliary table with positions and 
            flux densities of 6755 WFCS program sources detected in 
            each three solutions. Only first 11 columns out of 23 are shown.
           }
   \scriptsize\hspace{-10em}
   \begin{tabular}{l@{\quad}l@{\quad}l@{\quad}l@{\quad}l@{\quad}r@{\quad}r@{\quad}l@{\quad}l@{\quad}l@{\quad}l@{\quad}l}
      \hline
        \ntab{c}{(1)} &
        \ntab{c}{(2)} &
        \ntab{c}{(3)} &
        \ntab{c}{(4)} &
        \ntab{c}{(5)} &
        \ntab{c}{(6)} &
        \ntab{c}{(7)} &
        \ntab{c}{\hspace{-1em}(8)} &
        \ntab{c}{(9)} &
        \ntab{c}{(10)} &
        \ntab{c}{(11)} &
        \ldots \\
      \hline
          WFCS J0000$+$0248 & 2357$+$025 & 00 00 19.282714 & $+$02 48 14.69019 &   1.20 &   2.57 & 0.021  & 29 &  00 00 19.282530 & $+$02 48 14.68956  &  0.56 & \ldots \\
          RFCS J0000$+$0307 & 2357$+$028 & 00 00 27.022719 & $+$03 07 15.64335 &   1.17 &   2.68 & 0.101  & 36 &  00 00 27.022615 & $+$03 07 15.64531  &  0.52 & \ldots \\
          WFCS J0000$+$1139 & 2357$+$113 & 00 00 19.564227 & $+$11 39 20.72629 &   1.18 &   2.54 & 0.009  & 29 &  00 00 19.564105 & $+$11 39 20.72772  &  1.01 & \ldots \\
          WFCS J0000$+$3918 & 2358$+$390 & 00 00 41.527499 & $+$39 18 04.14617 &   1.37 &   2.18 &-0.040  & 45 &  00 00 41.527528 & $+$39 18 04.14665  &  0.49 & \ldots \\
          WFCS J0000$+$5157 & 2358$+$516 & 00 00 51.385221 & $+$51 57 19.89483 &   7.30 &   7.88 & 0.216  & 11 &  00 00 51.385344 & $+$51 57 19.89833  & 14.43 & \ldots \\
          WFCS J0000$-$1352 & 2357$-$141 & 00 00 03.124573 & $-$13 52 00.76274 &   3.01 &   6.45 & 0.586  & 19 &  00 00 03.124493 & $-$13 52 00.75819  &  1.71 & \ldots \\
          WFCS J0000$-$3738 & 2357$-$379 & 00 00 08.414694 & $-$37 38 20.70320 &   8.89 &  18.68 & 0.812  & 32 &  00 00 08.414234 & $-$37 38 20.68723  &  3.41 & \ldots \\
          WFCS J0001$+$1456 & 2358$+$146 & 00 01 32.830898 & $+$14 56 08.07612 &   1.44 &   2.83 & 0.058  & 28 &  00 01 32.830859 & $+$14 56 08.07853  &  0.53 & \ldots \\
          \ldots & &  &  &  &  &  &  &  &  &  &  \\
      \hline
   \end{tabular}
   \label{t:det_ext}
\end{table*}

\note{
  Extended Table~\ref{t:ndt} lists 6399 target sources that were observed 
but have not been detected. It contains their a~priori coordinates, the 
flux densities for matching sources from NVSS and either GB6 or PMN catalogues, 
and the spectral index. Among 6399 sources not detected in the WFCS campaign, 
51 have been detected with other VLBI programs that ran after WFCS. 
These sources are marked with a flag. The entire table can be found in the 
electronic attachment under name datafile4.txt.
}

\begin{table*}[h]
   \caption{The first 8 rows of the auxiliary table with positions and flux 
            densities of 6399 WFCS program sources that were observed, 
            but not detected.
           }
   \begin{tabular}{llllllll}
      \hline
        Source name & IVS name & Right ascension & Declination & $F_{1.4}$ & $F_{4.8}$ & Sp. ind & Flag \\
      \hline
          WFCScand J0000$+09$57 & 2357$+$096 & 00 00 02.87  & $+$09 57 06.6 & 0.3014 & 0.1030 & $-$0.86 & \\
          WFCScand J0000$+11$14 & 2358$+$109 & 00 00 47.60  & $+$11 14 12.0 & 0.2019 & 0.1070 & $-$0.51 & \\
          WFCScand J0000$+12$14 & 2358$+$119 & 00 00 37.80  & $+$12 13 57.0 & $-$9.9 & 0.0500 & $-$9.9  & \\
          WFCScand J0000$+29$47 & 2358$+$295 & 00 00 57.60  & $+$29 47 58.0 & 0.1298 & 0.0530 & $-$0.72 & \\
          WFCScand J0000$+55$39 & 2357$+$553 & 00 00 19.40  & $+$55 39 03.0 & 1.5178 & 0.4650 & $-$0.95 & \\
          WFCScand J0000$+60$20 & 2357$+$600 & 00 00 32.40  & $+$60 20 55.0 & 0.3741 & 0.1570 & $-$0.70 & \\
          WFCScand J0000$+61$26 & 2358$+$611 & 00 00 58.30  & $+$61 26 10.0 & 0.4360 & 0.1740 & $-$0.74 & \\
          WFCScand J0000$-14$23 & 2358$-$146 & 00 00 40.22  & $-$14 23 47.2 & 0.3088 & $-$9.9 & $-$9.9  & \\
      \hline
   \end{tabular}
   \label{t:ndt}
\end{table*}

\bibliographystyle{aasjournal}
\bibliography{wfcs}

\end{document}